\documentclass[prb,aps,twocolumn,superscriptaddress,showpacs]{revtex4-1}
\usepackage[colorlinks=true,linkcolor=blue, citecolor=blue, urlcolor=blue, 
    unicode=true,breaklinks]{hyperref}

\usepackage{graphicx}
\usepackage{amsmath,amssymb} 
\usepackage{physics}
\usepackage{chemformula}
\usepackage{siunitx}
\usepackage{tabularx}
\usepackage{multirow}
\usepackage{paralist} 
\usepackage[capitalise]{cleveref} 
\crefrangelabelformat{equation}
{(#3#1#4--#5\crefstripprefix{#1}{#2}#6)}
\crefrangelabelformat{subequation}
{(#3#1#4--#5\crefstripprefix{#1}{#2}#6)}
\usepackage[overload]{empheq}

\newcommand\ut[1]{_\mathrm{#1}} 
\newcommand\upt[1]{^\mathrm{#1}} 

\newcommand\et{\ensuremath{\epsilon\ut{T}}}
\newcommand\mnsb{\ch{MnSb2O6}}
\newcommand\tn{\ensuremath{T\ut{N}}}
\newcommand\Q{\ensuremath{\vb*{Q}}}
\newcommand\Qn{\abs{\Q}}

\newcommand\Ef{\ensuremath{E\ut{f}}}
\newcommand\sqe{S(\Q,E)}

\newcommand\ddsigma{\frac{\dd[2]{\sigma}}{\dd{\Omega}\dd{\Ef}}}

\newcommand\gme{\ensuremath{\gamma\ut{e}}}
\newcommand\gmi{\ensuremath{\gamma\ut{i}}}

\begin{document}

\title{Neutron scattering sum rules, symmetric exchanges, and helicoidal magnetism in \mnsb}

\author{E. Chan}
\affiliation{Institut Laue-Langevin, 71 avenue des Martyrs, CS 20156, 38042 Grenoble Cedex 9, France}
\affiliation{School of Physics and Astronomy, University of Edinburgh, Edinburgh EH9 3JZ, United Kingdom}

\author{H. Lane}
\affiliation{School of Physics and Astronomy, University of Edinburgh, Edinburgh EH9 3JZ, United Kingdom}
\affiliation{School of Chemistry and Centre for Science at Extreme Conditions, University of Edinburgh, Edinburgh, EH9 3FJ, United Kingdom}
\affiliation{ISIS Pulsed Neutron and Muon Source, STFC Rutherford Appleton Laboratory, Harwell Campus, Didcot, Oxon, OX11 0QX, United Kingdom}
\affiliation{School of Physics, Georgia Institute of Technology, Atlanta, Georgia 30332, USA}

\author{J. P\'{a}sztorov\'{a}}
\affiliation{School of Physics and Astronomy, University of Edinburgh, Edinburgh EH9 3JZ, United Kingdom}

\author{M. Songvilay}
\affiliation{School of Physics and Astronomy, University of Edinburgh, Edinburgh EH9 3JZ, United Kingdom}

\author{R. D. Johnson}
\affiliation{Department of Physics and Astronomy, University College London, Gower Street, London WC1E 6BT}

\author{R. Downie}
\affiliation{Institute of Chemical Sciences and Centre for Advanced Energy Storage and Recovery, School of Engineering and Physical Sciences, Heriot-Watt University, Edinburgh EH14 4AS, United Kingdom}

\author{J-W. G. Bos}
\affiliation{Institute of Chemical Sciences and Centre for Advanced Energy Storage and Recovery, School of Engineering and Physical Sciences, Heriot-Watt University, Edinburgh EH14 4AS, United Kingdom}

\author{J. A. Rodriguez-Rivera}
\affiliation{NIST Center for Neutron Research, National Institute of Standards  and Technology, Gaithersburg, Maryland 20899-6100, USA}
\affiliation{Department of Materials Science, University of Maryland, College Park, MD  20742}

\author{S.-W. Cheong}
\affiliation{Rutgers Center for Emergent Materials and Department of Physics and Astronomy, Rutgers University, 136 Frelinghuysen Road, Piscataway, New Jersey 08854, USA}

\author{R. A. Ewings}
\affiliation{ISIS Pulsed Neutron and Muon Source, STFC Rutherford Appleton Laboratory, Harwell Campus, Didcot, Oxon, OX11 0QX, United Kingdom}

\author{N. Qureshi}
\affiliation{Institut Laue-Langevin, 71 avenue des Martyrs, CS 20156, 38042 Grenoble Cedex 9, France}

\author{C. Stock}
\affiliation{School of Physics and Astronomy, University of Edinburgh, Edinburgh EH9 3JZ, United Kingdom}

\date{\today}

\begin{abstract}

MnSb$_{2}$O$_{6}$ is based on the noncentrosymmetric $P321$ space group with magnetic Mn$^{2+}$ ($S={5/2}$, $L\approx 0$) spins ordering below $T\ut{N}=\SI{12}{K}$ in a cycloidal structure. The spin rotation plane was found to be tilted away from the $c$-axis [M. Kinoshita \textit{et al.} Phys. Rev. Lett. {\bf{117}}, 047201 (2016)] resulting as a helicoidal ground state which we refer as the tilted structure. In our previous diffraction work [E. Chan \textit{et al.} Phys. Rev. B {\bf{106}}, 064403 (2022)] we
found no evidence that this tilted structure is favored over the pure cycloidal order (referred as the untilted structure). The ground state magnetic structure, expected to be built and originate from 7 nearest neighbor Heisenberg exchange constants, has been shown to be coupled to the underlying crystallographic chirality with polar domain switching being reported.  We apply neutron spectroscopy to extract these symmetric exchange constants.  Given the high complexity of the magnetic exchange network, crystallographic structure and complications fitting many parameter linear spin-wave models, we take advantage of multiplexed neutron instrumentation to use the first moment sum rule of neutron scattering to estimate these symmetric exchange constants. The first moment of neutron scattering provides a way of deriving the Heisenberg exchange constant between two neighboring spins if the relative angle and distance of the two ordered spins is known.  We show that the first moment sum rule combined with the known magnetic ordering wavevector fixes 6 of the 7 exchange constants.  The remaining exchange constant is not determined by this analysis because of the equal spatial bond distances present for different chiral exchange interactions.  However, we find this parameter is fixed by the magnon dispersion near the magnetic zone boundary which is not sensitive to the tilting of the global magnetic structure.  We then use these parameters to calculate the low-energy spin-waves in the N\'eel state to reproduce the neutron response without strong antisymmetric coupling.  Using Green's response functions, the stability of long-wavelength excitations in the context of our proposed untilted magnetic structures is then discussed.  The results show the presence of strong symmetric exchange constants for the chiral exchange pathways and illustrate an underlying coupling between crystallographic and magnetic ``chirality" through predominantely symmetric exchange.	We further argue that the excitations can be consistently modelled in terms of an untilted magnetic structure in the presence of symmetric-only exchange constants.

\end{abstract}


\maketitle

\section{Introduction}

Magnetic materials that lack an inversion center potentially host coupled magnetic and ferroelectric order parameters while also providing a framework for unusual magnetic excitations like directionally anisotropic (or nonreciprocal) spin-waves.\cite{cheon201898,stock2019100} Such materials often consist of magnetic ions in a low-symmetry environment with a complex set of magnetic interactions causing the coupling between structural (e.g. ferroelectricity) and magnetic orders.~\cite{cheong20076,Eerenstein06:442,fiebig16:1,spaldin2005309,spaldin201063}  Determining these magnetic interactions that provide the basis for coupled structural and magnetic properties is often complicated and based on many parameter fits from complex magnetic ground states.~\cite{mostovoy200696,johnson201444}  In this paper we investigate the magnetic excitations in powder and in an array of single crystals of the helicoidal magnet MnSb$_{2}$O$_{6}$ with the goal of extracting the symmetric exchange constants. Given the complexity of the excitation spectrum, the number of predicted exchange constants, and ambiguities of the magnetic structure (tilted versus untilted ground state), we apply a first moment sum rule~\cite{hohenberg197410} analysis to extract the symmetric exchange constants and compare the results to the excitation spectrum from mean field linear spin-wave theory.  This approach only depends on the relative orientation of neighboring magnetic moments and does not depend on whether the overall magnetic structure is tilted or untilted as discussed below. We also demonstrate a generalized methodology for obtaining symmetric Heisenberg exchange constants from multiplexed neutron scattering where extensive regions of momentum and energy transfers are sampled.

Iron-langasite (Ba$_{3}$NbFe$_{3}$Si$_{2}$O$_{14}$)~\cite{marty2008101,marty10:81,loire2011106,stock201183} and MnSb$_{2}$O$_{6}$~\cite{reimers1989,johnson2013111,kinoshita2016117,werner201694,chan2022106a} are two examples of magnetic compounds that are based on the noncentrosymmetric $P$321 (\#150) space group.  The magnetic order in these compounds is different with iron-langasite being described by a simple helix that can be quantified by a time-even pseudoscalar.\cite{johnson2013111} The magnetic structure in MnSb$_{2}$O$_{6}$, in contrast was first found cycloidal and quantified by a time-even polar vector.\cite{johnson2013111,chan2022106a}  Given the fact that magnetic Mn$^{2+}$ ($S=5/2$, $L\approx 0$) is not expected to have an orbital degeneracy that would enhance any anisotropy in the magnetic Hamiltonian,~\cite{Yosida:book} like antisymmetric terms, it is expected that such terms are small compared to symmetric exchange terms in the magnetic Hamiltonian. Furthermore, diagonal symmetric exchange interactions are coupled to the chirality of the underlying lattice.   

\begin{figure}[h]
    \begin{center}
    \includegraphics[scale=0.3]{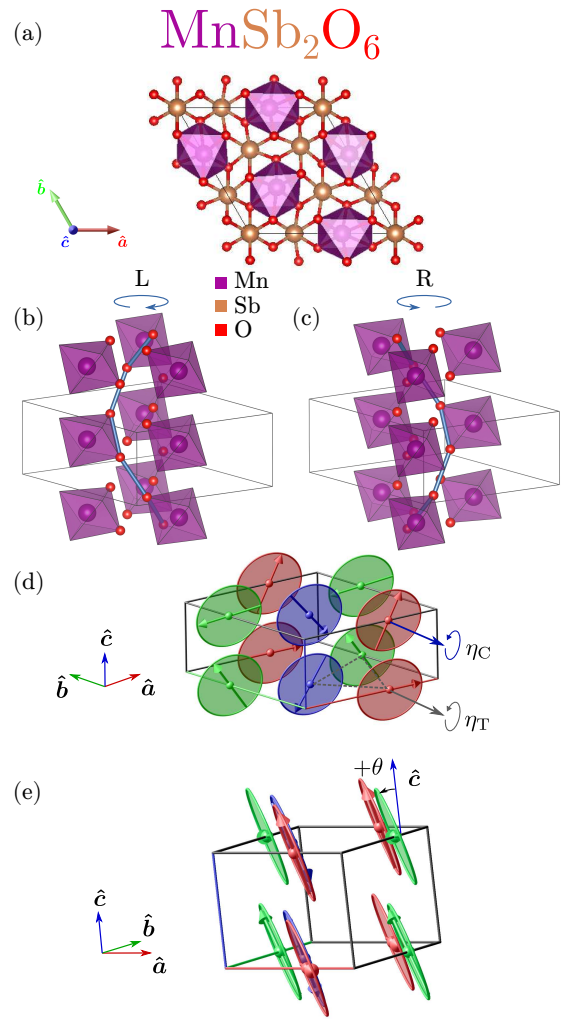}
    \end{center}
    \caption{(a) Nuclear structure of lattice-chiral \mnsb. The structural chirality can be defined as the helical winding of the Mn-O-O-Mn super-super-exchange pathway with respect to the $c$-axis: it is clockwise for left-handed structure (b), and anti-clockwise for right-handed structure (c). Figures made on \textsc{Vesta}.\cite{momma201144} (d) Cycloidal magnetic structure with magnetic parameters $\eta\ut{c}$ and $\eta\ut{T}$ describing sense of rotations of the spins. (e) Tilted magnetic structure where the spin rotation plane is tilted from the $c$-axis by an angle $\theta$. Our previous diffraction work found no evidence of this tilted magnetic structure over the cycloidal one, we discuss this below in the context of a model with symmetric-only exchange constants.  Figures made on \textsc{Mag2Pol}.\cite{qureshi201952}}
    \label{fig:MSO_struc}
\end{figure}

The nuclear structure of \mnsb , based on interlaying \ch{MnO6} and \ch{SbO6} octahedra is shown in \cref{fig:MSO_struc}(a). The only magnetic ions Mn$^{2+}$ arrange in a triangular motif. Magnetic interactions occur in these isolated \ch{MnO6} octahedra through super-super-exchange (SSE) pathways (Mn-O-O-Mn). In particular, chiral SSE pathways along the $c$-axis, shown in \cref{fig:MSO_struc}(b)-(c), define the structural chirality of the compound.

Below $\tn\approx\SI{12}{K}$, the magnetic ground state was found to follow a cycloidal order with a propagation vector $\vb*{k} = (0,0,0.182)$.\cite{johnson2013111} Within each triangle of Mn in the $(ab)$-plane, shown in dashed gray lines in \cref{fig:MSO_struc}(d), the moments are dephased by 120°. The sense of rotation of the spins along the $c$-axis and within a basal triangle can be described by magnetic parameters $\eta\ut{C}$ and $\eta\ut{T}$, often called magnetic ``chiralities'', which directly couple to the crystal chirality $\sigma$ through an energy invariant.\cite{chan2022106a} Later on, the cycloids were reported to be tilted away from the $c$-axis, with one of the main axes of the spin envelop parallel to $[1\bar{1}0]$, as shown in \cref{fig:MSO_struc}(e). This ground state was presented to be necessary to explain the electric polarization measured by pyroelectric current in the $(ab)$-plane in Ref.~\onlinecite{kinoshita2016117}. The magnetic structure was further investigated by complementary neutron diffraction techniques in Ref.~\onlinecite{chan2022106a}, showing no evidence of this tilted magnetic ground state. Furthermore, a mechanism based on the coupled structural and magnetic chiralities is proposed for the ferroelectric switching, which does not require a tilted cycloid ground state.

\begin{figure}[h]
    \begin{center}
    \includegraphics[]{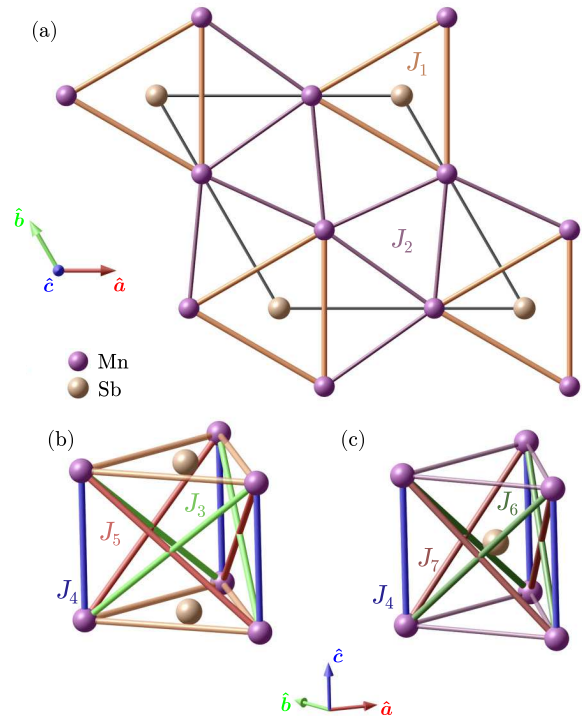}
    \end{center}
    \caption{Drawing of the seven nearest neighbors interactions in \mnsb. (a) Intraplane interactions $J_1$ connecting triangles of Mn centered at the lattice origin, and $J_2$ connecting between these triangles. (b) Interplane interactions based on the $J_1$ triangle, $J_4$ is the straight interplane interaction, while $J_3$ and $J_5$ are diagonal chiral interactions. (c) Interplane interactions based on the $J_2$ triangle, with $J_6$ and $J_7$ as chiral exchange interactions. Oxygen atoms are omitted here for clarity. Figure made on \textsc{Mag2Pol}.\cite{qureshi201952}} 
    \label{fig:MSO_exchanges}
\end{figure}

The magnetic interactions are described by a dominant Heisenberg Hamiltonian $\mathcal{\hat{H}}=\sum_{ij}J_{ij}\vb*{\hat{S}}_i\cdot\vb*{\hat{S}}_j$ with the symmetric exchange constants corresponding to the seven SSE pathways in \mnsb.\cite{johnson2013111} The nearest neighbor exchange paths are shown in \cref{fig:MSO_exchanges}, where the oxygen atoms are omitted for clarity. Each manganese and antimony atom is surrounded by six oxygen atoms forming edge-sharing octahedra. In a minimalist model considering only interactions between neighboring Mn$^{2+}$ ions, there are therefore 7 exchange constants which need to be considered.  Intraplane interactions are shown in \cref{fig:MSO_exchanges}(a) where $J_1$ connects a triangle of \ch{MnO6} octahedra through a \ch{SbO6} octahedra centered at the origin, and $J_2$ connects \ch{MnO6} octahedra between these triangles, through an interplane \ch{SbO6} octahedron shown in \cref{fig:MSO_exchanges}(c). Interplane interactions within a Mn triangle connected by $J_1$ are shown in \cref{fig:MSO_exchanges}(b), where $J_4$ is the straight interplane exchange interaction, and $J_3$ and $J_5$ are diagonal exchange interactions. Similarly, Figure~\ref{fig:MSO_exchanges}(c) shows $J_6$ and $J_7$, the diagonal exchange interactions connecting a Mn triangle linked by $J_2$. Interestingly, $J_3$ and $J_6$ are related to the right-handed helical winding of the Mn-O-O-Mn SSE pathways (shown in \cref{fig:MSO_struc}(c) for $J_3$), while $J_5$ and $J_7$ are related to left-handed SSE pathways (shown in \cref{fig:MSO_struc}(b) for $J_5$). Thus, these chiral exchange paths are interchanged by inversion symmetry between structurally left- and right-handed crystals.\cite{johnson2013111} We note that only the five first exchange constants were necessary to describe the SSE interactions in iron-langasite, due to structural differences with \mnsb. Indeed, in \ch{Ba3NbFe3Si2O14}, the bond distance $d_2=\SI{5.652}{\angstrom}$ associated with intertriangle interaction $J_2$ is significantly larger than the bond distance $d_1=\SI{3.692}{\angstrom}$ tied to intratriangle interaction $J_1$.\cite{stock201183} On the contrary, in \mnsb , $d_2=\SI{4.845}{\angstrom}$ is smaller than $d_1=\SI{5.596}{\angstrom}$, as a result the related interplane interactions $J_6$ and $J_7$ are expected to be more significant as they link magnetic Mn$^{2+}$ ions through SSE pathways.

In this paper, we present our inelastic neutron scattering data from both powder and single crystals of \mnsb. We apply the first moment (Hohenberg-Brinkman) sum rule of neutron scattering to extract the exchange constants from the Heisenberg model, therefore characterizing the magnetic Hamiltonian.  Then, we apply Green's functions on a rotating frame to generate spin-wave spectra based on our derived exchange constants. Using the values of the symmetric exchange constants from sum rules of neutron scattering, we refine the parameters to obtain a good description of the neutron inelastic spectra. Based on the Green's functions neutron response, the stability of spin-wave excitations is further tested for the proposed magnetic structures.

\section{Experimental details}

\subsection{Materials preparation}

Materials preparation followed the procedure outlined in Ref.~\onlinecite{nakua1995154}. Powders of MnSb$_{2}$O$_{6}$ were prepared by mixing stoichiometric amounts of pure MnCO$_{3}$ and Sb$_{2}$O$_{3}$. After mixing through grinding, the powder was pressed into a pellet and heated up to 1000$^{\circ}$C with the process repeated with intermediate grinding. It was found that heating the pellet to higher temperatures introduced the impurity Mn$_{2}$Sb$_{2}$O$_{7}$. Single crystals of MnSb$_{2}$O$_{6}$ were prepared using the flux method. Starting ratios for single-crystal growth were (by weight) 73\% of flux V$_{2}$O$_{5}$, 20\% of polycrystalline MnSb$_{2}$O$_{6}$ and 7\% of B$_{2}$O$_{3}$. The powder was ground and pressed into a pellet and flame sealed in a quartz ampoule under vacuum (less than 1e$^{-4}$ Torr). B$_{2}$O$_{3}$ was used to lower the melting temperature of the V$_{2}$O$_{5}$ flux. Back filling the ampoules with $\approx$ 200 mTorr of Argon gas was found to noticeably improve crystal sizes.  Quartz ampoules were then heated to 1000$^{\circ}$C at a rate of 60$^{\circ}$C/hour and soaked at this temperature for 24 hours. The furnace was then cooled to 700$^{\circ}$C at a rate of 2$^{\circ}$C/hour and held for 24 hours, before it was switched off and allowed to cool to room temperature. Crystal sizes in the range from a few millimeters to nearly a centimeter were obtained through this procedure.

\subsection{Neutron spectroscopy}

\begin{figure}[h]
    \begin{center}
    \includegraphics[scale=0.3]{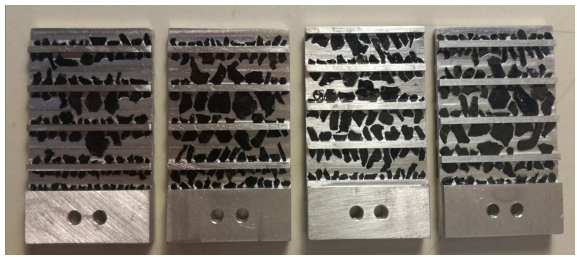}
    \end{center}
    \caption{\SI{1.3}{g} of single crystals of \mnsb\ aligned on four Al plates, and coated with Fomblin oil for inelastic neutron scattering.}
    \label{fig:MSO_plate}
\end{figure}

To investigate the magnetic dynamics, neutron spectroscopy was performed on the MACS (NIST, Gaithersburg) triple-axis spectrometer~\cite{Rodriguez200819} on both single crystals and powder samples. $\SI{1.3}{g}$ of single crystals were aligned in the $(HHL)$ scattering plane on both sides of four aluminium plates and coated with viscous hydrogen free Fomblin oil, as shown in \cref{fig:MSO_plate}. A select fraction of the crystals were aligned with Laue diffraction and the remainder were aligned using polarized optical microscopy based on the crystal morphology. These single crystals were synthesized the same way as the samples measured in our previous studies in Ref.~\onlinecite{chan2022106a}, where we have performed Schwinger scattering and transmission polarized optical microscopy and found only a small imbalance of chiral structural domains in the single crystals. This small imbalance distinguishes MnSb$_{2}$O$_{6}$ from the enantiopure single crystals of iron based langasite previously studied.\cite{marty2008101,loire2011106,qureshi201952,qureshi201952} During the coalignment of the single crystals used here for spectroscopy, great care was taken to align the relative $a$ and $b$ inplane axes, the choice of what constituted $\pm$ [001] was done at random. For the purposes here we consider the average crystal structure to be an equal mixture of the differing chiral domains. We will show in \cref{ssec:1stmoment} that our analysis holds no matter the proportion of chiral structural domains. To probe the dynamics in our array of single crystals, the final energy was fixed to either $E\ut{f}$=2.4 meV or 3.7 meV with BeO and Be filters, respectively, being used on the scattered side to filter out higher order neutrons from the monochromator. For all results presented here the pyrolytic graphite PG(002) monochromator was focused both horizontally and vertically.  The lattice parameters were measured to be $a=b=\SI{8.733}{\angstrom}$ and $c=\SI{4.697}{\angstrom}$.  For powder measurements, a 16.3 g sample was used with $E\ut{f}$=3.7 meV and a BeO filter on the scattered side.

\section{Results and discussion}

In this section, we will first present the neutron scattering data for both powders and single crystals of \mnsb, before detailing our absolute normalization process. Then, zeroth and first moment sum rules are applied to our inelastic data allowing the extraction of the symmetric exchange constants. We will finally use Green's functions on a rotating frame to compare the resulting spin-wave spectra to the experimental ones and to test the stability of proposed magnetic structures.

\begin{figure}[h]
    \begin{center}
    \includegraphics[scale=1]{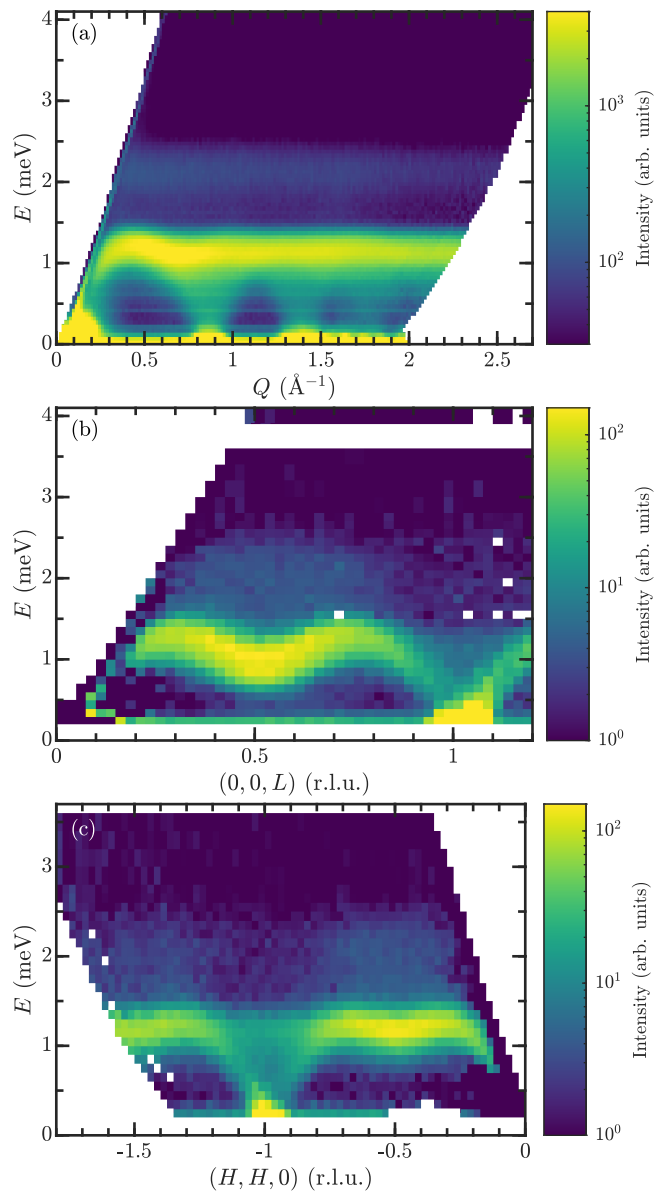}
    \end{center}
    \caption{(a) Powder averaged inelastic neutron scattering spectrum taken on MACS at $T=\SI{1.4}{K}$. (b)-(c) Single crystal inelastic neutron scattering spectrum from the $\Ef=\SI{3.7}{meV}$ dataset at $T=\SI{1.4}{K}$. The logarithmic intensity scales are chosen to show the two components to the scattering and in particular the higher energy weak scattering displayed at $\sim$ 2 meV.}
    \label{fig:two_magnon}
\end{figure}

\subsection{Excitation spectra}

\subsubsection{Total excitation spectra}

The excitation spectra of both powders and single crystals of \mnsb\ at $T=\SI{1.4}{K}$ are shown in Fig.~\ref{fig:two_magnon}, with the $E\ut{f}=\SI{3.7}{meV}$ MACS setup. The powder data in Fig.~\ref{fig:two_magnon}(a) display intense low energy magnetic scattering extending from the elastic line to $\sim$ 1 meV, and a weaker band of excitations at approximately twice this value at $\sim$ 2 meV. The single crystal data displayed in Fig.~\ref{fig:two_magnon}(b)-(c) illustrate two different types of scattering: one with intense dispersive fluctuations that are well defined both in momentum and energy at low energies, and the other with a weaker momentum and energy broadened continuum of scattering extending to larger energy transfers.  This continuum of scattering is most apparent at the zone boundaries in the single crystal data.  Given the kinematics of these two types of scattering, we associate the lower energy dispersive fluctuations with one-magnon scattering and the higher energy continuum with two-magnon scattering.  While two-magnon scattering is expected to be most prominent in $S={1/2}$ magnets,~\cite{Cloizeaux1962128,Muller198124,Endoh197432,coldea200368,Tennant199370,Nagler199144,Lake2013111,Stone200391,Mourigal20139} it is a direct result of the uncertainty associated with non-commuting observables and has been studied extensively in other large-$S$ magnets.~\cite{heilmann198124,huberman200572,songvilay2018}  We discuss this cross section later in the paper in the context of the zeroth moment sum rule and show indeed that these two components of scattering originate from single and multi magnon processes.

\subsubsection{Powder low-energy spectrum}\label{sec:macspowder}

\begin{figure}[h]
    \begin{center}
    \includegraphics[scale=1]{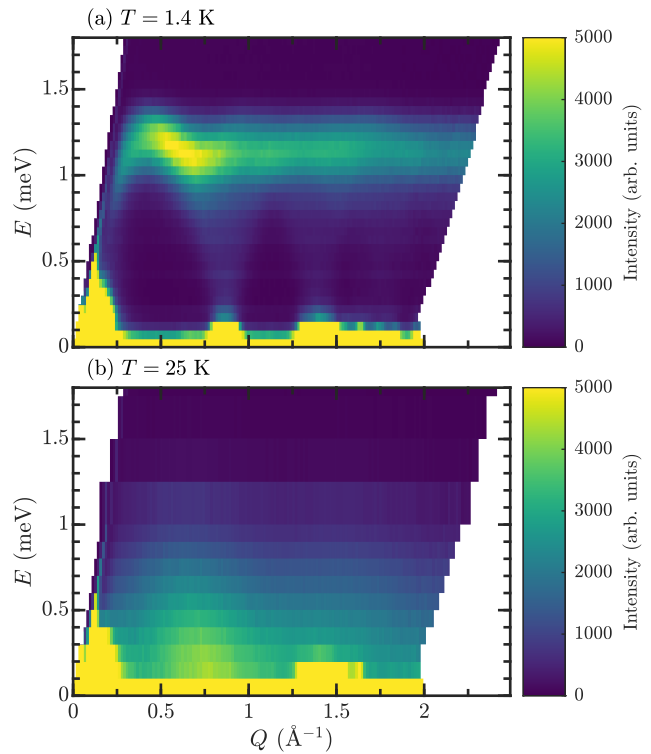}
    \end{center}
    \caption{Powder inelastic neutron scattering spectrum of the one-magnon cross section at (a) $T=\SI{1.4}{K}$ (below \tn) and (b) $T=\SI{25}{K}$ (above \tn).}
    \label{fig:pow_spec}
\end{figure}

Results of the low-energy powder inelastic neutron scattering experiment performed on MACS, with fixed final energy $\Ef=\SI{3.7}{meV}$ are shown in \cref{fig:pow_spec}. The powder averaged spin-wave dispersion at $T=\SI{1.4}{K}$, below the N\'eel magnetic ordering transition, is presented in \cref{fig:pow_spec}(a), showing low-energy spin dynamics below $E\approx\SI{1.4}{meV}$. These dynamics are highly dispersive from the magnetic ordering wavevector and are gapless within experimental resolution ($\Delta E\approx\SI{0.15}{meV}$). In contrast, above $\tn\approx\SI{12}{K}$, the magnetic scattering is considerably broadened both in momentum and energy indicative of spatially and temporally short-range correlations.  This paramagnetic scattering is very strong due to high spin $S=5/2$ of Mn$^{2+}$ magnetic ions, as shown in \cref{fig:pow_spec}(b) with the spectrum measured at $T=\SI{25}{K}$.  Both experimental datasets below and above the magnetic ordering temperature also display a decay in intensity with increasing momentum transfer, characteristic of magnetic scattering.  The powder averaged spectra establish the presence of dispersive magnetic dynamics and the energy scale of the spin excitations.

\subsubsection{Single crystal low-energy spectrum}

\begin{figure}[h]
    \begin{center}
    \includegraphics[scale=1]{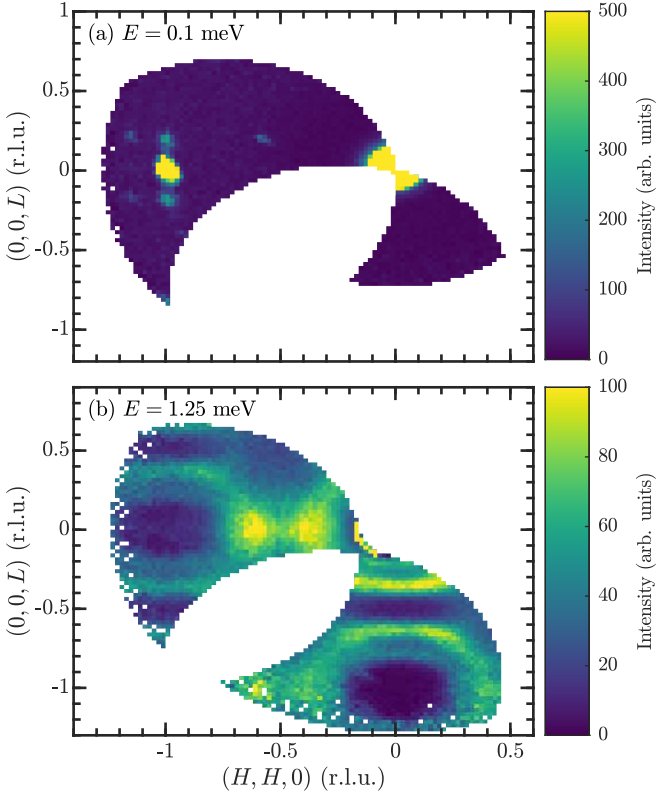}
    \end{center}
    \caption{MACS single crystal inelastic neutron scattering spectra at $T=\SI{1.4}{K}$: constant energy slices for (a) $E=\SI{0.1}{meV}$ and (b) $E=\SI{1.25}{meV}$.  The weak scattering in (a) at $(H,H)$ $\sim$ -0.5 and displaced at $(H,H)$ $\sim$ -1.1 originate from some crystals miss-aligned by $\sim$ 60$^{\circ}$ in the multi crystal mount.}
    \label{fig:MACS_Eslice}
\end{figure}

\begin{figure}[h]
    \begin{center}
    \includegraphics[scale=1]{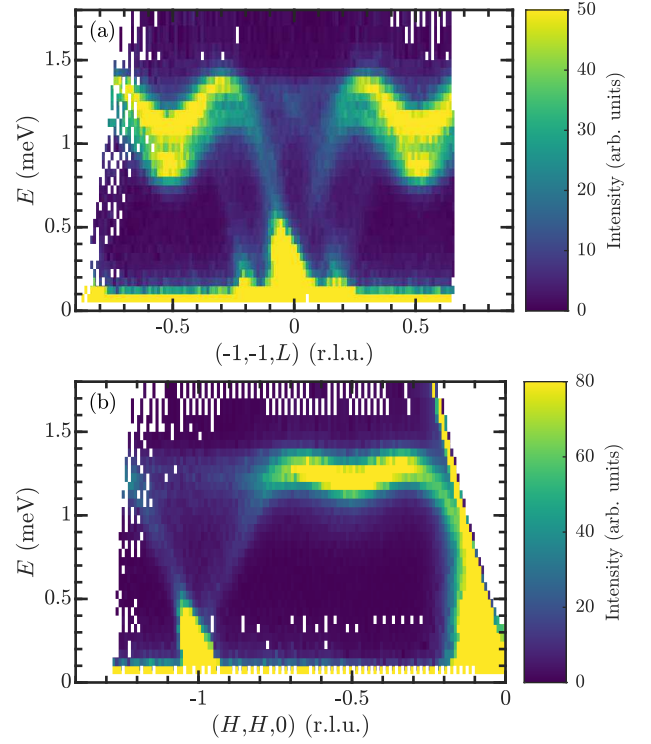}
    \end{center}
    \caption{MACS single crystal inelastic neutron scattering spectra at $T=\SI{1.4}{K}$: spin-wave dispersion along (a) $(-1,-1,L)$ and (b) $(H,H,0)$.}
    \label{fig:MACS_spec}
\end{figure}

Results of single crystal inelastic neutron scattering performed on MACS with a fixed final energy $E\ut{f}=\SI{2.4}{meV}$ are displayed in \cref{fig:MACS_Eslice} and \cref{fig:MACS_spec} at $T=\SI{1.4}{K}$ below $T\ut{N}$.  The data are illustrative of dispersive dynamics originating from the magnetic ordering wavevector.  Constant energy slices at $E=\SI{0.1}{meV}$ and $E=\SI{1.25}{meV}$ are shown in \cref{fig:MACS_Eslice}(a) and (b). Spin-wave dispersion along $(-1,-1,L)$ and $(H,H,0)$ are respectively shown in \cref{fig:MACS_spec}(a) and \cref{fig:MACS_spec}(b). Spin-wave branches emerging from nuclear Bragg peak (-1,-1,0) and also its magnetic satellites (-1,-1,0)$\pm\vb*{k}$ are visible in \cref{fig:MACS_Eslice}(a) and \cref{fig:MACS_spec}(a). Within the instrumental resolution ($\Delta E\approx\SI{0.1}{meV}$), all modes seem gapless, which is consistent with the low anisotropy measured from electron spin resonance,\cite{werner201694} and observed from the tunability of the magnetic structure by small magnetic fields.\cite{kinoshita2016117,chan2022106a} 

As already presented in \cref{fig:two_magnon}(b, c), inelastic neutron scattering data were also obtained on MACS with the same array of single crystals, but with a fixed final energy $E\ut{f}=\SI{3.7}{meV}$. In the following of the paper, the dataset used for each analysis will be mentioned.

\subsection{Absolute normalization of magnetic cross section}\label{sec:normalization}

In order to straightly compare the magnetic scattering intensities from the different datasets, they have to be converted into absolute units. This is particularly important given our goal of applying sum rules of neutron scattering to obtain the magnetic exchange constants in absolute units of energy.  Through this we will apply the zeroth moment sum rule to demonstrate that all of the magnetic spectral weight is measured in the experiments discussed above.  We then apply the first moment sum rule to obtain the symmetric exchange constants. In this section, we describe our normalization process, adapted from Ref.~\onlinecite{hammar199857} and Ref.~\onlinecite{xu201384} and introduce our definition for the dynamical structure factor $\sqe$.

The intensity measured during the experiment $I(\Q,E)$ (in counts) is related to the differential cross section via a convolution with an instrumental-dependent resolution function $R$:

\begin{equation}
I(\Q,E) = \int\dd{\Q_0}\dd{E_0}\ddsigma(\Q_0,E_0)R(\Q_0,E_0,\Q,E)
\end{equation}

\noindent By assuming a slow variation of this resolution function in the region of study (over the narrow energy range probed in this study), it can be approximated by a constant $R_0$, which allows us to decouple the intensity into:

\begin{equation}\label{eq:IQE}
I(\Q,E) \approx R_0\ddsigma(\Q_0,E_0)
\end{equation}

\noindent During the data reduction, the intensity is normalized to the monitor counts based on a low efficiency detector placed in the incident beam after the monochromator and before the sample. The efficiency of which is inversely dependent to the speed of the incident neutrons, which is proportional to $k\ut{i}$, giving the normalized intensity (in counts/mon):

\begin{equation}\label{eq:ItQE}
\bar{I}(\Q,E) = k\ut{i} I(\Q,E) = k\ut{i}R_0\ddsigma(\Q,E)
\end{equation}

\noindent Having related the measured scattering intensity to the cross section, we now focus on the magnetic differential cross section for unpolarized neutrons and identical magnetic ions. Assuming isotropic spin excitations, we can define the dynamic structure factor $\sqe=S^{xx}=S^{yy}=S^{zz}$, where $S^{\alpha\beta}$ is the dynamic spin correlation function related to the Fourier transform of the spin-spin correlation function. Neglecting the Debye-Waller factor gives the following double differential cross section:

\begin{equation}\label{eq:magsigmasimple}
\ddsigma(\Q,E)=N\frac{k\ut{f}}{k\ut{i}}\left(\frac{\gamma r_0}{2}\right)^2(g\abs{f(\Q)})^22S(\Q,E) 
\end{equation}

\noindent where $N$ is the number of unit cells, ${\gamma r_0}/{2}\approx \SI{0.2695E-12}{cm}$ is the typical magnetic scattering length, $g$ is the Land\'e factor and $f(\Q)$ the magnetic form factor. Combining Eq.~\ref{eq:ItQE} and \ref{eq:magsigmasimple} we get the dynamical structure factor (in $\SI{}{meV^{-1}}$) from the measured intensity by:

\begin{equation}\label{eq:SQE}
S(\Q,E)=\frac{\bar{I}(\Q,E)}{\abs{gf(\Q)}^2(\frac{\gamma r_0}{2})^22Nk\ut{f}R_0}
\end{equation}

\noindent we can write directly the numerical values of the magnetic cross section $(\gamma r_0/2)^2$ into the equation:

\begin{equation}
S(\Q,E)=\frac{13.8(\SI{}{b^{-1}})\bar{I}(\Q,E)}{\abs{gf(\Q)}^22Nk\ut{f}R_0}
\end{equation}

\noindent The key for normalizing the magnetic intensity is thus to evaluate this instrumental-dependent factor $Nk\ut{f}R_0$ expressed in (meV)(counts/mon)(b$^{-1}$).  

There are several ways reported in the literature for obtaining this instrument calibration factor. One possibility is to evaluate the incoherent scattering from the elastic line of a known standard compound (for example as done in Ref. \onlinecite{Nakatsuji2012339}). By energy integrating the measured intensity close to elastic energy transfer, far from any magnetic or nuclear Bragg peak, we obtain, as $k\ut{i}=k\ut{f}$ for elastic scattering:

\begin{equation}
{\int_{-\epsilon}^{+\epsilon}{\dd{E} \bar{I}(\Q,E)}}=Nk\ut{f}R_0{\sum_{i} (b\upt{inc}_i)^2}
\end{equation}

\noindent where $b\upt{inc}_i$ is the incoherent scattering length of atom $i$, and the sum is over the unit cell. Vanadium having a large incoherent scattering cross section compared to its coherent one, it is usually used as a standard sample to normalize inelastic neutron scattering data. We have measured the Vanadium sample in the same geometry and instrumental configuration as our MnSb$_{2}$O$_{6}$ powder sample. With $N\ut{V}$ the number of Vanadium atoms and its incoherent scattering length $b\upt{inc}\ut{V}=\SI{6.35}{fm}$,\cite{sears19923} we can write: 

\begin{equation}
N\ut{V}k\ut{f}R_0 = \frac{\int_{-\epsilon}^{+\epsilon}{\dd{E} \bar{I}\ut{V}(\Q,E)}}{(b\upt{inc}\ut{V})^2}
\end{equation}

\noindent By writing $N\ut{V}=m\ut{V}/(A\ut{r}(V)m\ut{u})$ with $m\ut{V}$ the mass of the Vanadium sample, $A\ut{r}(V)$ the relative atomic mass of Vanadium, and $m\ut{u}$ the atomic mass constant, we can write the ratio $N/N\ut{v}=\frac{m/A\ut{r}(\mnsb)\ut{cell}}{m\ut{V}/A\ut{r}(V)}$ with $m$ the mass of the \mnsb\ sample, and $A\ut{r}(\mnsb)\ut{cell}$ the relative mass of a unit cell (three formula units of \mnsb\ per unit cell), the normalization factor becomes: 

\begin{equation}
Nk\ut{f}R_0 = \frac{m/A\ut{r}(\mnsb)\ut{cell}}{m\ut{V}/A\ut{r}(V)}\frac{\int_{-\epsilon}^{+\epsilon}{\dd{E} \bar{I}\ut{V}(\Q,E)}}{\SI{0.403}{b}}
\end{equation}

\noindent This equation allows us to obtain the instrumental calibration factor from the incoherent cross section centered at the elastic ($E=0$) position.  We note that an alternate way to obtain this calibration constant is to measure the elastic incoherent cross section from the sample given Manganese has a comparatively large incoherent cross section. We did not take this approach in this experiment as we found the elastic line where incoherent scattering is present in our single crystal geometry was contaminated by scattering from hydrogen free (yet fluorine based) Fomblin oil.  Fomblin, while having a comparatively small incoherent cross section in comparison to hydrogen, has a non-negligible coherent liquid-like cross section.  This cross section is difficult to disentangle from the purely Mn$^{2+}$ incoherent cross section and therefore we relied on a separate Vanadium standard of known mass. 

\subsection{Total moment sum rule}\label{sec:totalmoment}

Having established the procedure for calibration of the instrument, we now discuss the sum rules of neutron scattering.  Magnetic neutron scattering is governed by sum rules which are satisfied by integrating the dynamical spin correlation function $S^{\alpha\beta}(\Q,E)$ over energy and momentum transfer.\cite{hohenberg197410} In particular the energy moments, $\int_{-\infty}^{+\infty}E^n S^{\alpha\beta}(\Q,E)\dd{E}$ are given theoretically,\cite{hohenberg197410,zaliznyak2005,sarte2019a} with $n=0,1$ the zeroth and first moment.

\begin{figure}[h]
    \begin{center}
    \includegraphics[scale=1]{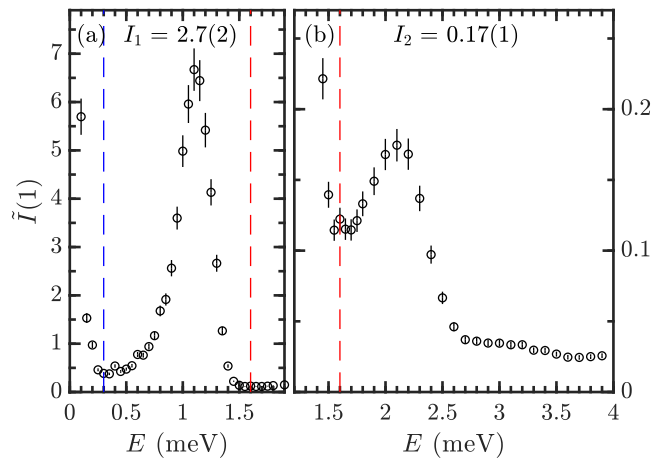}
    \end{center}
    \caption{Momentum integrated intensities as a function of the energy, for (a) $E\in {[0,1.9]}$ meV, and (b) $E\in[1.3,4]$ meV. The intensities are integrated between the dashed blue (0.4 meV) and red (1.6 meV) lines to get the one-magnon spectral weight $I_1$, and above the red lines to 4 meV to get the two-magnon spectral weight $I_2$.}
    \label{fig:zeroth_IE}
\end{figure}

The zeroth moment sum rule is often referred to as the total moment sum rule and corresponds to the integral of all the magnetic spectral weights:\cite{sarte201898,stock2009103,hammar199857,stone200265}

\begin{equation}
\frac{3\int\dd[3]{\Q}\int\dd{E}\sqe}{\int\dd[3]{\Q}}=N\ut{m}S(S+1)
\end{equation}

\noindent where $N\ut{m}=3$ is the number of magnetic ions per unit cell. This quantity can be considered as a conservation rule and allows us to confirm whether we have experimentally measured all of the spectral weight.  This rule has become particularly important in itinerant compounds near potential critical points.~\cite{Matsumoto21:90}  We will apply this zeroth moment sum rule to our powder data, which was normalized using a vanadium standard sample, following the process described above. In this case, the total moment can be written as:

\begin{equation}
I=\frac{\int\dd{Q}Q^2\int\dd{E}S(Q,E)}{\int\dd{Q}Q^2}=S(S+1)
\end{equation}

\noindent with $Q=\Qn$. In order to estimate the spectral contributions from one-magnon and two-magnon scattering, we can introduce the momentum integrated intensity:

\begin{equation}
\tilde{I}(E)=\frac{3\int\dd{Q}Q^2S(Q,E)}{\int\dd{Q}Q^2}
\end{equation}

\noindent which measures the magnetic density of states.\cite{stone200265,stock2009103} Then the integral $\int_{E\ut{min}}^{E\ut{max}}\dd{E}\tilde{I}(E)$ gives the spectral weight for the energy interval $[E\ut{min},E\ut{max}]$. Figure \ref{fig:zeroth_IE} shows the momentum integrated intensities as a function of the energy. As discussed above, the magnetic intensity consists of two components with a low-energy component which consists of harmonic excitations well defined in momentum and energy and a second considerably weaker component which is broadened in momentum and energy transfer.  These correspond to single [Fig.~\ref{fig:zeroth_IE}(a)] and two-magnon [Fig.~\ref{fig:zeroth_IE}(b)] dynamics and are separated in the powder averaged data. We can see that the one- and two-magnon contributions crossover around $\SI{1.6}{meV}$ (red dashed line), but since the intensities are quite low at this energy we consider 1.6 meV as the upper bound of the one-magnon scattering, and 0.3 meV as its lower bound (blue dashed line).

\begin{figure}[h]
    \begin{center}
    \includegraphics[scale=1]{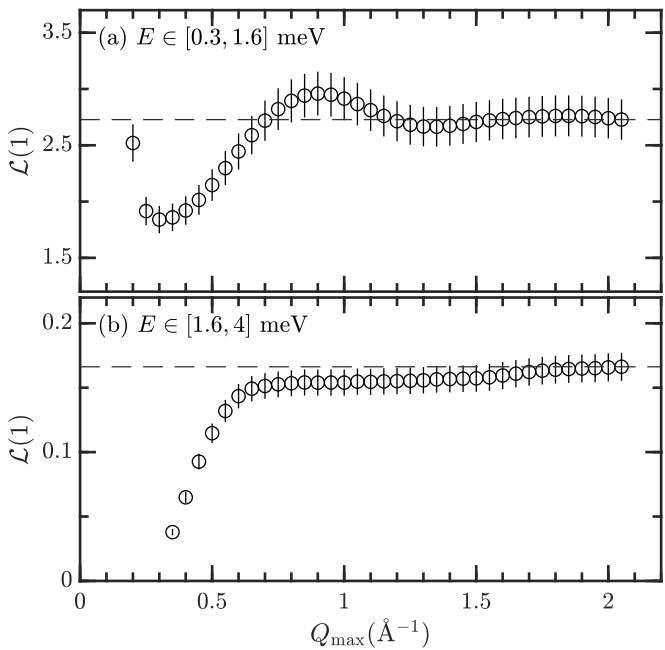}
    \end{center}
    \caption{Integrated intensities as a function of $Q\ut{max}$ the momentum integration upper bound, for (a) one-magnon and (b) two-magnon scattering. The dashed lines indicate the final values for $Q\ut{max}=\SI{2.05}{\angstrom}$.}
    \label{fig:zeroth_LQ}
\end{figure}

To extract numerical values for the integrated zeroth moments from our powder data we average the data in momentum. Accounting from the momentum powder average, the $Q$-dependence of the integrated intensity is given by:\cite{sarte201898,songvilay2020102}

\begin{equation}
\mathcal{L}(Q\ut{max})=\frac{\int_0^{Q\ut{max}}\dd{Q}Q^2\int\dd{E}S(Q,E)}{\int_0^{Q\ut{max}}\dd{Q}Q^2}
\end{equation}

\noindent and is shown in \cref{fig:zeroth_LQ} for both (a) one-magnon and (b) two-magnon contributions discussed above. The momentum average in this plot allows us to account for limited kinematic coverage of the detectors at low momentum transfers (see low momentum transfers in \cref{fig:pow_spec}).  From \cref{fig:zeroth_LQ}, we can see that $\mathcal{L}(Q\ut{max})$ approximately fully saturates close to $\SI{2}{\AA}^{-1}$ thereby illustrating that approximately all of the spectral weight has been sampled.   

Based on this momentum average of the powder data, the spectral weight $I_1=2.7(2)$ for one-magnon scattering is then calculated by integrating the intensity between 0.3 meV [dashed blue line in \cref{fig:zeroth_IE}(a)], and 1.6 meV (dashed red line in \cref{fig:zeroth_IE}). The two-magnon spectral weight is obtained by integrating between 1.6 and 4 meV, leading to $I_2=0.17(1)$.

\begin{table}[h]
    \centering
    \begin{tabular}{c c c}\\ \hline\hline
        & Theory & Experiment \\ \hline
         Total & $S(S+1)=8.75$ & 8.2(2)\\
         Elastic & \multicolumn{2}{c}{$\langle S_z\rangle^2=5.3$} \\
         One-magnon & $(S-\Delta S)(1+2\Delta S)$ = 3.2 & 2.7(2)\\
         Two-magnon & $\Delta S(\Delta S + 1)$ = 0.2 & 0.17(1) \\ \hline\hline
    \end{tabular}
    \caption{Contributions of the different components of the scattering for $S=5/2$ and $\Delta S=0.2$ deduced from neutron powder diffraction.}
    \label{tab:total}
\end{table}

The elastic (static) scattering contribution to the total moment is $\langle S_z\rangle^2$ where $z$ indicates the direction of the Mn$^{2+}$ spin in the rotated local frame. From our neutron powder diffraction (previously outlined in Ref.~\onlinecite{chan2022106a}) the ordered moment is $g\langle S_z\rangle=4.6\,\mu\ut{B}$ at $\SI{2.6}{K}$ leading to $\langle S_z\rangle^2=5.3$, and a spin reduction from the expected full saturated moment corresponding to $S={5/2}$ of $\Delta S = S-\langle S_z\rangle=0.2$. This missing component from the experimental $\langle S_z\rangle$ by conservation of spectral weight is expected to reside in the multimagnon component of the neutron dynamics corresponding to longitudinal fluctuations.  

Based on this elastic spectral weight, the theoretical total, one-magnon, and two-magnon contributions can be computed.\cite{huberman200572,songvilay2021126} They are compared with those obtained experimentally in \cref{tab:total}. The experimental total moment is 8.2(2), which is to be compared to the expected value of 8.75 for $S=5/2$. The discrepancies can be due to the relatively small $Q$-range measured during this experiment and experimental systematic issues such as the use of an external Vanadium standard or small variations in the resolution function over the energy range probed here.  Given the small energy and momentum ranges, and that we have integrated the intensity over all momentum and energy, we do not expect that changes in the resolution to be important.  However, the results are in good agreement illustrating the relative weights of one- and two-magnon cross sections and the energy range over which the magnetic dynamics are present in MnSb$_{2}$O$_{6}$.  This also confirms our assignment of the higher energy component to longitudinal two-magnon scattering and also illustrates all of the spectral weight is sampled in the dynamic range of our experiments.

\subsection{First moment sum rule}\label{ssec:1stmoment}

The previous discussion of the zeroth moment sum rule has established several points relevant for the rest of the paper.  First, we established the energy range of the magnetic dynamics in MnSb$_{2}$O$_{6}$.  Second, we have established the relative spectral weights of the single and two-magnon cross sections and found these to be in good agreement with missing spectral weight observed in diffraction experiments.  Third, we have established and verified a calibration procedure for the powder data.  

\subsubsection{Theory}

In this section, we discuss the first moment sum rule and how it can be applied to extract symmetric exchange constants.  The first moment is defined for general dynamic spin correlation function $S^{\alpha \beta} ({\vb*{Q}},E)$ as:

\begin{align}
\langle E\rangle(\Q)&\equiv \int_{-\infty}^{\infty} \dd{E}E \ S^{\alpha \beta} ({\Q},E) \\ \nonumber
&=\int_{-\infty}^{\infty}\dd{E} \langle [\hat{S}^{\alpha}({\Q},E),\mathcal{\hat{H}}]\hat{S}^{\beta}(-\Q,0) \rangle \\ \nonumber
&=\langle [\hat{S}^{\alpha}(\Q),\mathcal{\hat{H}}]\hat{S}^{\beta}(-\Q) \rangle
\end{align}

\noindent For nuclear scattering from a monotonic system, this reduces to ${\hbar^{2} Q^{2} \over {2 M}}$, where $M$ is the mass of the scattering nucleus.~\cite{cowley200315, Stock201081} For magnetic systems and in the case for symmetric-only exchange where the Hamiltonian has the form $\mathcal{\hat{H}}=\sum_{i,j} J_{ij} {\vu*{S}_{i}} \cdot {\vu*{S}_{j}}$, the Hohenberg-Brinkman first moment sum rule is given by:\cite{hohenberg197410,hammar199857,sarte201898,stock2009103,stone200265}

\begin{align}\label{eq:firstmomentxtal}
\langle E\rangle(\vb*{Q})&= \int \dd{E}E \ \sqe\\ \nonumber
&=-\frac{2}{3}\sum_{i,j}n_{ij}J_{ij}\langle\vu*{S}_i\cdot\vu*{S}_j\rangle[1-\cos(\Q\cdot\vb*{d}_{ij})]
\end{align}

\noindent where $\langle\vu*{S}_i\cdot\vu*{S}_j\rangle$ is the ground-state equal-time correlation function of spins $\vu*{S}_i$ and $\vu*{S}_j$ at sites $i$ and $j$, $n_{ij}$ is the multiplicity of $J_{ij}$, the exchange constant associated to the bond vector $\vb*{d}_{ij}$.  This equation assumes symmetric-only exchange as we anticipate is dominant for $3d$ magnetic transition metal ions in the absence of spin-orbit coupling. Anisotropic terms in the magnetic Hamiltonian appear as constants to this equation for the first moment, however, given the lack of an orbital degree of freedom in Mn$^{2+}$ in an octahedra, we expect such terms to be small in comparison to the symmetric Heisenberg exchange and therefore neglect them here.

Knowing the nuclear and magnetic structure of a compound gives the bond vectors $\vb*{d}_{ij}$ and the correlators $\langle\vu*{S}_i\cdot\vu*{S}_j\rangle$. Then, measuring the first moment for different $\Q$ values allows to fit the exchange constants, which correspond to the amplitudes of the sinusoidal oscillations.  We note that Eqn. \ref{eq:firstmomentxtal} only depends on the relative orientation of neighboring spins which has been modelled previously using neutron diffraction.   For the following, in terms of notation, the spin component $S(S+1)$ will be included in the exchange constants instead of the correlators and the exchange constants are in units of meV.

In \mnsb, 7 nearest neighbors exchange interactions are considered and expected to be relevant, as shown in \cref{fig:MSO_exchanges}, related to a total of 30 Mn-Mn bonds per unit cell. The first thing to evaluate is the ground-state correlation functions $\langle\vu*{S}_i\cdot\vu*{S}_j\rangle$ for each of the bonds. The magnetic ground state of \mnsb\ is unclear, rather reported as a pure cycloid in Ref.~\onlinecite{johnson2013111} or tilted from the $c$-axis in Ref.~\onlinecite{kinoshita2016117}. But in both cases, the spin structure is helicoidal with the spins co-rotating in the same plane.\cite{chan2022106a} Thus, the scalar product can be simply evaluated by $\cos{\Delta\theta_{ij}}$, with $\Delta\theta_{ij}$, the angle difference between the spins in the same rotation plane. The exchange interactions are listed in \cref{tab:exchange} with their associated multiplicities, bond distances, and ground-state correlators, with $k=0.182$ the propagation vector component along the $c$-axis.  We emphasize that this method only depends on relative orientation of neighboring spins and not on details for the tilted and non tilted helicoidal structures.  Indeed, the $\langle\vu*{S}_i\cdot\vu*{S}_j\rangle$ correlators are the same in both models.  Therefore, this method allows us an independent means of measuring the exchange constants without details of the long-range magnetic structure that is relevant for spin-wave calculations. We discuss this point later in the context of stability of the long-wavelength excitations once we have obtained the exchange constants from the first moment analysis.

Furthermore, we note that the correlators for diagonal paths actually depend on the sense of rotations of the spins, and thus on the magnetic parameters $\eta\ut{C}$ and $\eta\ut{T}$. From the energy invariant, these magnetic parameters are related to the structural chirality by $\sigma=\eta\ut{C}\eta\ut{T}$.\cite{chan2022106a} Thus the correlators for the diagonal exchange paths are
$\cos(2\pi(\eta\ut{C}k\pm\eta\ut{T}/3))=\cos(2\pi(k\pm\sigma/3))$ for left- $J_5$, $J_7$ ($+$) and right-handed $J_3$, $J_6$ ($-$) exchange interactions. The diagonal exchange interactions are interchanged by inversion symmetry, which corresponds to an inversion of $\sigma$. Thus, ground-state correlators are invariant for a given exchange constant. Thus the analysis holds independently of the structural and magnetic domains populations. This is convenient as a mixture of structural and magnetic domains was previously measured in a single crystal of \mnsb.\cite{chan2022106a}

For a fixed scattering vector \Q, the cosine frequency will only depend on the bond distances. We can therefore define the parameters $\gamma$ associated to each of the five distinct bond lengths, which are functions of the exchange constants and ground-state correlation functions:

\begin{subequations}\label{eq:gamma}
\begin{align}[left = \empheqlbrace\,]
\gamma_1 &= J_1c_1\label{eq:g1}\\
\gamma_2 &= J_2c_1\label{eq:g2}\\
\gamma_4 &= J_4c_4\label{eq:g4}\\
\gmi&= J_3 c\ut{R} + J_5 c\ut{L}\label{eq:gi}\\
\gme&= J_6 c\ut{R} + J_7 c\ut{L}\label{eq:ge}
\end{align}
\end{subequations}

\noindent where the $c_i$ are calculated from the co-rotating helicoidal magnetic structure\cite{chan2022106a} and displayed in \cref{tab:exchange}.

\begin{table}[]
\centering
\begin{tabular}{ccccc}
\hline
\hline
$J_i$   & $n_i$ & $d_i$ (\AA)               & $\Delta\theta_{ij}$       & $c_{ij}=\langle\vu*{S}_i\cdot\vu*{S}_j\rangle=\cos\Delta\theta_{ij}$ \\ \hline
$J_1$ & 3     & $d_1=5.5961$                  & \multirow{2}{*}{$2\pi/3$} & \multirow{2}{*}{$c_1 = -0.5$}                                        \\
$J_2$ & 6     & $d_2=4.8445$                  &                           &                                                                      \\
$J_3$ & 3     & $d\ut{i}=7.3235$                  & $2\pi(k+\et/3)$           & $c\ut{R} = -0.995$                                                   \\
$J_4$ & 3     & $d_4=4.7241$                  & $2\pi k$                & $c_4 = 0.414$                                                        \\
$J_5$ & 3     & $d\ut{i}=7.3235$               & $2\pi(k-\et/3)$           & $c\ut{L} = 0.58$                                                     \\
$J_6$ & 6     & \multirow{2}{*}{$d\ut{e}=6.7666$} & $2\pi(k+\et/3)$           & $c\ut{R} = -0.995$                                                   \\
$J_7$ & 6     &                         & $2\pi(k-\et/3)$           & $c\ut{L} = 0.58$                                                    \\
\hline
\hline
\end{tabular}
\caption{Summary of the exchange interactions $J_i$, with their multiplicity in the unit cell $n_i$, the related bond distance $d_i$, the spin angle difference $\Delta\theta_{ij}$ and the associated ground-state correlation functions $c_{ij}$. Subindices i and e refer to the diagonal bond distances internal and external to the triangle of Mn interconnected by $J_1$. Subindices L and R refer to left- and right-handed correlation functions.}
\label{tab:exchange}
\end{table}

\subsubsection{Single-crystal data}\label{sec:xtal}

Having discussed the equations and theory for the first moment sum rule applied to MnSb$_{2}$O$_{6}$, we now apply this to our single crystal sample aligned in the $(HHL)$ scattering plane. We can simplify the calculation of the first moment by fixing $H=H_0$ and varying $L$ ($L$-scan), or fixing $L=L_0$ and varying $H$ ($H$-scan). This leads to two different analyses. The $L$-scan analysis will be detailed in the following section, while the $H$-scan analysis is presented in \cref{sec:annexHscan}.

\begin{figure}[h]
    \begin{center}
    \includegraphics[]{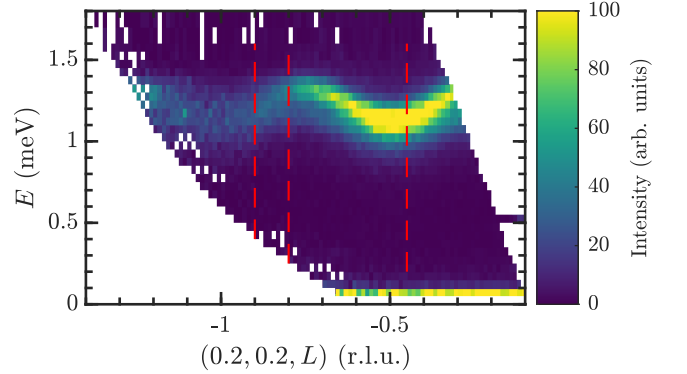}
    \end{center}
    \caption{MACS single crystal inelastic neutron scattering spectrum: spin-wave dispersion along $(0.2,0.2,L)$. The red dashed lines indicate constant-$\Q$ scans shown in \cref{fig:L24_1}(a)-(c).}
    \label{fig:MACS_02}
\end{figure}

\begin{figure*}[!]
    \begin{center}
    \includegraphics[]{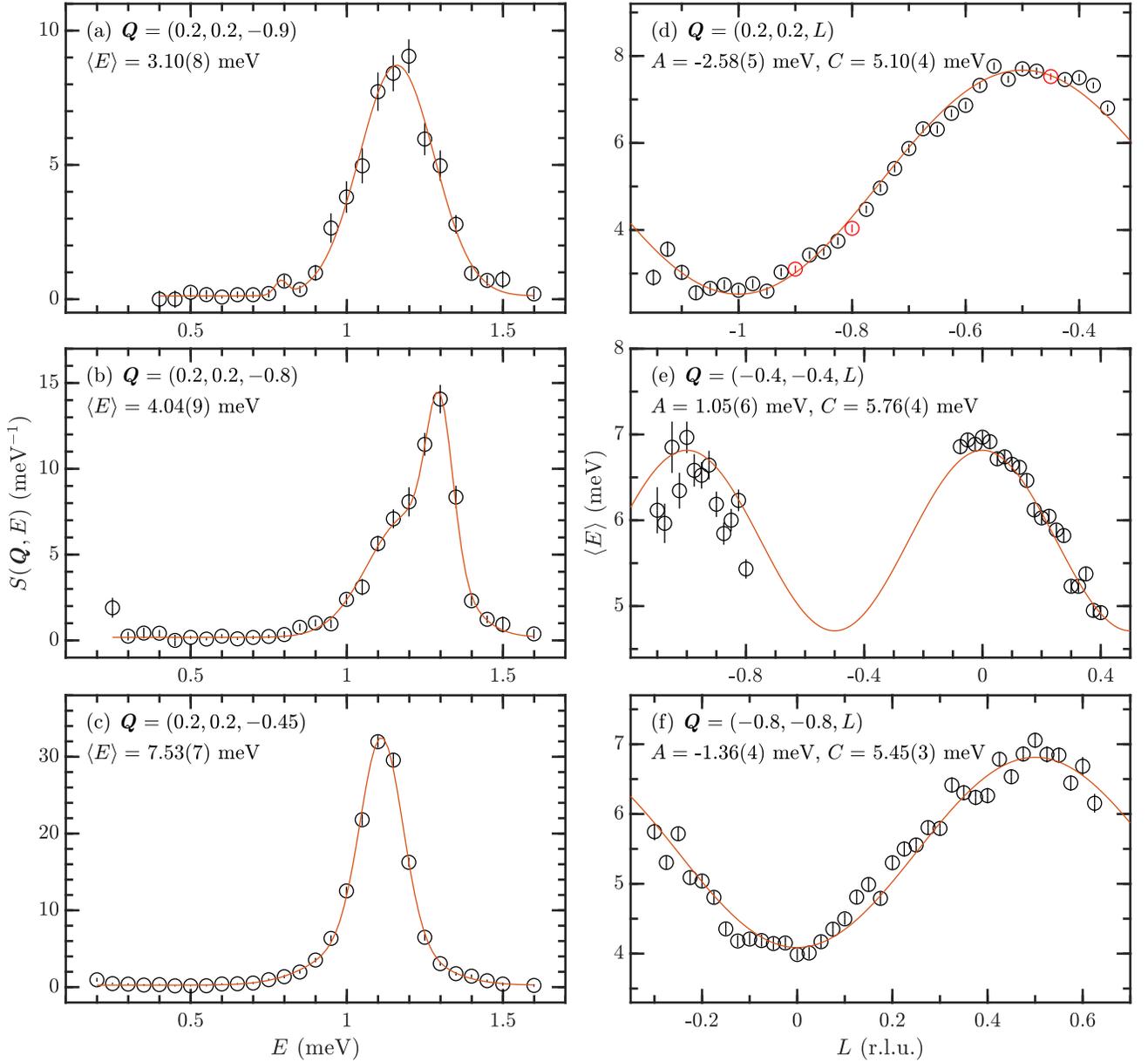}
    \end{center}
    \caption{(a)-(c) Constant-$\Q$ scans for different $\Q=(0.2, 0.2, L)$, indicated with dashed red lines in \cref{fig:MACS_02}. A fit to a double gaussian is shown in red, and the first moment is calculated from trapezoidal integration where the background is removed from the gaussian fit. (d) First moment as a function of $L$ for $H_0=0.2$, fitted to its theoretical expression (red curve). The red data points corresponds to the first moments calculated in the cuts plotted in (a)-(c). (e)-(f) First moment as a function of $L$ for (e) $H_0=-0.4$ and (f) $H_0=-0.8$, fitted to theoretical expression in red.}
    \label{fig:L24_1}
\end{figure*}

\begin{figure*}[!]
    \begin{center}
    \includegraphics[]{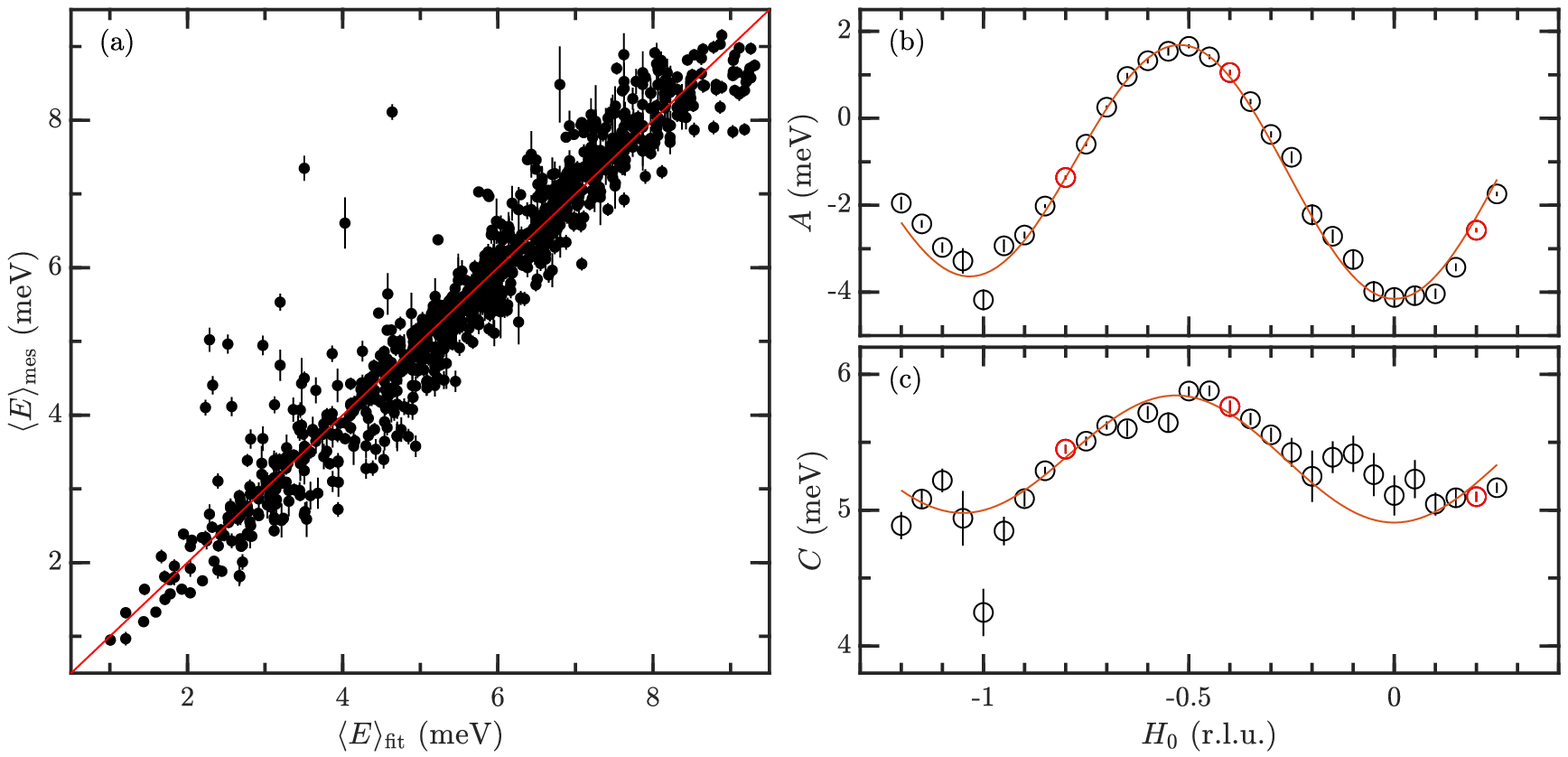}
    \end{center}
    \caption{(a) Measured first moments versus fitted first moments for $L$-scan analysis, for the $E\ut{f}=\SI{2.4}{meV}$ dataset. A total of 969 $\langle E\rangle(\Q)$ were taken into account. (b)-(c) Fitting of coefficients (b) $A$ and (c) $C$ giving the $\gamma$ parameters. The red data points show the values calculated in \cref{fig:L24_1}(d)-(f).}
    \label{fig:L24_2}
\end{figure*}

The data is extracted along an $L$-scan, considering $\Q = (H_0,H_0,L)$  with $L$ varying and a given $H_0$. In the following we will consider the $\Ef=\SI{2.4}{meV}$ dataset, as an example, we take $H_0=0.2$. The spin-wave dispersion along $(0.2, 0.2,L)$ is shown in \cref{fig:MACS_02}. For each interaction indexed by spins $i$ and $j$, the corresponding term in the first moment cosine from \cref{eq:firstmomentxtal} can be written as:

\begin{equation}
\Q\cdot\vb*{d}_{ij} = 2\pi H_0(d_{ij,x}+d_{ij,y})+2\pi Ld_{ij,z}
\end{equation}

\noindent where the distances $d_{ij}$ are expressed in lattice units, and the scattering vector in reciprocal lattice units. Using trigonometric identities to expand the cosine term, and summing \cref{eq:firstmomentxtal} over the 30 bonds in the unit cell, a general formula for the first moment is derived, for a fixed $H_0$:

\begin{equation}\label{eq:Lscan}
\langle E\rangle(H_0,L)= A(H_0)\cos(2\pi L)+C(H_0)
\end{equation}

\noindent where $A$ and $C$ are two $H_0$-dependent functions of the $\gamma$ parameters, given by:

\begin{align}
A(H_0) &= \frac{2}{3}[(1+2c(H_0))\gamma_i+3\gamma_4+2\Sigma_c(H_0)\gamma_e]\label{eq:Lscan_A}\\
C(H_0) &= -\frac{2}{3}[2(1-c(H_0))\gamma_1+... \label{eq:Lscan_C}\\
& 2(3-\Sigma_c(H_0))\gamma_2+3\gamma_i+3\gamma_4+6\gamma_e] \nonumber
\end{align}

\noindent where,

\begin{align}
c(H_0) &= \cos(2\pi H_0\delta_1) \nonumber \\
\Sigma_c(H_0) &= \cos(2\pi H_0\delta_2)+...\nonumber \\
& \cos(2\pi H_0\delta_3)+\cos(2\pi H_0\delta_4) \nonumber
\end{align}

\noindent are $H_{0}$-dependent harmonic oscillations, and 

\begin{align}
    \delta_1 &= 3(1-r_x) \nonumber \\
    \delta_2 &= 1 \nonumber\\
    \delta_3 &= 2-3r_x \nonumber \\
    \delta_4 &= 3r_x-1 \nonumber
\end{align}

\noindent are Mn-Mn interatomic distances (in r.l.u.) projected in the $(ab)$-plane. $r_x=0.6329$ is the $a$-axis coordinate of the Mn atom at Wyckoff site 3$e$, taken from the single crystal neutron diffraction refinement at $T=\SI{2}{K}$ in Ref.~\onlinecite{chan2022106a}.

From \cref{eq:Lscan}, for a specific $H_0$, we can compute the first moment as a function of $L$, and fit the coefficients $A(H_0)$ and $C(H_0)$ for a scan along $(H_0,H_0,L)$. The next step is to repeat the same process for several $H_0$, and fit the $\gamma$ parameters in coefficients $A$ and $C$ with \cref{eq:Lscan_A} and \cref{eq:Lscan_C}.

Examples of calculations of the first moment for different $L$, for $\Q=(0.2, 0.2, L)$ are shown in \cref{fig:L24_1}(a)-(c). These constant-\Q\ scans are indicated in red dashed lines in \cref{fig:MACS_02}. Most of the $\sqe$ are well fitted by two gaussians, shown in red in the figures, but to take into account any deviation from a two-mode spectrum, the numerical integration of the first moment from \cref{eq:firstmomentxtal} was performed using a trapezoidal integration, with the background removed from these two-gaussian fits. The calculation is performed above 0.2 meV to get rid of any contribution from elastic scattering, and below 1.6 meV to only capture contribution from one-magnon scattering. This criterion is arbitrary, and low-energy scattering can be miscalculated. Actually, due to \cref{eq:firstmomentxtal}, lowest energy points contribute less to the first moment (given a low magnetic intensity at low energy), so the differences are not significant within uncertainties. More information concerning the numerical integration and the differences between the methods of integration are given in \cref{sec:error}.

These first moments are calculated for a range of $L$, as shown in \cref{fig:L24_1}(d) where first moments computed in \cref{fig:L24_1}(a)-(c) are highlighted in red. For this specific $H_0=0.2$, the $A$ and $C$ parameters are obtained from the fit (red curve) to \cref{eq:firstmomentxtal}. The $H_0$-dependence of $A$ and $C$ is then obtained by repeating the same procedure for different $H_0$, as illustrated in \cref{fig:L24_1}(e)-(f) for $H_0=-0.4$ and $H_0=-0.8$.

\begin{figure*}[!]
    \begin{center}
    \includegraphics[]{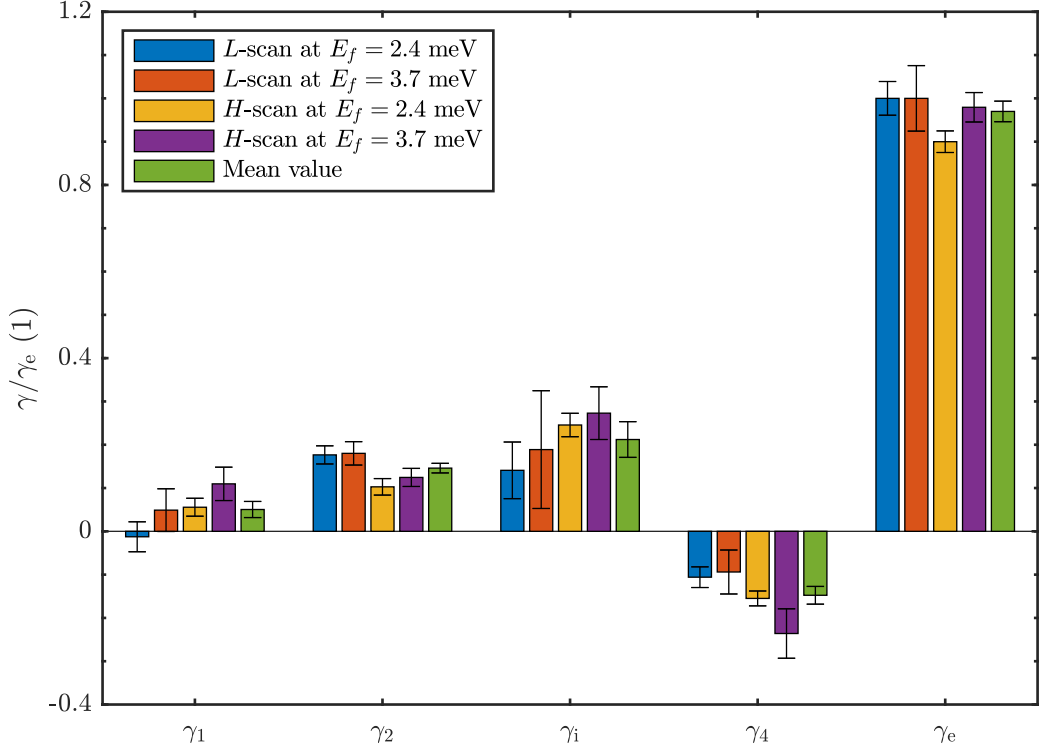}
    \end{center}
    \caption{Fitted parameters for the different analysis and dataset, normalized to $\gme$ obtained in the $L$-scan analysis from the $E\ut{f}=\SI{2.4}{meV}$ dataset. Mean values (green bars) are calculated averaging over the four analysis.}
    \label{fig:param}
\end{figure*}

Finally, a total of 969 first moments $\langle E\rangle(\Q)$ were calculated from the MACS $E\ut{f}=\SI{2.4}{meV}$ dataset for this analysis and are shown as a function of the fitted first moment in \cref{fig:L24_2}(a). Finally the $\gamma$ parameters are obtained by fitting $A$ and $C$ to \cref{eq:Lscan_A} and \cref{eq:Lscan_C} as shown in \cref{fig:L24_2}(b)-(c), where the red data points are the coefficients calculated in \cref{fig:L24_1}(d)-(f). We note from \cref{eq:Lscan} that some remaining background can be included in the computation of $C$, as well as small contributions from anisotropic terms in the magnetic Hamiltonian, as discussed above. For this reason, the $H_0$-independent part of \cref{eq:Lscan_C} is not fitted to get the parameters $\gamma_4$, $\gmi$ and $\gme$, which are rather fitted with \cref{eq:Lscan_A}, where $A$ represents the amplitude of the first moment cosine variation.

A similar analysis can be performed by considering a fixed $L_0$ and varying along $H$ and is detailed in \cref{sec:annexHscan}, giving another set of fitted $\gamma$ parameters. Then, these two analyses were performed again with the second single crystal dataset, with $\Ef=\SI{3.7}{meV}$, giving two other sets of $\gamma$ parameters. This is detailed in \cref{sec:annexsecond}. These fitted $\gamma$ parameters are shown in \cref{fig:param}, where they have been normalized to $\gme$ obtained from the $L$-scan analysis for each dataset, in order to get rid of any scale issue coming from the absolute normalization process and to directly compare the fitted parameters. We discuss below how we obtain the overall scaling factor to obtain units of meV. 

\subsubsection{Powder data}

As described in \cref{sec:macspowder}, powder inelastic neutron scattering was also performed on MACS and first moment sum rule can also be applied to these data.

For polycrystalline samples, the intensity measured is related to the powder averaged $S(\Qn,E)=\int\dd{\Omega_{\Q}}\sqe/4\pi$ of the dynamic structure factor. This gives the powder averaged first moment sum rule:\cite{stone200265,sarte201898}

\begin{align}\label{eq:firstmomentpowder}
\langle E\rangle(\Qn)&= \int \dd{E}E S(\Qn,E)\\ \nonumber
&=-\frac{2}{3}\sum_{i,j}n_{ij}J_{ij}\langle\vu*{S}_i\cdot\vu*{S}_j\rangle\left\{1-\frac{\sin(\abs{\Q}\abs{\vb*{d}_{ij}})}{\abs{\Q}\abs{\vb*{d}_{ij}}}\right\}
\end{align}

\noindent As for the single crystal analysis, for a fixed $Q=\abs{\Q}$, the sine frequency only depends on the bond lengths, which are the same for diagonal exchange paths as listed in \cref{tab:exchange}, resulting in five distinct bond distances. We can further simplify the first moment by summing over these distinct bond distances:

\begin{equation}
\langle E\rangle(Q)=-\frac{2}{3}\sum_{i}n_i \gamma_i\left\{1-\frac{\sin(Q\abs{\vb*{d}_{i}})}{Q\abs{\vb*{d}_{i}}}\right\}
\end{equation}

\noindent where $i\in[1,5]$ is related to the $i$-th bond length and the $\gamma_i$ are defined in \cref{eq:gamma}. Due to the very close bond distances (especially $d_2=\SI{4.8445}{\AA}$ and $d_4=\SI{4.7241}{\AA}$), and the relatively small $Q$-range probed in the experiment (from 0.3 to 2.05 \AA$^{-1}$), we were not able to conveniently fit the $\gamma$ parameters, because of high correlations in the fitting process. However, we can compare the first moment extracted from the powder inelastic neutron scattering with the theoretical one calculated using the $\gamma$ parameters obtained from the single crystal analysis described above. 

The first step for extracting the first moment from the experimental data is to define the region of integration for the energy. For the powder, the first moment was integrated for $E\in [0.3,1.6]$~meV to get rid of the elastic and two-magnon scattering. This is justified by the spectral weight calculated in the total moment sum rule analysis described in \cref{sec:totalmoment}. Due to gapless modes in the one-magnon spectrum, around $\SI{0.8}{\AA^{-1}}$ and $\SI{1.4}{\AA^{-1}}$, as shown in \cref{fig:pow_spec}(a), the contribution from elastic scattering and one-magnon can be mixed. However, this mixture happens at low energies and low intensities, so that deviations from the actual first moment are small. As for the single crystal analysis, the data were integrated numerically using a trapezoid integration, and the background was removed by fitting with two gaussians. The theoretical $\gamma$ parameters calculated from the single crystal first moment sum rule analysis were rescaled to match the scale of the first moment observed in the powder experiment, as we know the powder data have been fairly normalized as it captures all the magnetic spectral weight as detailed in \cref{sec:totalmoment}. The magnetic form factor is also taken into account during this rescaling process.

\begin{figure}[h]
    \begin{center}
    \includegraphics[]{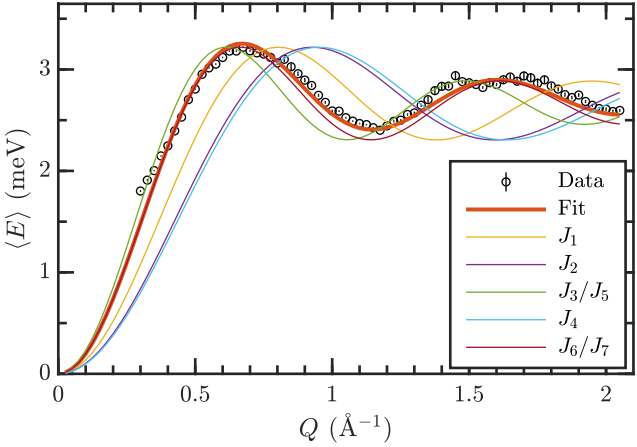}
    \end{center}
    \caption{(data points) First moment computed from the powder data, as a function of the scattering vector amplitude. (red thick curve) First moment calculated from the $\gamma$ parameters fitted in the single crystal first moment sum rule analysis. (thin curves) Contributions to the first moment from the different exchange paths, normalized to the powder computed first moment.}
    \label{fig:powder}
\end{figure}

The theoretical first moment calculated from the $\gamma$ parameters obtained from the single crystal sum rules analysis is shown in red in \cref{fig:powder}, and matches well the first moment computed from the powder experiment. The contribution from each exchange constant associated to their bond distance is shown in thin lines (normalized to the powder computed first moment). From this, we can see how the contributions from $J_2$ and $J_4$ to the first moment are close, which makes the fit difficult within this small wavevector range probed during this experiment.

\subsection{Determination of exchange constants}

In the first moment sum rules analysis, we have used the five $\gamma$ parameters which are related to the seven exchange constants. $\gamma_1$, $\gamma_2$  and $\gamma_4$ are uniquely related to $J_1$, $J_2$ and $J_4$, and can be deduced from  \cref{eq:g1,eq:g2,eq:g4}, leaving $J_3$, $J_5$, $J_6$ and $J_7$.   $\gamma\ut{i}$ and $\gamma\ut{e}$ are related in \cref{eq:gi,eq:ge} to these four chiral exchange constants. Considering the energy minimization using the experimental propagation vector from diffraction,\cite{chan2022106a} these four unknown exchange constants can be written into three linearly independent equations:

\begin{subequations}
\label{eq:system}
\begin{align}[left = \empheqlbrace\,]
\tan 2\pi k &= \sqrt{3}\frac{J_3-J_5+2(J_6-J_7)}{J_3+J_5+2(J_6+J_7-J_4)}\label{eq:tank}\\
\gamma\ut{i}&= J_3 c\ut{R} + J_5 c\ut{L}\\
\gamma\ut{e}&= J_6 c\ut{R} + J_7 c\ut{L}
\end{align}
\end{subequations}

\noindent This analysis presents an ambiguity given the presence of three equations and four unknown exchange constants.   This ambiguity is intrinsic originating from many of the exchange parameters corresponding to the same bond distances which is the the basis of the first moment sum rule analysis discussed above.  In particular, the exchange constants $J_{3}$ ($J_{6}$) and $J_{5}$ ($J_{7}$) correspond to the same bond distance and only differ by the SSE pathway defined by the crystal chirality.  We therefore need further information to close this set of equations and seek this through a comparison between calculated and measured single crystal excitation spectra, focusing on the overall bandwidth and excitations near the zone boundary.  

\begin{figure*}[!]
    \begin{center}
    \includegraphics[]{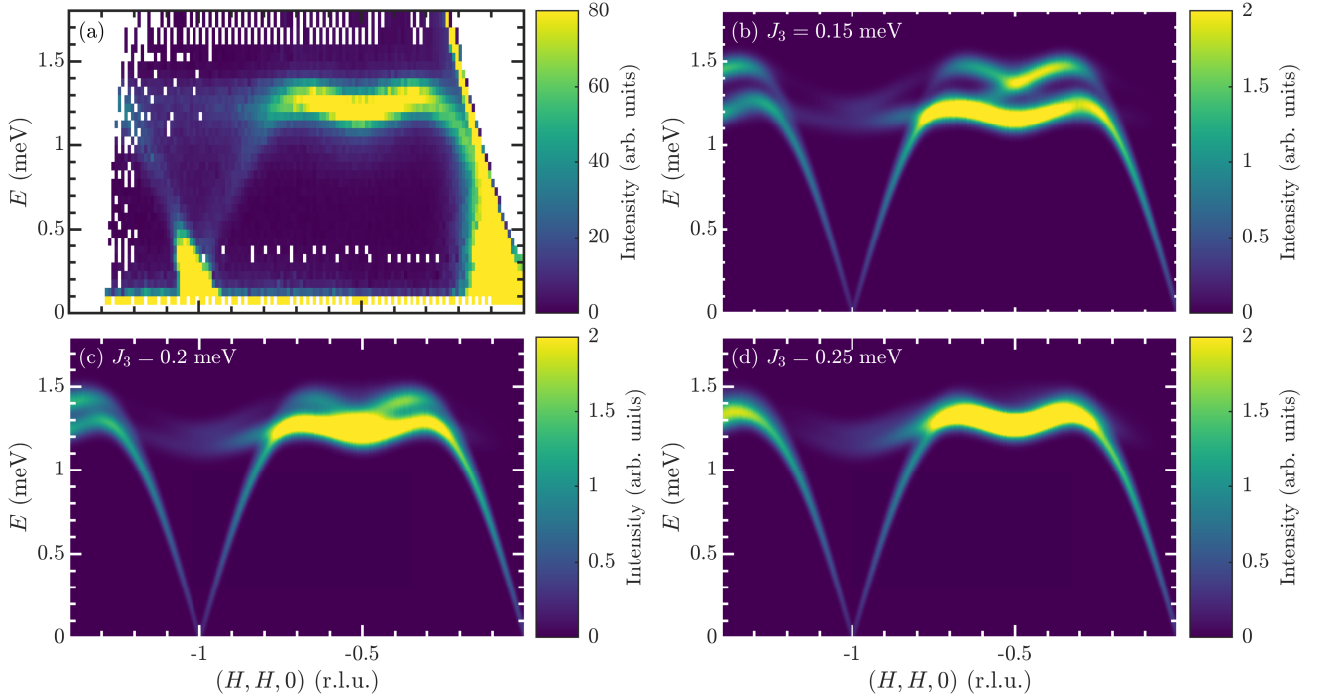}
    \end{center}
    \caption{Spin-wave dispersion along  $(H,H,0)$ for: (a) MACS single crystal inelastic neutron scattering spectrum. (b)-(d) Inelastic neutron scattering spectrum calculated from linear spin-wave theory by fixing different $J_3$ values.  The other parameters for these calculations are listed in \cref{tab:J_param}.} 
    \label{fig:guessJ3}
\end{figure*}

By calculating the excitation spectra using linear spin-wave theory software \textsc{SpinW}\cite{toth201527} with an simulated instrumental resolution $\Delta E\approx\SI{0.1}{meV}$, we can see that the upper magnon branch along $(H,H,0)$ is largely affected by a change of the $J_{3}$ exchange parameter. We note that the calculation was done assuming an untilted structure [cycloidal ground state shown in \cref{fig:MSO_struc}(d)], however, the scattering near the top of the single magnon branch was found not to be sensitive to the tilting of the magnetic moments.  Analyzing the scattering near the top of the single magnon branch near the magnetic zone boundary therefore provides an independent means of fixing $J_{3}$.  The experimental spectrum from MACS $E\ut{f}=\SI{2.4}{meV}$ dataset is shown in \cref{fig:guessJ3}(a), and compared to calculated spectra for different values of $J_3$ in \cref{fig:guessJ3}(b)-(d), where we can observe a significant change of the position and structure of the upper mode.  In particular, tuning $J_{3}$ affects the maximum energy of the one-magnon band and also the splitting of multiple bands at the maximum energy of the single magnon bands as observed in the $H$-scans.  Given our experimental data [Fig. \ref{fig:guessJ3}(a)] and to close off the set of Eqns. \ref{eq:system}, we assume no observable splitting of bands in the $H$-scans and a maximum single-magnon energy excitation given by experiment.  These two observations fix both the absolute value of $J_{3}$ and also an overall scaling factor taking the data to absolute units of meV.  For these calculations, $J_5$, $J_6$ and $J_7$ are obtained by fixing $J_3$ in \cref{eq:system} resulting in a system of three equations and three unknowns with $\gmi$ and $\gme$ the mean values obtained in the single crystal sum rules analysis shown in \cref{fig:param}. We have chosen to fix $J_3$ as it has the lesser influence on the ordering wavevector which is seen by partially differentiating \cref{eq:tank}. Finally, the exchange constants obtained by fixing $J_3$ with the best agreement are listed in \cref{tab:J_calc}. The uncertainty associated to $J_3$ is an estimation based on the instrumental resolution of how far from $J_3=\SI{0.25}{meV}$ we can observe the band splitting. From this estimated error, and the least-square refinement of \gmi\ and \gme, we subsequently compute the uncertainties associated to $J_{5,6,7}$. The obtained exchange constants are compared with the values calculated from DFT from Ref.~\onlinecite{johnson2013111}. First we can see that the interactions are overall lower in energy than expected from the DFT calculations. Then, the left-handed interactions $J_3$ and $J_6$ are dominant in comparison to right-handed $J_5$ and $J_7$, as expected to impose the structural chirality of \mnsb.


\begin{table*}[!]
    \centering
    \begin{tabular}{cccccccc}\hline\hline
          & $J_1$ & $J_2$ & $J_3$ & $J_4$ & $J_5$ & $J_6$ & $J_7$\\ \hline
         DFT\cite{johnson2013111} & 0.77 & 1.47 & 2.2 & 1.16 & 0.4 & 1.94 & 0.4\\ 
         Sum rules & 0.10(4) & 0.29(2) & 0.25(2) & 0.35(5) & 0.07(8) & 0.97(3) & 0.03(5)\\ 
         Refined & 0.10 & 0.29 & 0.25 & \color{red} \bf{0.25} & 0.07 & 0.97 & \color{red} \bf{-0.023}\\\hline \hline
    \end{tabular}
    \caption{Symmetric $J$ exchange constants obtained by DFT calculations\cite{johnson2013111} and the mean values from the four single crystal sum rules analyses (normalized to $\gme$ and then rescaled to experimental data, in meV, note that all values of $J$ in the table are multiplied by $S(S+1)$ with $S=5/2$).  The refined parameters using Green's function approach are highlighted in \color{red} red.}
    \label{tab:J_calc}
\end{table*}

From mean field theory, the Curie-Weiss temperature can be estimated by summing the exchange constants over the nearest neighbors of a Mn$^{2+}$ ion:\cite{smart1966}

\begin{align}
\Theta\ut{CW}&=-\frac{S(S+1)}{3k\ut{B}} \left[ 2(J_1+J_3+J_4+J_5)+\right.\nonumber\\& \left.4(J_2+J_6+J_7)\right]
\end{align}

\noindent We note that this equation is not linearly independent from the system in \cref{eq:system}, and thus cannot be used to uniquely determine the four chiral exchange constants $J_3$, $J_5$, $J_6$, and $J_7$. Furthermore, the Curie-Weiss temperature obtained from magnetic susceptibility on \mnsb\ powder, $\Theta\ut{CW}=\SI{-19.6}{K}$ in Ref.~\onlinecite{johnson2013111} and $\Theta\ut{CW}=\SI{-23}{K}$ in Ref.~\onlinecite{werner201694} have a difference $\Delta T=\SI{3.4}{K}$ corresponding to an energy difference of $\Delta E \approx \SI{0.3}{meV}$, which is significant given the low energy scale of the exchange constants in \mnsb\ (see \cref{tab:exchange}). This variation in experimentally reported results is justifiable given the choice of the linear regime when fitting mean-field Curie Weiss law and reflects the experimental uncertainty.  For these reasons, we have not used the experimental Curie-Weiss temperatures as a hard constraint for the exchange constants. On the contrary, we can compute afterwards $\Theta\ut{CW}=\SI{-26(1)}{K}$, which reasonably agrees with the measured ones, given the experimental variations.

\subsection{Comparison to spin-wave theory}

In the previous sections we have applied the first moment sum rule to extract the complex series of Heisenberg exchange constants in MnSb$_{2}$O$_{6}$.  In this section we compare these results to a mean-field linear spin-wave theory to compare results and also to test for stability of the ground state magnetic structure.  We use the Green's function formalism for this.  While this technique for calculating magnetic excitations is more versatile in cases where the low-energy response is determined by a series of single-ion states (such as the case in rare-earths or in the presence of spin-orbit coupling like in, for example, Co$^{2+}$~[\onlinecite{sarte2019100}] or V$^{3+}$~[\onlinecite{Lane2021a}] based compounds), it is also useful to test for stability of harmonic long-wavelength magnetic excitations with changes in the local magnetic environment.  In this section we first briefly outline the use of the Green's function technique and then we apply it to calculate the spin excitation spectrum, comparing sum rule results presented above to experiment, then refining results.  We then test stability of the proposed magnetic structure and interactions based on the series of exchange constants extracted with the first moment sum rule and refined values.  In particular, we discuss the stability of long-wavelength magnetic fluctuations for tilted helicoidal structures.

\subsubsection{Green's functions on a rotating frame}

The basic technique for applying the Green's function approach has been outlined in several previous papers by us.  The application of the technique to collinear systems CoO,~\cite{sarte2019100} in the presence of spin-orbit coupling with Co$^{2+}$ ($S={3\over2}$, $l\ut{eff}$=1) ions, and CaFe$_{2}$O$_{4}$,~\cite{Lane2021} based on a spin-only ground state of Fe$^{3+}$ ($S={5\over2}$) ions.  We then recently extended this methodology to the noncollinear magnetic structure of RbFe$^{2+}$Fe$^{3+}$F$_{6}$ which involved coupled spin-only Fe$^{3+}$ ($S={5\over2}$) and orbitally degenerate Fe$^{2+}$ ($S=2$, $l\ut{eff}$=1) ions. In terms of MnSb$_{2}$O$_{6}$ where only a spin-degree of freedom exists (Mn$^{2+}$ with $S={5\over2}$), we quote only the key results here and refer the reader to Ref. \onlinecite{Lane22:106} for further details.  The methodology here is to use the Green's functions results from the collinear cases and transform to a local rotating frame of reference for use in incommensurate magnets like MnSb$_{2}$O$_{6}$.

The neutron scattering cross section is proportional to the dynamical structure factor $S(\mathbf{Q}, \omega)$ which is related to the Green's response function, $G(\mathbf{Q}, \omega)$ via the fluctuation-dissipation theorem.

\begin{equation}
    S(\mathbf{Q}, \omega)=-{1\over \pi} [n(\omega)+1] \Im G(\mathbf{Q}, \omega)
    \label{fluctdiss:eq}
\end{equation}  

\noindent where $n(\omega)$ is the Bose factor.  The Green's function, in the laboratory frame, is defined here as

\begin{equation}
G^{\alpha\beta}_{\tilde{\gamma}\tilde{\gamma}'}(i'j',t)=-i\Theta(t)\langle [\hat{S}^{\alpha}_{i'\tilde{\gamma}}(t),\hat{S}^{\beta}_{j'\tilde{\gamma}'}(0)]\rangle.
\nonumber
\end{equation}

\noindent The three sets of indices in this definition of the Green's function and used throughout the remaining discussion in this paper are summarized in Table \ref{Table:indices}.

\begin{table}[h]
\caption{\label{Table:indices} Summary of labeling convention for indices.}
\begin{ruledtabular}
\begin{tabular}{cc}
Index & Description \\ 
\hline
$\gamma$, $\gamma'$ & sites within unit cell  \\ 
$i$, $j$ & unit cell  \\ 
$\alpha$, $\beta$, $\mu$, $\nu$ & Cartesian coordinates  \\ 
\end{tabular}
\end{ruledtabular}
\end{table}

\begin{figure*}[!]
    \begin{center}
    \includegraphics[width=\linewidth]{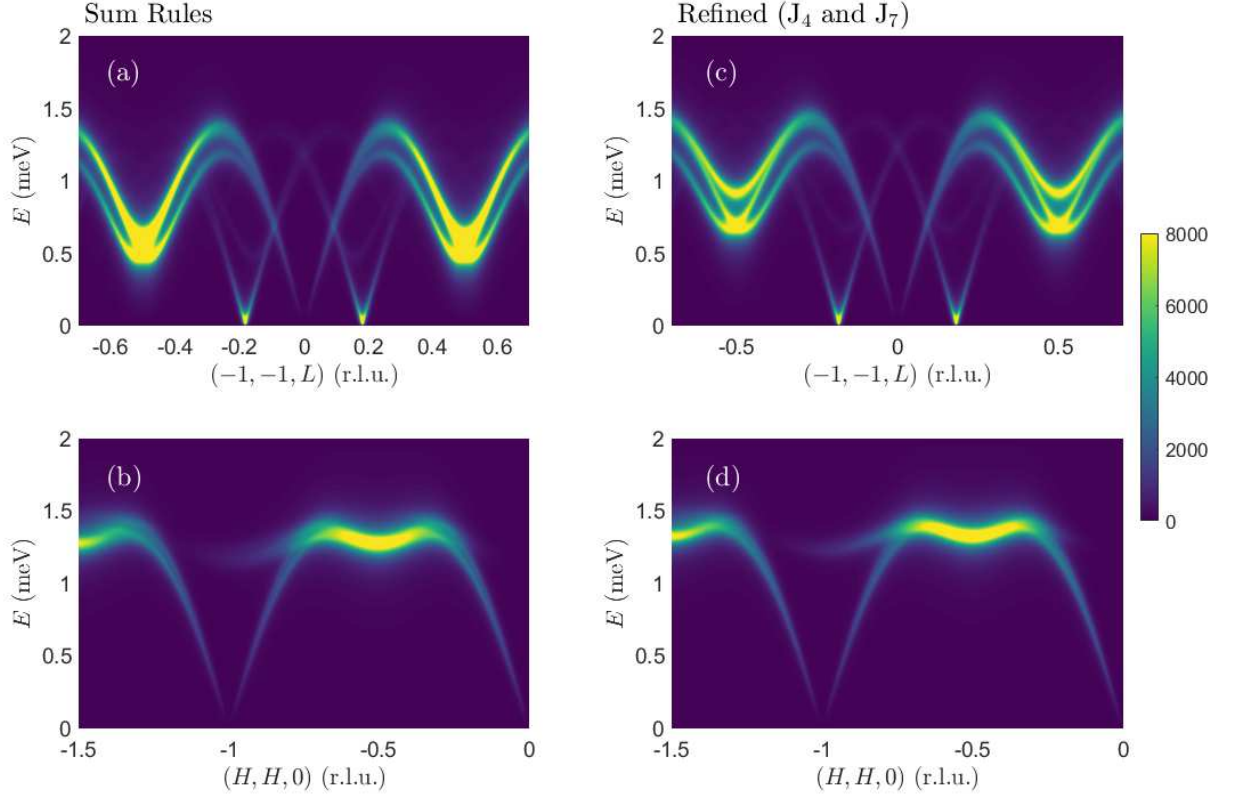}
    \end{center}
    \caption{The theoretical dispersive neutron scattering results based on our theoretical calculations using Green's functions taking an untilted magnetic ground state [see \cref{fig:MSO_struc}(d)].  (a)-(b) display calculations with the exchange parameters fixed from those derived using the first-moment sum rule described in the main text. (c)-(d) show calculations but refining $J_{4}$ to give agreement with experiment at the zone boundary and $J_{7}$ refined to keep the ordering wavevector consistent with experiment. }
    \label{Theory}
\end{figure*}

Following previous methods applying the RPA (random phase approximation),~\cite{Buyers197511,Cooke19737} we take an interaction Hamiltonian between Mn$^{2+}$ ($S={5\over2}$) ions of the form $\mathcal{H}\ut{int}=\frac{1}{2}\sum_{ij}^{\gamma\gamma'}\mathcal{J}_{ij}^{\gamma\gamma'}\mathbf{S}_{i\gamma}\cdot\mathbf{S}_{j\gamma'}$, where $\mathcal{J}_{ij}^{\gamma\gamma'}$ is a symmetric Heisenberg exchange parameter.  Note that we have changed notation here from Eqn.\ref{eq:firstmomentxtal} and written the symmetric exchange $J_{1\rightarrow7}$, discussed above in the context of the first moment sum rule, as a diagonal matrix $\mathcal{J}_{ij}^{\gamma\gamma'}$ which we use below when moving to a rotating frame as required for incommensurate magnets.  Note also that the factor of ${1\over2}$ in $\mathcal{H}\ut{int}$ originates from the application of mean field theory as discussed previously in Refs. \onlinecite{Buyers197511,sarte2019100,Sarte2020102,Lane2021,Lane22:106}. As shown in Ref. \onlinecite{Lane22:106}, applying mean field decoupling and converting to a local rotating frame, where we define rotation matrices, 

\begin{equation}
\mathbf{S}_{i\gamma}=R_{i\gamma}\mathbf{\tilde{S}}_{i\gamma},
\nonumber
\end{equation}    

\noindent with ${\vb*{\tilde{S}}}_{i\gamma}$ being the spin operators in the rotating frame.  As discussed in Ref. \onlinecite{Lane22:106} the Green's function equation of motion becomes after transforming to $\mathbf{Q}$ and $\omega$ space 
\begin{equation}
\begin{split}
&\tilde{G}_{\tilde{\gamma}\tilde{\gamma}'}^{\alpha\beta}(\mathbf{Q},\omega)=g_{\tilde{\gamma}\tilde{\gamma}'}^{\alpha\beta}(\omega)\delta_{\tilde{\gamma}\tilde{\gamma}'}\\ &\qquad+\sum_{\gamma'}^{\mu\nu}g_{\tilde{\gamma}\tilde{\gamma}}^{\alpha \mu}(\omega)\tilde{\mathcal{J}}^{\mu\nu}_{\tilde{\gamma}\gamma'}(\mathbf{Q})\tilde{G}_{\gamma'\tilde{\gamma}'}^{\nu\beta}(\mathbf{Q},\omega)
\end{split}
\label{fulleqRF}
\nonumber
\end{equation}
\noindent where the Fourier transform of the exchange interaction in the rotating frame is

\begin{subequations}
\begin{gather}
\begin{split}
&\underline{\underline{\tilde{\mathcal{J}}}}(\mathbf{Q})=X'\Big[\underline{\underline{\mathcal{J}}}(\mathbf{Q}+\mathbf{\tilde{q}})T_{3N}\\&+\underline{\underline{\mathcal{J}}}(\mathbf{Q}-\mathbf{\tilde{q}})T_{3N}^{*}+\underline{\underline{\mathcal{J}}}(\mathbf{Q})(\mathbb{I}_{3}\otimes\mathbf{n}\textbf{n}^{T} )\Big]X
\end{split}\\
\left[\underline{\underline{\mathcal{J}}}(\mathbf{Q})\right]_{\gamma\gamma'}=\sum_{ij}\underline{\underline{\mathcal{J}}}_{ij}^{\gamma\gamma'}e^{-i\mathbf{Q}\cdot(\mathbf{r}_{i}-\mathbf{r}_{j})}\label{Jmat:eq}\\
X=\mathrm{diag}\left(R_{1},R_{2},...,R_{N}\right)\\
X'=\mathrm{diag}\left(R^{T}_{1},R^{T}_{2},...,R^{T}_{N}\right)
\end{gather}
\end{subequations}
\noindent with $\mathbf{\tilde{q}}$ the ordering wavevector, $\mathbf{n}$ is the normal to the spin rotation plane and $T_{3N}=\mathbb{I}_{3}\otimes\frac{1}{2}\left(\mathbf{1}-\mathbf{n}\mathbf{n}^{T}-i[\mathbf{n}]_{\times}\right)$. We note the use of the notation ${\left[[\mathbf{n}]_{\times}\right]^{i}}_{j}={\epsilon_{i}}^{jk}n_{k}$ where we have made use of the Levi-Civita symbol for the antisymmetric tensor.  The matrices, $R$, rotate each of the $N$ spins in the unit cell onto a common axis. The single-ion Green's function is given by

\begin{equation}
g_{\tilde{\gamma}\tilde{\gamma}'}^{\alpha\beta}(\omega)=\sum_{qp}\frac{S^{\tilde{\gamma}}_{\alpha qp}S^{\tilde{\gamma}'}_{\beta pq}\phi_{qp}}{\omega-(\omega_{p}-\omega_{q})},
\label{SingleionGreensFunction}
\end{equation}
\noindent which has poles corresponding to the transitions between the eigenvalues of the single-ion Hamiltonian, $\omega_{p}$. We will sum over transitions to and from the ground state, as appropriate for magnon excitations at zero temperature. The rotation back to the lab frame can be achieved by

\begin{equation}
    \begin{split}
    \underline{\underline{G}}(\mathbf{Q},\omega)=D_{\mathbf{Q}}(\mathbb{I}_{3}\otimes\mathbf{n}\textbf{n}^{T})X\underline{\underline{\tilde{G}}}(\mathbf{Q},\omega)X^{\prime}(\mathbb{I}_{3}\otimes\mathbf{n}\textbf{n}^{T})D_{-\mathbf{Q}}\\+D_{\mathbf{Q}}T_{3N}^{*}X\underline{\underline{\tilde{G}}}(\mathbf{Q}+\mathbf{\tilde{q}},\omega)X^{\prime}T_{3N}^{\prime}D_{-\mathbf{Q}}\\+D_{\mathbf{Q}}T_{3N}X\underline{\underline{\tilde{G}}}(\mathbf{Q}-\mathbf{\tilde{q}},\omega)X^{\prime}T_{3N}^{*\prime}D_{\mathbf{-Q}}
    \end{split}
    \nonumber
\end{equation}
\noindent where the matrix $D_{\mathbf{Q}}=\delta_{\gamma\gamma'}e^{i\mathbf{Q}\cdot\delta_{\gamma}}\otimes \mathbb{I}_{3}$ accounts for the interference between ions in the unit cell. 

Finally the neutron scattering cross section is 
\begin{equation*}
S({\bf{Q}},\omega)=g_{L}^{2}f^2({\bf{Q}})\sum_{\alpha \beta} (\delta_{\alpha \beta}-\hat{q}_{\alpha}\hat{q}_{\beta}) S^{\alpha \beta}({\bf{Q}},\omega), 
\end{equation*}
\noindent where the partial dynamical structure factor, $S^{\alpha \beta}(\bf{Q},\omega)$ is proportional to the imaginary part of the Green's function [\cref{fluctdiss:eq}], $g_{L}$ is the Land{\'e} g-factor, $f(\bf{Q})$ is the Mn$^{2+}$ magnetic form factor and the polarization factor selects the component perpendicular to the momentum transfer.

We now apply this theory to \mnsb, which comprises a triangular motif of coupled Mn$^{2+}$ ($3d^{5}$) ions. In an intermediate octahedral field, the single-ion ground state of Mn$^{2+}$ is $^{6}$S ($S=5/2$, $L\approx0$) and the orbital moment is quenched. As a result, the effect of spin-orbit coupling and crystallographic distortions are small and may be neglected. The single-ion Hamiltonian is thus remarkably simple and consists solely of the molecular mean field created by the magnetic coupling to neighboring ions, which breaks time reversal symmetry, $\mathcal{H}\ut{SI}=h\ut{MF}\hat{S}_{z}$. This ``Zeeman-like" term acts to split the 6-fold degenerate $|S=5/2, m \rangle$ states.  At low temperatures (as illustrated in Fig. 7 of Ref. \onlinecite{Lane2021}) when only the ground state is populated, only one transition is allowed under the constraints of dipole selection rules of neutron scattering.  We note that this approach is equivalent to semi-classical linear spin-wave theory.

\subsubsection{Comparison to Experiment}

Inputting the symmetric exchange constants derived from the first moment sum rule into the Green's function calculation with an untilted magnetic structure, we derive the predicted neutron scattering excitation spectrum in Fig.~\ref{Theory}(a)-(b). This calculation is done with no anisotropic terms.  Symmetric exchange is expected to be dominant here owing to the lack of an orbital degree of freedom for Mn$^{2+}$.  The general results are in good qualitative agreement with experiment, however the calculated zone-boundary excitations are clearly in disagreement with experiment with the calculation predicting lower energy excitations than observed in experiment at the zone boundary.  

To address this, there are two noteworthy points of our first moment sum rule analysis.  First, on inspection of Fig. \ref{fig:param}, the values of $\gamma_{4}$, which fixes $J_{4}$ maybe dominated by the $H$-scan experiment performed with $\Ef=\SI{3.7}{meV}$. In comparison to iron based langasite, this value for $J_{4}$ is also considerably larger in \mnsb.\cite{stock201183}  We therefore consider a case when this value is lowered in Fig. \ref{Theory}(c)-(d). To ensure the same ordering wavevector we correspondingly tune $J_{7}$ given the relatively large error bar in our analysis and also the large sensitivity of the magnetic ordering wavevector to this exchange constant (Eqn. \ref{eq:tank}).  After refining $J_{4,7}$ (to within one-two sigma of the calculated error bar from the first moment sum rule analysis) we obtain a good description of the data (both along the $L$ and $H$ directions) with sum rule and refined exchange parameters illustrated in Table \ref{tab:J_calc} (refined values from this step highlighted in red).   

\subsubsection{Stability analysis}

Having derived a set of symmetric exchange constants from the first moment sum rule and written down a response function theory for the spin waves in terms of Green's functions, we discuss stability of the ground state fixed by the magnetic structure.  There have been two magnetic structures proposed in the literature involving a tilting of the plane of the helicoid at an angle away from the $c$-axis [Ref. \onlinecite{kinoshita2016117} and \cref{fig:MSO_struc}(e)] and one without tilting [Ref. \onlinecite{johnson2013111} and \cref{fig:MSO_struc}(d)].  While initially it was proposed that the observed polar domain switching in MnSb$_{2}$O$_{6}$ requires a tilted structure, other work based on neutron diffraction has suggested that it is not a requirement.  While in a previous paper we have argued for the existence of an untilted structure, the goodness of fit to the diffraction data was not markedly worse for the tilted case making the results arguably ambiguous.\cite{chan2022106a}  Here we evaluate the stability of the long-wavelength magnetic excitations as a function of tilting the vertical axis of the spin rotation plane given our exchange constants derived from the first moment rule.  We emphasize that the exchange constants derived above from the first moment sum rule depend only on the relative orientation of neighboring spins and is independent of the static magnetic structure being tilted or not.  Given the good description of the data to a symmetric-only exchange model, we test here how stable these excitations are when the static magnetic structure is gradually tilted. 

\begin{figure}[!]
    \begin{center}
    \includegraphics[width=\linewidth]{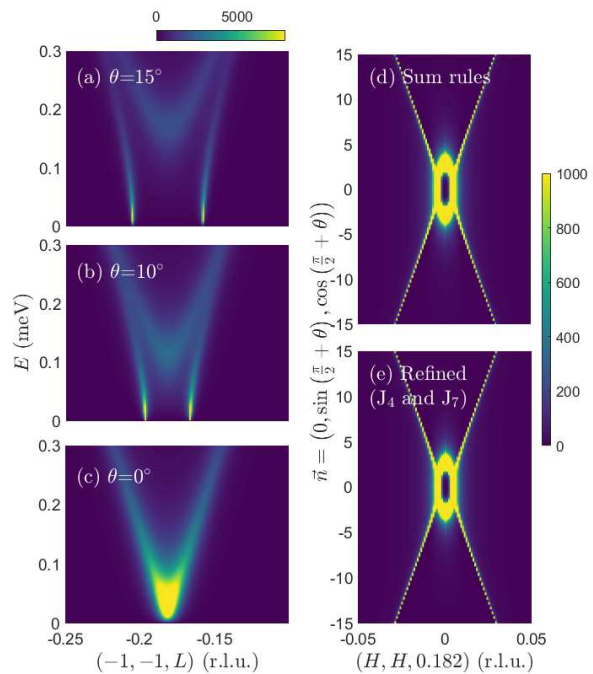}
    \end{center}
    \caption{Calculations investigating the stability of long-wavelength spin-waves as a function of tilting the spin rotation plane away from the $c$-axis.  Calculations of the neutron response for tilts of $\theta=15^{\circ}$ (a), 10$^{\circ}$ (b), and 0$^{\circ}$ (c) are displayed with low-energy, long-wavelength excitations only stable for tilts of $\theta \sim 0^{\circ}$.  This is further illustrated in panels (d)-(e) that display the response at low energies as a function of tilt-angle of the spin rotation plane away from the $c$-axis. We emphasize that these calculations are done for a magnetic Hamiltonian with \textit{symmetric-only} exchange constants.  No anisotropic terms are included in the magnetic Hamiltonian as discussed in the main text.}
    \label{Stability}
\end{figure}

The Green's function calculation predicts the energy and momentum values of stable harmonic excitations through the imaginary part of the response, given a magnetic ground state and a set of symmetric exchange constants.  In the first moment analysis presented above, the exchange constants are derived based on relative orientation of the magnetic moments, and does not depend on global details like tilting of the overall magnetic structure.  Our Green's function analysis, however, does require this tilting as the magnetic ground state determines the local molecular field on each site.  

Given that the Green's function approach predicts stable harmonic excitations as a function of momentum and energy, in this section we search for stable long-wavelength excitations as a function of tilting of the spin rotation plane given our derived exchange parameters based on the first moment rule. We focus on $L$-scans as calculations of the excitation spectrum along $H$ were found to not noticeably change with tilting the spin rotation plane away from the $c$-axis over the range of 0-15$^{\circ}$.  We note that such $H$-scans were used above to fix one of the exchange parameters and the overall calibration constant to take the data to absolute units of meV.  The two assumptions behind that step, namely the energy value of the top of the single-magnon band and the splitting, are not found to observably change with tilting in our calculations. 

In Fig.~\ref{Stability}, we search for long-wavelength excitations given our sum rule exchange constants as a function of tilting of the vertical main axis of the spin rotation plane away from the $c$-axis at an angle $\theta$. The long-wavelength excitations ($q \rightarrow 0$) are calculated for several tilt angles and shown in Fig.~\ref{Stability}(a)-(c), based on the set of parameters derived from the sum rule analysis. Given that the sum rules and the fixing of the value of $J_{3}$ described above is independent of the tilting of the static magnetic moments, in our stability calculations described here we fix the exchange constants to these determined values and vary the long-range static magnetic structure.   On increased tilting, the exchange parameters derived from sum rules show no stable long-wavelength excitations, indicative that the derived exchange parameters combined with a tilted helicoid is unstable.  This is further displayed in Fig.~\ref{Stability}(d)-(e) which plot calculated constant energy cuts (integrating calculated data below 0.02 meV) as a function of tilting of the cycloid away from the $c$-axis for both the cases of exchange constants derived from sum rules, and refined values discussed above.  In both cases, increased tilting of the helicoid results in unstable long-wavelength excitations.  Based on this analysis, we suggest that the derived exchange constants are consistent with an untilted ($\theta=0$) magnetic structure. However, we emphasize that this analysis is based only a Hamiltonian with \textit{symmetric-only} exchange constants as expected based on the high-spin value of Mn$^{2+}$. We cannot rule out the possibility of small anisotropic or more complex magnetic exchange terms that may arise from the distorted framework surrounding the magnetic ions. In Ref.~\onlinecite{chan2022106a} we have shown with diffraction under magnetic field the possibility to manipulate the spin structure in \mnsb\, and that the appearance of electric polarization does not require a tilted structure as raised in Ref.~\onlinecite{kinoshita2016117}. Therefore, the stability analysis above is consistent with our neutron diffraction analysis.  The elastic scattering outlined in our previous paper and the spin excitations can be modeled and understood in terms of a symmetric-only exchange model on an untilted structure.

\section{Conclusions}

In this paper, we have studied structurally chiral polar magnet \mnsb, with magnetic interactions being described by seven symmetric Heisenberg exchanges in the magnetic Hamiltonian. We have presented a method using the first moment sum rule, and have applied this to extract the exchange constants from multiplexed neutron data. This method only depends on the correlators (angles) between neighboring spins and not the tilting of the overall spin rotation plane.  Using Green's functions on a rotating frame, we have reproduced the spin-wave spectra, which are in good agreement with the measured ones and discussed refined values.  Finally, we investigated the stability of the magnetic structure in terms of long-wavelength magnetic excitations present at low energies and suggest that the pure cycloid is favored in terms of stability given the derived exchange constants from the first moment sum rule.

\begin{acknowledgements}
We would like to thank M. Georgopoulou for helpful discussions. We thank the Carnegie Trust for the Universities of Scotland, the EPSRC, and the STFC for financial support.  S.W.C. was supported by the DOE under Grant No. DOE: DE-FG02-07ER46382.  Access to MACS was provided by the Center for High Resolution Neutron Scattering, a partnership between the National Institute of Standards and Technology and the National Science Foundation under Agreement No. DMR-1508249
\end{acknowledgements}

\appendix

\section{Single-crystal sum rules analysis}

\subsection{Integration methods for first moment}\label{sec:error}

\begin{figure*}[!]
    \begin{center}
    \includegraphics[]{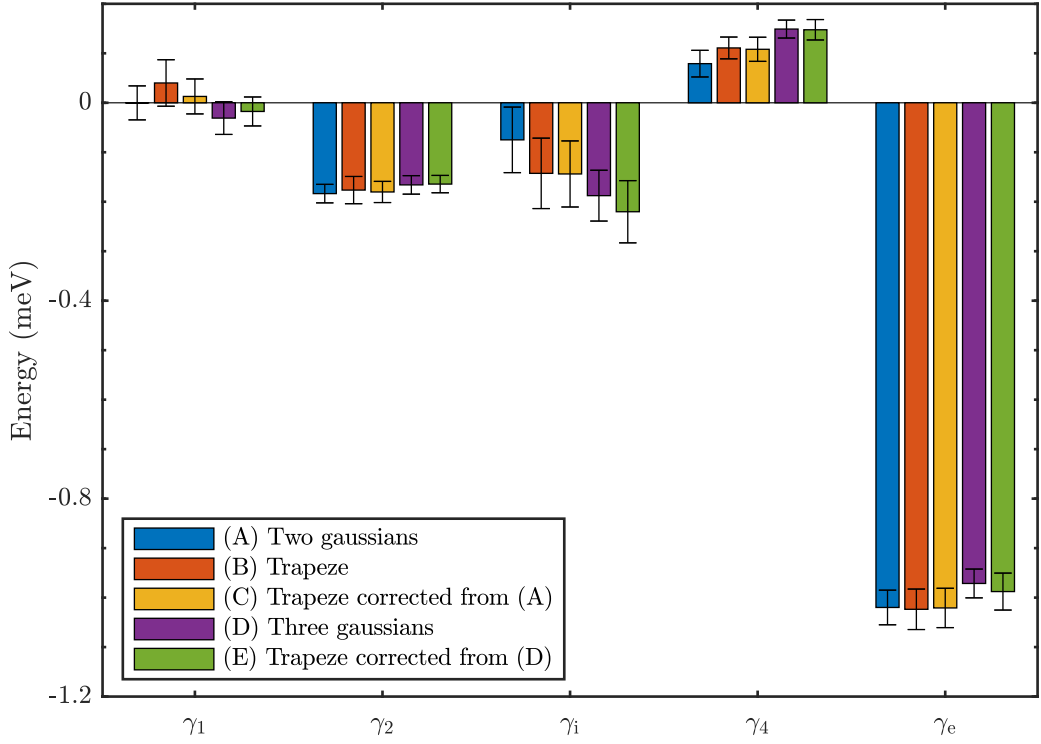}
    \end{center}
    \caption{Fitted parameters for different integrations methods to compute the first moments, from the $E\ut{f}=\SI{2.4}{meV}$ dataset.}
    \label{fig:gamma_methods}
\end{figure*}

As the first moments are computed by a numerical integration, it is important to make sure that the integration methods do not have a significant impact on the results of the analysis. This section outlines five integration methods, and the resulting $\gamma$ parameters are compared in \cref{fig:gamma_methods}, following a $L$-scan analysis on the $\Ef=\SI{2.4}{meV}$ dataset.

In \cref{sec:xtal}, the constant-$\Q$ scans are fitted to two gaussians as shown in \cref{fig:L24_1}, and then the first moments were calculated by numerically integrating with a trapezoidal rule with the background removed from the fit to a two-gaussian model. The results are shown with bars (C). Of course, the first moments can also be computed without removing the background, resulting with bars (B). Then, they can be computed analytically using the fit parameters of the two-gaussian model, shown with the bars (A) in \cref{fig:gamma_methods}. In order to avoid the mixture of elastic scattering and one-magnon scattering, the elastic line can be fitted to a third gaussian, while the actual data above $E=\SI{0.2}{meV}$ are fitted to two gaussians. Then, the first moments can be again calculated analytically with the fitted parameters of these two gaussians in the good energy range. This is shown in bars (D). Finally, the trapezoidal integration can be performed, removing the background from this three-gaussian model, as shown in bars (E). Finally, it can be seen that all the fitted parameters agree within uncertainties. We have rather chosen to adopt trapezoidal integration, removing the background from the two-gaussian fit, to deal with any deviation from a two-mode spin-wave spectrum.
 
\subsection{$H$-scan}\label{sec:annexHscan}

In \cref{sec:xtal}, we have described the first moment sum rule analysis of the single crystal data, by fixing some $H_0$ and calculating the first moment as a function of $L$.
We can perform the same analysis considering $\Q = (H,H,L_0)$ with $H$ varying for a chosen $L_0$ ($H$-scan). For each interaction indexed by spins $i$ and $j$, the corresponding term in the cosine in \cref{eq:firstmomentxtal} can be written now:

\begin{equation}
\Q\cdot\vb*{d}_{ij} = 2\pi H(d_{ij,x}+d_{ij,y})+2\pi L_0d_{ij,z}
\end{equation}

where the distances are expressed in lattice units, and the scattering vector in reciprocal lattice units. Similarly as in \cref{eq:Lscan}, a general formula for the first moment can be derived for a fixed $L_0$, using trigonometric identities:

\begin{align}\label{eq:Hscan}
\langle E\rangle(H,L_0) &= A\ut{i}(L_0)\cos(2\pi \delta_1 H)+... \\
& A\ut{e}(L_0)[\cos(2\pi \delta_2 H)+... \nonumber \\ 
& \cos(2\pi \delta_3 H)+\cos(2\pi \delta_4 H)]+C(L_0) \nonumber
\end{align}

where we have now three functions $A\ut{i}$, $A\ut{e}$ and $C$ which are $L_0$-dependent, expressed by:

\begin{align}
A\ut{i}(L_0) &= \frac{4}{3}[\gamma_1+\gamma_i\cos(2\pi L_0)] \\
A\ut{e}(L_0) &= \frac{4}{3}[\gamma_2+\gamma_e\cos(2\pi L_0)] \\
C(L_0) &= -\frac{2}{3}[2\gamma_1+6\gamma_2+3\gamma_i+3\gamma_4+6\gamma_e]+...\label{eq:H_scan_C}\\
& \frac{2}{3}\cos(2\pi L_0)(\gamma_i+3\gamma_4)\nonumber
\end{align}

\begin{figure*}[!]
    \begin{center}
    \includegraphics[]{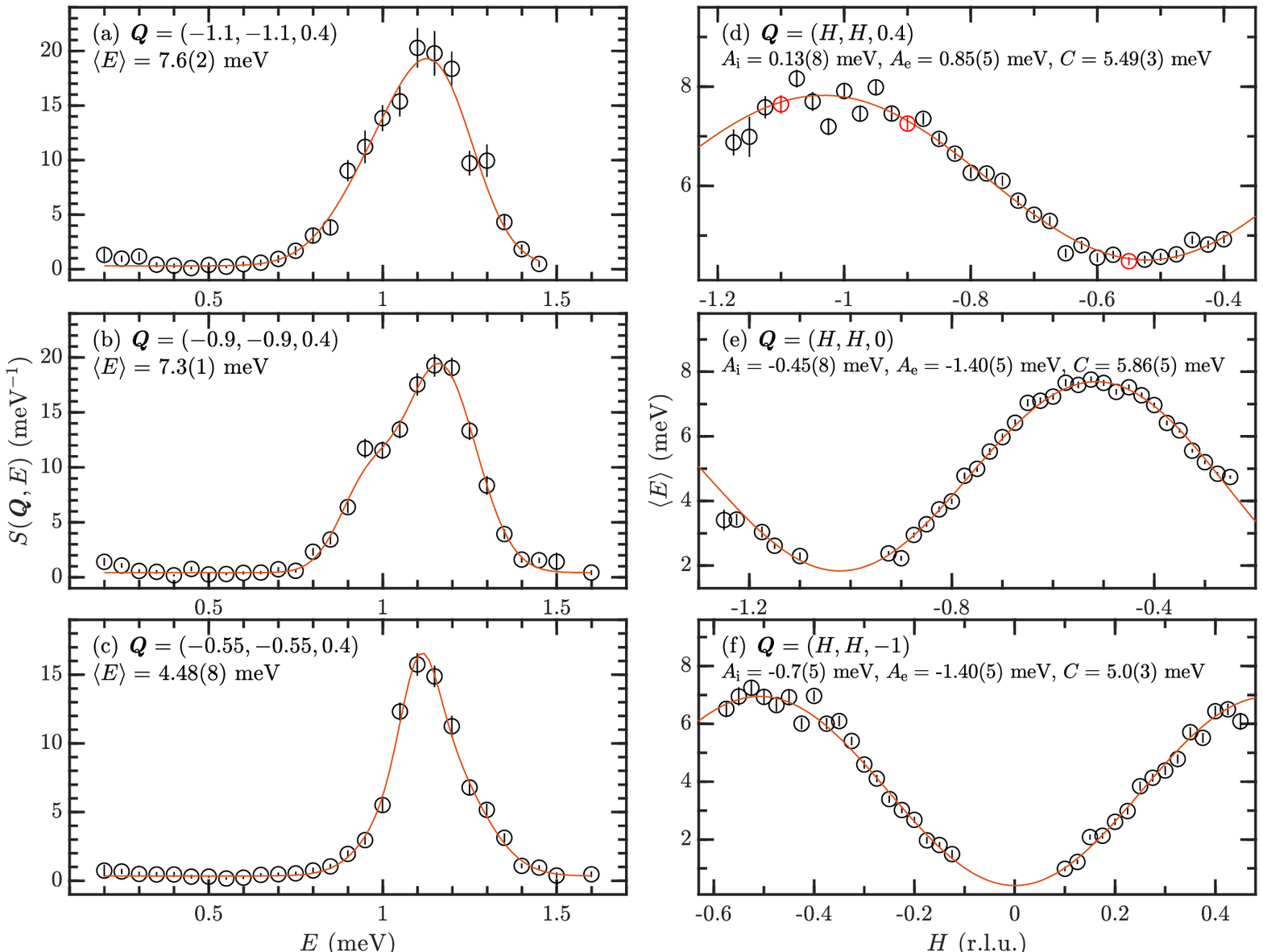}
    \end{center}
	\caption{(a)-(c) Constant-$\Q$ scans for different $\Q=(H, H, L_0=0.4)$. A fit to a double gaussian is shown in red, and the first moment is calculated from trapezoidal integration where the background is removed from the gaussian fit. (d) First moment as a function of $H$ for $L_0=0.4$, fitted to its theoretical expression (red curve). The red data points correspond to the first moments calculated in the cuts plotted in (a)-(c). (e)-(f) First moment as a function of $H$ for (e) $L_0=0$ and (f) $L_0=-1$, fit to theoretical expression in red.}
    \label{fig:H24}
\end{figure*}

\cref{fig:H24}(a)-(c) shows some constant-\Q\ cuts for $(H,H,L_0=0.4)$ and their fit to two gaussians. The first moments are again calculated numerically using trapezoidal integration and the background is removed from the two-gaussian fit. These computed first moments are the red data points in \cref{fig:H24}(d), along with the $H$-dependence of the computed first moment, and the fit to \cref{eq:Hscan}, to extract $A\ut{i}$, $A\ut{e}$ and $C$. This operation is repeated for several $L_0$, as shown in \cref{fig:H24}(e)-(f).

\begin{figure*}[!]
    \begin{center}
    \includegraphics[]{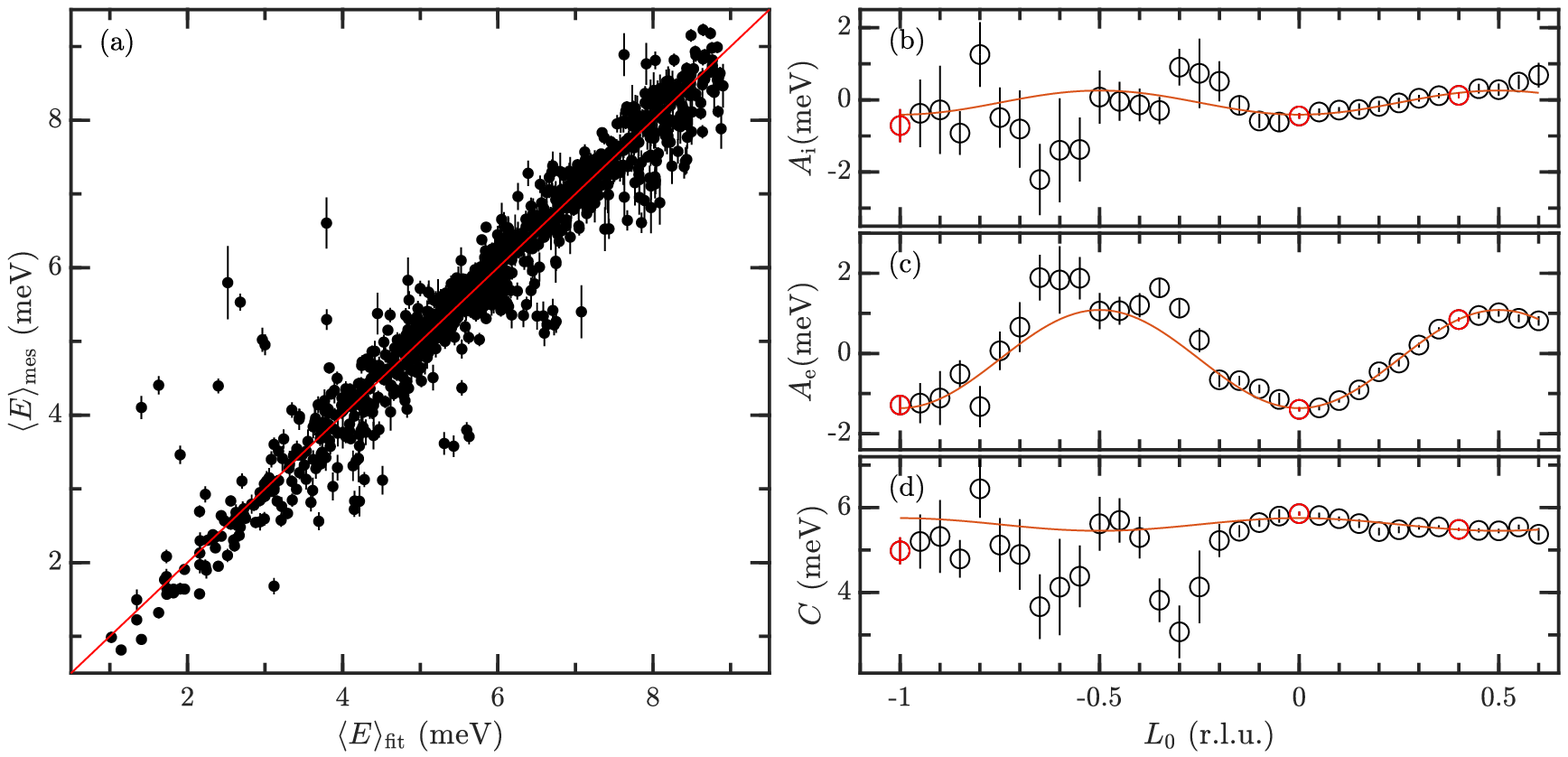}
    \end{center}
    \caption{(a) Measured first moments versus fitted first moments for $H$-scan analysis, for the $E\ut{f}=\SI{2.4}{meV}$ dataset. A total of 999 $\langle E\rangle(\Q)$ were taken into account. (b)-(d) Fitting of coefficients (b) $A\ut{i}$, (c) $A\ut{e}$ and (d) $C$ giving the $\gamma$ parameters. The red data points show the values calculated in \cref{fig:H24}(d)-(f).}
    \label{fig:H24_2}
\end{figure*}

Finally, a total of 999 first moments $\langle E\rangle(\Q)$ are computed for this analysis on this $\Ef=\SI{2.4}{meV}$ dataset, and plotted against the fitted first moments in \cref{fig:H24_2}(a). The $\gamma$ parameters are then obtained by fitting the measured $A\ut{i}$, $A\ut{e}$ and $C$ to their theoretical values, as shown in \cref{fig:H24_2}(b)-(d), where the red data points are the coefficients calculated in \cref{fig:H24}(d)-(f). As for the $L$-scan analysis, some remaining background can be included in the computation of $C$. For this reason, the $L_0$-independent part of \cref{eq:H_scan_C}, which corresponds to an overall constant to the first moment sum rule, is not used to get the $\gamma$ parameters and hence the exchange constants $J_{i}$.

\subsection{Second dataset results}\label{sec:annexsecond}

\begin{figure*}[!]
    \begin{center}
    \includegraphics[]{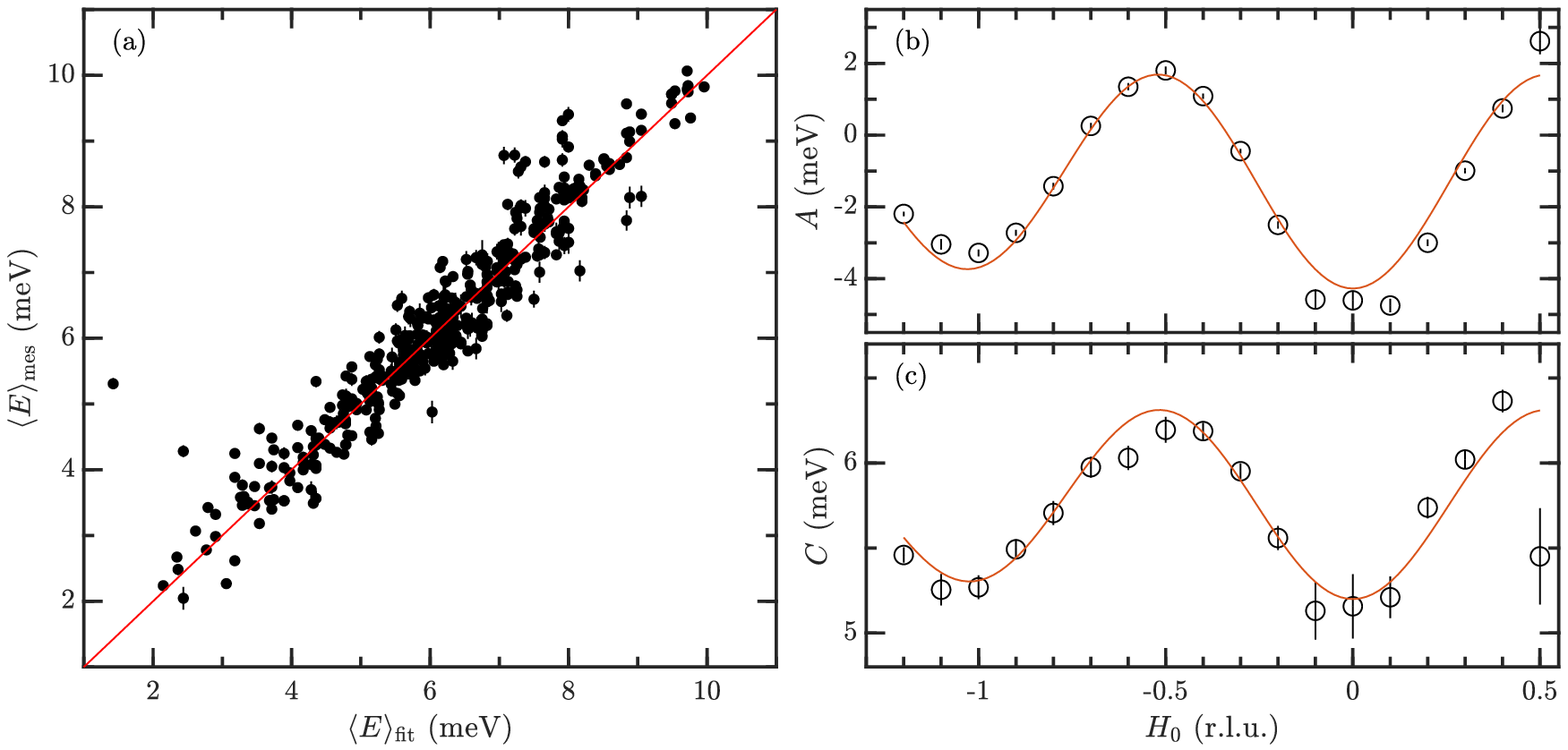}
    \end{center}
    \caption{(a) Measured first moments versus fitted first moments for $L$-scan analysis, for the $E\ut{f}=\SI{3.7}{meV}$ dataset. A total of 469 $\langle E\rangle(\Q)$ were taken into account. (b)-(c) Fitting of coefficients (b) $A$ and (c) $C$ giving the $\gamma$ parameters.}
    \label{fig:L37}
\end{figure*}

\begin{figure*}[!]
    \begin{center}
    \includegraphics[]{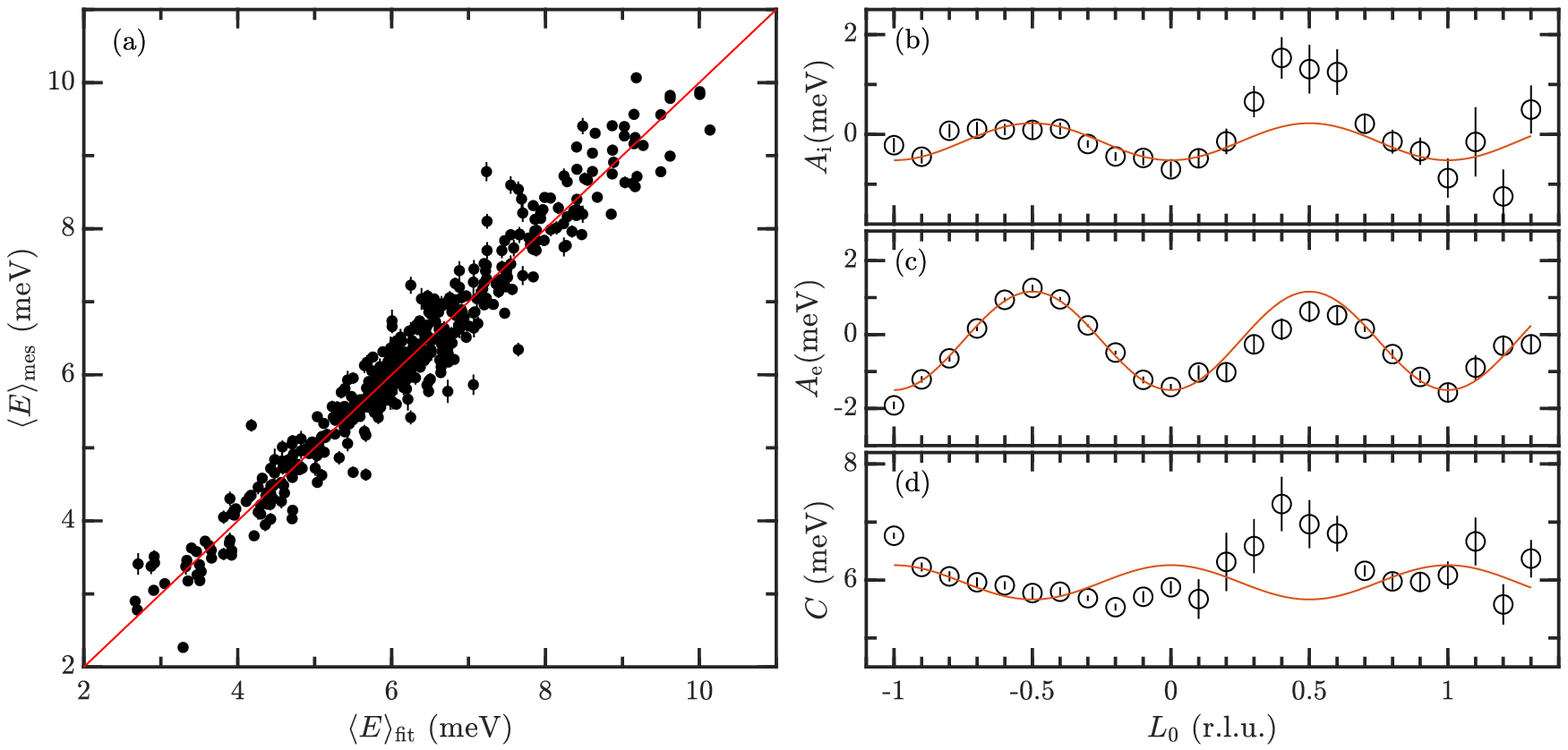}
    \end{center}
    \caption{(a) Measured first moments versus fitted first moments for $H$-scan analysis, for the $E\ut{f}=\SI{3.7}{meV}$ dataset. A total of 487 $\langle E\rangle(\Q)$ were taken into account. (b)-(d) Fitting of coefficients (b) $A\ut{i}$, (c) $A\ut{e}$ and (d) $C$ giving the $\gamma$ parameters.}
    \label{fig:H37}
\end{figure*}

The single crystal first moment sum rule analysis was repeated on the second dataset measured on MACS with $\Ef=\SI{3.7}{meV}$. The results of the $L$-scan (469 computed first moments) and $H$-scan (487 computed first moments) analyses are respectively shown in \cref{fig:L37} and \cref{fig:H37}.

\subsection{Parameters for Figure 14}

\begin{table*}[!]
    \centering
    \begin{tabular}{cccccccc}\hline\hline
          & $J_1$ & $J_2$ & $J_3$ & $J_4$ & $J_5$ & $J_6$ & $J_7$\\ \hline
         Fig. 14 (b) & 0.0988 & 0.2859 & \color{blue} 0.1500 & 0.3491 & \color{blue} -0.1011 & \color{blue} 1.0222 & \color{blue} 0.1161\\ 
         Fig. 14 (c) & 0.0988 & 0.2859 & \color{blue} 0.2000 & 0.3491 & \color{blue} -0.0155 & \color{blue} 0.9972 & \color{blue} 0.0732\\ 
         Fig. 14 (d) & 0.0988 & 0.2859 & \color{blue} 0.2500 & 0.3491 & \color{blue} 0.0702 & \color{blue} 0.9722 & \color{blue} 0.0304\\\hline \hline
    \end{tabular}
    \caption{The parameters for the calculations performed in \textsc{SpinW} displayed in \cref{fig:guessJ3}(b)-(d).  Parameters varied for the three calculations are highlighted in \color{blue} blue.}
    \label{tab:J_param}
\end{table*}

The sum rule analysis had an ambiguity in the set of equations resulting from the fact that several exchange constants corresponded to the same bond distance.  We therefore needed to fix one exchange constant through a comparison to the single crystal dispersion as discussed in the main text.  This qualitative analysis is described in Fig. \ref{fig:guessJ3}.  The parameters for the calculations are listed in Table \ref{tab:J_param}.


\begin{thebibliography}{59}%
\makeatletter
\providecommand \@ifxundefined [1]{%
 \@ifx{#1\undefined}
}%
\providecommand \@ifnum [1]{%
 \ifnum #1\expandafter \@firstoftwo
 \else \expandafter \@secondoftwo
 \fi
}%
\providecommand \@ifx [1]{%
 \ifx #1\expandafter \@firstoftwo
 \else \expandafter \@secondoftwo
 \fi
}%
\providecommand \natexlab [1]{#1}%
\providecommand \enquote  [1]{``#1''}%
\providecommand \bibnamefont  [1]{#1}%
\providecommand \bibfnamefont [1]{#1}%
\providecommand \citenamefont [1]{#1}%
\providecommand \href@noop [0]{\@secondoftwo}%
\providecommand \href [0]{\begingroup \@sanitize@url \@href}%
\providecommand \@href[1]{\@@startlink{#1}\@@href}%
\providecommand \@@href[1]{\endgroup#1\@@endlink}%
\providecommand \@sanitize@url [0]{\catcode `\\12\catcode `\$12\catcode
  `\&12\catcode `\#12\catcode `\^12\catcode `\_12\catcode `\%12\relax}%
\providecommand \@@startlink[1]{}%
\providecommand \@@endlink[0]{}%
\providecommand \url  [0]{\begingroup\@sanitize@url \@url }%
\providecommand \@url [1]{\endgroup\@href {#1}{\urlprefix }}%
\providecommand \urlprefix  [0]{URL }%
\providecommand \Eprint [0]{\href }%
\providecommand \doibase [0]{http://dx.doi.org/}%
\providecommand \selectlanguage [0]{\@gobble}%
\providecommand \bibinfo  [0]{\@secondoftwo}%
\providecommand \bibfield  [0]{\@secondoftwo}%
\providecommand \translation [1]{[#1]}%
\providecommand \BibitemOpen [0]{}%
\providecommand \bibitemStop [0]{}%
\providecommand \bibitemNoStop [0]{.\EOS\space}%
\providecommand \EOS [0]{\spacefactor3000\relax}%
\providecommand \BibitemShut  [1]{\csname bibitem#1\endcsname}%
\let\auto@bib@innerbib\@empty
\bibitem [{\citenamefont {Cheon}\ \emph {et~al.}(2018)\citenamefont {Cheon},
  \citenamefont {Lee},\ and\ \citenamefont {Cheong}}]{cheon201898}%
  \BibitemOpen
  \bibfield  {author} {\bibinfo {author} {\bibfnamefont {S.}~\bibnamefont
  {Cheon}}, \bibinfo {author} {\bibfnamefont {H.-W.}\ \bibnamefont {Lee}}, \
  and\ \bibinfo {author} {\bibfnamefont {S.-W.}\ \bibnamefont {Cheong}},\
  }\href {\doibase 10.1103/PhysRevB.98.184405} {\bibfield  {journal} {\bibinfo
  {journal} {Phys. Rev. B}\ }\textbf {\bibinfo {volume} {98}},\ \bibinfo
  {pages} {184405} (\bibinfo {year} {2018})}\BibitemShut {NoStop}%
\bibitem [{\citenamefont {Stock}\ \emph {et~al.}(2019)\citenamefont {Stock},
  \citenamefont {Johnson}, \citenamefont {{Giles-Donovan}}, \citenamefont
  {Songvilay}, \citenamefont {{Rodriguez-Rivera}}, \citenamefont {Lee},
  \citenamefont {Xu}, \citenamefont {Radaelli}, \citenamefont {Chapon},
  \citenamefont {Bombardi}, \citenamefont {Cochran}, \citenamefont
  {Niedermayer}, \citenamefont {Schneidewind}, \citenamefont {Husges},
  \citenamefont {Lu}, \citenamefont {Meng},\ and\ \citenamefont
  {Cheong}}]{stock2019100}%
  \BibitemOpen
  \bibfield  {author} {\bibinfo {author} {\bibfnamefont {C.}~\bibnamefont
  {Stock}}, \bibinfo {author} {\bibfnamefont {R.~D.}\ \bibnamefont {Johnson}},
  \bibinfo {author} {\bibfnamefont {N.}~\bibnamefont {{Giles-Donovan}}},
  \bibinfo {author} {\bibfnamefont {M.}~\bibnamefont {Songvilay}}, \bibinfo
  {author} {\bibfnamefont {J.~A.}\ \bibnamefont {{Rodriguez-Rivera}}}, \bibinfo
  {author} {\bibfnamefont {N.}~\bibnamefont {Lee}}, \bibinfo {author}
  {\bibfnamefont {X.}~\bibnamefont {Xu}}, \bibinfo {author} {\bibfnamefont
  {P.~G.}\ \bibnamefont {Radaelli}}, \bibinfo {author} {\bibfnamefont {L.~C.}\
  \bibnamefont {Chapon}}, \bibinfo {author} {\bibfnamefont {A.}~\bibnamefont
  {Bombardi}}, \bibinfo {author} {\bibfnamefont {S.}~\bibnamefont {Cochran}},
  \bibinfo {author} {\bibfnamefont {C.}~\bibnamefont {Niedermayer}}, \bibinfo
  {author} {\bibfnamefont {A.}~\bibnamefont {Schneidewind}}, \bibinfo {author}
  {\bibfnamefont {Z.}~\bibnamefont {Husges}}, \bibinfo {author} {\bibfnamefont
  {Z.}~\bibnamefont {Lu}}, \bibinfo {author} {\bibfnamefont {S.}~\bibnamefont
  {Meng}}, \ and\ \bibinfo {author} {\bibfnamefont {S.-W.}\ \bibnamefont
  {Cheong}},\ }\href {\doibase 10.1103/PhysRevB.100.134429} {\bibfield
  {journal} {\bibinfo  {journal} {Phys. Rev. B}\ }\textbf {\bibinfo {volume}
  {100}},\ \bibinfo {pages} {134429} (\bibinfo {year} {2019})}\BibitemShut
  {NoStop}%
\bibitem [{\citenamefont {Cheong}\ and\ \citenamefont
  {Mostovoy}(2007)}]{cheong20076}%
  \BibitemOpen
  \bibfield  {author} {\bibinfo {author} {\bibfnamefont {S.-W.}\ \bibnamefont
  {Cheong}}\ and\ \bibinfo {author} {\bibfnamefont {M.}~\bibnamefont
  {Mostovoy}},\ }\href {\doibase 10.1038/nmat1804} {\bibfield  {journal}
  {\bibinfo  {journal} {Nat. Mater.}\ }\textbf {\bibinfo {volume} {6}},\
  \bibinfo {pages} {13} (\bibinfo {year} {2007})}\BibitemShut {NoStop}%
\bibitem [{\citenamefont {Eerenstein}\ \emph {et~al.}(2006)\citenamefont
  {Eerenstein}, \citenamefont {Mathur},\ and\ \citenamefont
  {Scott}}]{Eerenstein06:442}%
  \BibitemOpen
  \bibfield  {author} {\bibinfo {author} {\bibfnamefont {W.}~\bibnamefont
  {Eerenstein}}, \bibinfo {author} {\bibfnamefont {N.~D.}\ \bibnamefont
  {Mathur}}, \ and\ \bibinfo {author} {\bibfnamefont {J.~F.}\ \bibnamefont
  {Scott}},\ }\href {\doibase 10.1038/nature05023} {\bibfield  {journal}
  {\bibinfo  {journal} {Nature}\ }\textbf {\bibinfo {volume} {6442}},\ \bibinfo
  {pages} {759} (\bibinfo {year} {2006})}\BibitemShut {NoStop}%
\bibitem [{\citenamefont {Fiebig}\ \emph {et~al.}(2016)\citenamefont {Fiebig},
  \citenamefont {Lottermoser}, \citenamefont {Meier},\ and\ \citenamefont
  {Trassin}}]{fiebig16:1}%
  \BibitemOpen
  \bibfield  {author} {\bibinfo {author} {\bibfnamefont {M.}~\bibnamefont
  {Fiebig}}, \bibinfo {author} {\bibfnamefont {T.}~\bibnamefont {Lottermoser}},
  \bibinfo {author} {\bibfnamefont {D.}~\bibnamefont {Meier}}, \ and\ \bibinfo
  {author} {\bibfnamefont {M.}~\bibnamefont {Trassin}},\ }\href {\doibase
  10.1038/natrevmats.2016.46} {\bibfield  {journal} {\bibinfo  {journal} {Nat.
  Rev. Mater.}\ }\textbf {\bibinfo {volume} {1}},\ \bibinfo {pages} {1}
  (\bibinfo {year} {2016})}\BibitemShut {NoStop}%
\bibitem [{\citenamefont {Spaldin}\ and\ \citenamefont
  {Fiebig}(2005)}]{spaldin2005309}%
  \BibitemOpen
  \bibfield  {author} {\bibinfo {author} {\bibfnamefont {N.~A.}\ \bibnamefont
  {Spaldin}}\ and\ \bibinfo {author} {\bibfnamefont {M.}~\bibnamefont
  {Fiebig}},\ }\href {\doibase 10.1126/science.1113357} {\bibfield  {journal}
  {\bibinfo  {journal} {Science}\ }\textbf {\bibinfo {volume} {309}},\ \bibinfo
  {pages} {391} (\bibinfo {year} {2005})}\BibitemShut {NoStop}%
\bibitem [{\citenamefont {Spaldin}\ \emph {et~al.}(2010)\citenamefont
  {Spaldin}, \citenamefont {Cheong},\ and\ \citenamefont
  {Ramesh}}]{spaldin201063}%
  \BibitemOpen
  \bibfield  {author} {\bibinfo {author} {\bibfnamefont {N.~A.}\ \bibnamefont
  {Spaldin}}, \bibinfo {author} {\bibfnamefont {S.-W.}\ \bibnamefont {Cheong}},
  \ and\ \bibinfo {author} {\bibfnamefont {R.}~\bibnamefont {Ramesh}},\
  }\href@noop {} {\bibfield  {journal} {\bibinfo  {journal} {Physics Today}\
  }\textbf {\bibinfo {volume} {63}},\ \bibinfo {pages} {38} (\bibinfo {year}
  {2010})}\BibitemShut {NoStop}%
\bibitem [{\citenamefont {Mostovoy}(2006)}]{mostovoy200696}%
  \BibitemOpen
  \bibfield  {author} {\bibinfo {author} {\bibfnamefont {M.}~\bibnamefont
  {Mostovoy}},\ }\href {\doibase 10.1103/PhysRevLett.96.067601} {\bibfield
  {journal} {\bibinfo  {journal} {Phys. Rev. Lett.}\ }\textbf {\bibinfo
  {volume} {96}},\ \bibinfo {pages} {067601} (\bibinfo {year}
  {2006})}\BibitemShut {NoStop}%
\bibitem [{\citenamefont {Johnson}\ and\ \citenamefont
  {Radaelli}(2014)}]{johnson201444}%
  \BibitemOpen
  \bibfield  {author} {\bibinfo {author} {\bibfnamefont {R.~D.}\ \bibnamefont
  {Johnson}}\ and\ \bibinfo {author} {\bibfnamefont {P.~G.}\ \bibnamefont
  {Radaelli}},\ }\href {\doibase 10.1146/annurev-matsci-070813-113524}
  {\bibfield  {journal} {\bibinfo  {journal} {Annu. Rev. Mater. Res.}\ }\textbf
  {\bibinfo {volume} {44}},\ \bibinfo {pages} {269} (\bibinfo {year}
  {2014})}\BibitemShut {NoStop}%
\bibitem [{\citenamefont {Marty}\ \emph {et~al.}(2008)\citenamefont {Marty},
  \citenamefont {Simonet}, \citenamefont {Ressouche}, \citenamefont {Ballou},
  \citenamefont {Lejay},\ and\ \citenamefont {Bordet}}]{marty2008101}%
  \BibitemOpen
  \bibfield  {author} {\bibinfo {author} {\bibfnamefont {K.}~\bibnamefont
  {Marty}}, \bibinfo {author} {\bibfnamefont {V.}~\bibnamefont {Simonet}},
  \bibinfo {author} {\bibfnamefont {E.}~\bibnamefont {Ressouche}}, \bibinfo
  {author} {\bibfnamefont {R.}~\bibnamefont {Ballou}}, \bibinfo {author}
  {\bibfnamefont {P.}~\bibnamefont {Lejay}}, \ and\ \bibinfo {author}
  {\bibfnamefont {P.}~\bibnamefont {Bordet}},\ }\href {\doibase
  10.1103/PhysRevLett.101.247201} {\bibfield  {journal} {\bibinfo  {journal}
  {Phys. Rev. Lett.}\ }\textbf {\bibinfo {volume} {101}},\ \bibinfo {pages}
  {247201} (\bibinfo {year} {2008})}\BibitemShut {NoStop}%
\bibitem [{\citenamefont {Marty}\ \emph {et~al.}(2010)\citenamefont {Marty},
  \citenamefont {Bordet}, \citenamefont {Simonet}, \citenamefont {Loire},
  \citenamefont {Ballou}, \citenamefont {Darie}, \citenamefont {Kljun},
  \citenamefont {Bonville}, \citenamefont {Isnard}, \citenamefont {Lejay},
  \citenamefont {Zawilski},\ and\ \citenamefont {Simon}}]{marty10:81}%
  \BibitemOpen
  \bibfield  {author} {\bibinfo {author} {\bibfnamefont {K.}~\bibnamefont
  {Marty}}, \bibinfo {author} {\bibfnamefont {P.}~\bibnamefont {Bordet}},
  \bibinfo {author} {\bibfnamefont {V.}~\bibnamefont {Simonet}}, \bibinfo
  {author} {\bibfnamefont {M.}~\bibnamefont {Loire}}, \bibinfo {author}
  {\bibfnamefont {R.}~\bibnamefont {Ballou}}, \bibinfo {author} {\bibfnamefont
  {C.}~\bibnamefont {Darie}}, \bibinfo {author} {\bibfnamefont
  {J.}~\bibnamefont {Kljun}}, \bibinfo {author} {\bibfnamefont
  {P.}~\bibnamefont {Bonville}}, \bibinfo {author} {\bibfnamefont
  {O.}~\bibnamefont {Isnard}}, \bibinfo {author} {\bibfnamefont
  {P.}~\bibnamefont {Lejay}}, \bibinfo {author} {\bibfnamefont
  {B.}~\bibnamefont {Zawilski}}, \ and\ \bibinfo {author} {\bibfnamefont
  {C.}~\bibnamefont {Simon}},\ }\href {\doibase 10.1103/PhysRevB.81.054416}
  {\bibfield  {journal} {\bibinfo  {journal} {Phys. Rev. B}\ }\textbf {\bibinfo
  {volume} {81}},\ \bibinfo {pages} {054416} (\bibinfo {year}
  {2010})}\BibitemShut {NoStop}%
\bibitem [{\citenamefont {Loire}\ \emph {et~al.}(2011)\citenamefont {Loire},
  \citenamefont {Simonet}, \citenamefont {Petit}, \citenamefont {Marty},
  \citenamefont {Bordet}, \citenamefont {Lejay}, \citenamefont {Ollivier},
  \citenamefont {Enderle}, \citenamefont {Steffens}, \citenamefont {Ressouche},
  \citenamefont {Zorko},\ and\ \citenamefont {Ballou}}]{loire2011106}%
  \BibitemOpen
  \bibfield  {author} {\bibinfo {author} {\bibfnamefont {M.}~\bibnamefont
  {Loire}}, \bibinfo {author} {\bibfnamefont {V.}~\bibnamefont {Simonet}},
  \bibinfo {author} {\bibfnamefont {S.}~\bibnamefont {Petit}}, \bibinfo
  {author} {\bibfnamefont {K.}~\bibnamefont {Marty}}, \bibinfo {author}
  {\bibfnamefont {P.}~\bibnamefont {Bordet}}, \bibinfo {author} {\bibfnamefont
  {P.}~\bibnamefont {Lejay}}, \bibinfo {author} {\bibfnamefont
  {J.}~\bibnamefont {Ollivier}}, \bibinfo {author} {\bibfnamefont
  {M.}~\bibnamefont {Enderle}}, \bibinfo {author} {\bibfnamefont
  {P.}~\bibnamefont {Steffens}}, \bibinfo {author} {\bibfnamefont
  {E.}~\bibnamefont {Ressouche}}, \bibinfo {author} {\bibfnamefont
  {A.}~\bibnamefont {Zorko}}, \ and\ \bibinfo {author} {\bibfnamefont
  {R.}~\bibnamefont {Ballou}},\ }\href {\doibase
  10.1103/PhysRevLett.106.207201} {\bibfield  {journal} {\bibinfo  {journal}
  {Phys. Rev. Lett.}\ }\textbf {\bibinfo {volume} {106}},\ \bibinfo {pages}
  {207201} (\bibinfo {year} {2011})}\BibitemShut {NoStop}%
\bibitem [{\citenamefont {Stock}\ \emph {et~al.}(2011)\citenamefont {Stock},
  \citenamefont {Chapon}, \citenamefont {Schneidewind}, \citenamefont {Su},
  \citenamefont {Radaelli}, \citenamefont {McMorrow}, \citenamefont {Bombardi},
  \citenamefont {Lee},\ and\ \citenamefont {Cheong}}]{stock201183}%
  \BibitemOpen
  \bibfield  {author} {\bibinfo {author} {\bibfnamefont {C.}~\bibnamefont
  {Stock}}, \bibinfo {author} {\bibfnamefont {L.~C.}\ \bibnamefont {Chapon}},
  \bibinfo {author} {\bibfnamefont {A.}~\bibnamefont {Schneidewind}}, \bibinfo
  {author} {\bibfnamefont {Y.}~\bibnamefont {Su}}, \bibinfo {author}
  {\bibfnamefont {P.~G.}\ \bibnamefont {Radaelli}}, \bibinfo {author}
  {\bibfnamefont {D.~F.}\ \bibnamefont {McMorrow}}, \bibinfo {author}
  {\bibfnamefont {A.}~\bibnamefont {Bombardi}}, \bibinfo {author}
  {\bibfnamefont {N.}~\bibnamefont {Lee}}, \ and\ \bibinfo {author}
  {\bibfnamefont {S.-W.}\ \bibnamefont {Cheong}},\ }\href {\doibase
  10.1103/PhysRevB.83.104426} {\bibfield  {journal} {\bibinfo  {journal} {Phys.
  Rev. B}\ }\textbf {\bibinfo {volume} {83}},\ \bibinfo {pages} {104426}
  (\bibinfo {year} {2011})}\BibitemShut {NoStop}%
\bibitem [{\citenamefont {Reimers}\ and\ \citenamefont
  {Greedan}(1989)}]{reimers1989}%
  \BibitemOpen
  \bibfield  {author} {\bibinfo {author} {\bibfnamefont {J.~N.}\ \bibnamefont
  {Reimers}}\ and\ \bibinfo {author} {\bibfnamefont {J.~E.}\ \bibnamefont
  {Greedan}},\ }\href {\doibase 10.1016/0022-4596(89)90273-9} {\bibfield
  {journal} {\bibinfo  {journal} {J. Solid State Chem.}\ }\textbf {\bibinfo
  {volume} {79}},\ \bibinfo {pages} {263} (\bibinfo {year} {1989})}\BibitemShut
  {NoStop}%
\bibitem [{\citenamefont {Johnson}\ \emph {et~al.}(2013)\citenamefont
  {Johnson}, \citenamefont {Cao}, \citenamefont {Chapon}, \citenamefont
  {Fabrizi}, \citenamefont {Perks}, \citenamefont {Manuel}, \citenamefont
  {Yang}, \citenamefont {Oh}, \citenamefont {Cheong},\ and\ \citenamefont
  {Radaelli}}]{johnson2013111}%
  \BibitemOpen
  \bibfield  {author} {\bibinfo {author} {\bibfnamefont {R.~D.}\ \bibnamefont
  {Johnson}}, \bibinfo {author} {\bibfnamefont {K.}~\bibnamefont {Cao}},
  \bibinfo {author} {\bibfnamefont {L.~C.}\ \bibnamefont {Chapon}}, \bibinfo
  {author} {\bibfnamefont {F.}~\bibnamefont {Fabrizi}}, \bibinfo {author}
  {\bibfnamefont {N.}~\bibnamefont {Perks}}, \bibinfo {author} {\bibfnamefont
  {P.}~\bibnamefont {Manuel}}, \bibinfo {author} {\bibfnamefont {J.~J.}\
  \bibnamefont {Yang}}, \bibinfo {author} {\bibfnamefont {Y.~S.}\ \bibnamefont
  {Oh}}, \bibinfo {author} {\bibfnamefont {S.-W.}\ \bibnamefont {Cheong}}, \
  and\ \bibinfo {author} {\bibfnamefont {P.~G.}\ \bibnamefont {Radaelli}},\
  }\href {\doibase 10.1103/PhysRevLett.111.017202} {\bibfield  {journal}
  {\bibinfo  {journal} {Phys. Rev. Lett.}\ }\textbf {\bibinfo {volume} {111}},\
  \bibinfo {pages} {017202} (\bibinfo {year} {2013})}\BibitemShut {NoStop}%
\bibitem [{\citenamefont {Kinoshita}\ \emph {et~al.}(2016)\citenamefont
  {Kinoshita}, \citenamefont {Seki}, \citenamefont {Sato}, \citenamefont
  {Nambu}, \citenamefont {Hong}, \citenamefont {Matsuda}, \citenamefont {Cao},
  \citenamefont {Ishiwata},\ and\ \citenamefont {Tokura}}]{kinoshita2016117}%
  \BibitemOpen
  \bibfield  {author} {\bibinfo {author} {\bibfnamefont {M.}~\bibnamefont
  {Kinoshita}}, \bibinfo {author} {\bibfnamefont {S.}~\bibnamefont {Seki}},
  \bibinfo {author} {\bibfnamefont {T.~J.}\ \bibnamefont {Sato}}, \bibinfo
  {author} {\bibfnamefont {Y.}~\bibnamefont {Nambu}}, \bibinfo {author}
  {\bibfnamefont {T.}~\bibnamefont {Hong}}, \bibinfo {author} {\bibfnamefont
  {M.}~\bibnamefont {Matsuda}}, \bibinfo {author} {\bibfnamefont {H.~B.}\
  \bibnamefont {Cao}}, \bibinfo {author} {\bibfnamefont {S.}~\bibnamefont
  {Ishiwata}}, \ and\ \bibinfo {author} {\bibfnamefont {Y.}~\bibnamefont
  {Tokura}},\ }\href {\doibase 10.1103/PhysRevLett.117.047201} {\bibfield
  {journal} {\bibinfo  {journal} {Phys. Rev. Lett.}\ }\textbf {\bibinfo
  {volume} {117}},\ \bibinfo {pages} {047201} (\bibinfo {year}
  {2016})}\BibitemShut {NoStop}%
\bibitem [{\citenamefont {Werner}\ \emph {et~al.}(2016)\citenamefont {Werner},
  \citenamefont {Koo}, \citenamefont {Klingeler}, \citenamefont {Vasiliev},
  \citenamefont {Ovchenkov}, \citenamefont {Polovkova}, \citenamefont
  {Raganyan},\ and\ \citenamefont {Zvereva}}]{werner201694}%
  \BibitemOpen
  \bibfield  {author} {\bibinfo {author} {\bibfnamefont {J.}~\bibnamefont
  {Werner}}, \bibinfo {author} {\bibfnamefont {C.}~\bibnamefont {Koo}},
  \bibinfo {author} {\bibfnamefont {R.}~\bibnamefont {Klingeler}}, \bibinfo
  {author} {\bibfnamefont {A.~N.}\ \bibnamefont {Vasiliev}}, \bibinfo {author}
  {\bibfnamefont {Y.~A.}\ \bibnamefont {Ovchenkov}}, \bibinfo {author}
  {\bibfnamefont {A.~S.}\ \bibnamefont {Polovkova}}, \bibinfo {author}
  {\bibfnamefont {G.~V.}\ \bibnamefont {Raganyan}}, \ and\ \bibinfo {author}
  {\bibfnamefont {E.~A.}\ \bibnamefont {Zvereva}},\ }\href {\doibase
  10.1103/PhysRevB.94.104408} {\bibfield  {journal} {\bibinfo  {journal} {Phys.
  Rev. B}\ }\textbf {\bibinfo {volume} {94}},\ \bibinfo {pages} {104408}
  (\bibinfo {year} {2016})}\BibitemShut {NoStop}%
\bibitem [{\citenamefont {Chan}\ \emph {et~al.}(2022)\citenamefont {Chan},
  \citenamefont {Pásztorová}, \citenamefont {Johnson}, \citenamefont
  {Songvilay}, \citenamefont {Downie}, \citenamefont {Bos}, \citenamefont
  {Fabelo}, \citenamefont {Ritter}, \citenamefont {Beauvois}, \citenamefont
  {Niedermayer}, \citenamefont {Cheong}, \citenamefont {Qureshi},\ and\
  \citenamefont {Stock}}]{chan2022106a}%
  \BibitemOpen
  \bibfield  {author} {\bibinfo {author} {\bibfnamefont {E.}~\bibnamefont
  {Chan}}, \bibinfo {author} {\bibfnamefont {J.}~\bibnamefont {Pásztorová}},
  \bibinfo {author} {\bibfnamefont {R.~D.}\ \bibnamefont {Johnson}}, \bibinfo
  {author} {\bibfnamefont {M.}~\bibnamefont {Songvilay}}, \bibinfo {author}
  {\bibfnamefont {R.~A.}\ \bibnamefont {Downie}}, \bibinfo {author}
  {\bibfnamefont {J.-W.~G.}\ \bibnamefont {Bos}}, \bibinfo {author}
  {\bibfnamefont {O.}~\bibnamefont {Fabelo}}, \bibinfo {author} {\bibfnamefont
  {C.}~\bibnamefont {Ritter}}, \bibinfo {author} {\bibfnamefont
  {K.}~\bibnamefont {Beauvois}}, \bibinfo {author} {\bibfnamefont
  {C.}~\bibnamefont {Niedermayer}}, \bibinfo {author} {\bibfnamefont {S.-W.}\
  \bibnamefont {Cheong}}, \bibinfo {author} {\bibfnamefont {N.}~\bibnamefont
  {Qureshi}}, \ and\ \bibinfo {author} {\bibfnamefont {C.}~\bibnamefont
  {Stock}},\ }\href {\doibase 10.1103/PhysRevB.106.064403} {\bibfield
  {journal} {\bibinfo  {journal} {Phys. Rev. B}\ }\textbf {\bibinfo {volume}
  {106}},\ \bibinfo {pages} {064403} (\bibinfo {year} {2022})}\BibitemShut
  {NoStop}%
\bibitem [{\citenamefont {Yosida}(1996)}]{Yosida:book}%
  \BibitemOpen
  \bibfield  {author} {\bibinfo {author} {\bibfnamefont {K.}~\bibnamefont
  {Yosida}},\ }\href@noop {} {\emph {\bibinfo {title} {Theory of Magnetism}}}\
  (\bibinfo  {publisher} {Springer},\ \bibinfo {address} {New York},\ \bibinfo
  {year} {1996})\BibitemShut {NoStop}%
\bibitem [{\citenamefont {Hohenberg}\ and\ \citenamefont
  {Brinkman}(1974)}]{hohenberg197410}%
  \BibitemOpen
  \bibfield  {author} {\bibinfo {author} {\bibfnamefont {P.~C.}\ \bibnamefont
  {Hohenberg}}\ and\ \bibinfo {author} {\bibfnamefont {W.~F.}\ \bibnamefont
  {Brinkman}},\ }\href {\doibase 10.1103/PhysRevB.10.128} {\bibfield  {journal}
  {\bibinfo  {journal} {Phys. Rev. B}\ }\textbf {\bibinfo {volume} {10}},\
  \bibinfo {pages} {128} (\bibinfo {year} {1974})}\BibitemShut {NoStop}%
\bibitem [{\citenamefont {Momma}\ and\ \citenamefont
  {Izumi}(2011)}]{momma201144}%
  \BibitemOpen
  \bibfield  {author} {\bibinfo {author} {\bibfnamefont {K.}~\bibnamefont
  {Momma}}\ and\ \bibinfo {author} {\bibfnamefont {F.}~\bibnamefont {Izumi}},\
  }\href {\doibase 10.1107/S0021889811038970} {\bibfield  {journal} {\bibinfo
  {journal} {J. Appl. Crystallogr.}\ }\textbf {\bibinfo {volume} {44}},\
  \bibinfo {pages} {1272} (\bibinfo {year} {2011})}\BibitemShut {NoStop}%
\bibitem [{\citenamefont {Qureshi}(2019)}]{qureshi201952}%
  \BibitemOpen
  \bibfield  {author} {\bibinfo {author} {\bibfnamefont {N.}~\bibnamefont
  {Qureshi}},\ }\href {\doibase 10.1107/S1600576718016084} {\bibfield
  {journal} {\bibinfo  {journal} {J. Appl. Crystallogr.}\ }\textbf {\bibinfo
  {volume} {52}},\ \bibinfo {pages} {175} (\bibinfo {year} {2019})}\BibitemShut
  {NoStop}%
\bibitem [{\citenamefont {Nakua}\ and\ \citenamefont
  {Greedan}(1995)}]{nakua1995154}%
  \BibitemOpen
  \bibfield  {author} {\bibinfo {author} {\bibfnamefont {A.~M.}\ \bibnamefont
  {Nakua}}\ and\ \bibinfo {author} {\bibfnamefont {J.~E.}\ \bibnamefont
  {Greedan}},\ }\href {\doibase 10.1016/0022-0248(95)00217-0} {\bibfield
  {journal} {\bibinfo  {journal} {J. Cryst. Growth}\ }\textbf {\bibinfo
  {volume} {154}},\ \bibinfo {pages} {334} (\bibinfo {year}
  {1995})}\BibitemShut {NoStop}%
\bibitem [{\citenamefont {Rodriguez}\ \emph {et~al.}(2008)\citenamefont
  {Rodriguez}, \citenamefont {Adler}, \citenamefont {Brand}, \citenamefont
  {Broholm}, \citenamefont {Cook}, \citenamefont {Brocker}, \citenamefont
  {Hammond}, \citenamefont {Huang}, \citenamefont {Hundertmakr}, \citenamefont
  {Lynn}, \citenamefont {Maliszewskyj}, \citenamefont {Moyer}, \citenamefont
  {Orndorff}, \citenamefont {Pierce}, \citenamefont {Pike}, \citenamefont
  {Scharfstein}, \citenamefont {Smee},\ and\ \citenamefont
  {Vilaseca}}]{Rodriguez200819}%
  \BibitemOpen
  \bibfield  {author} {\bibinfo {author} {\bibfnamefont {J.~A.}\ \bibnamefont
  {Rodriguez}}, \bibinfo {author} {\bibfnamefont {D.~M.}\ \bibnamefont
  {Adler}}, \bibinfo {author} {\bibfnamefont {P.~C.}\ \bibnamefont {Brand}},
  \bibinfo {author} {\bibfnamefont {C.}~\bibnamefont {Broholm}}, \bibinfo
  {author} {\bibfnamefont {J.~C.}\ \bibnamefont {Cook}}, \bibinfo {author}
  {\bibfnamefont {C.}~\bibnamefont {Brocker}}, \bibinfo {author} {\bibfnamefont
  {R.}~\bibnamefont {Hammond}}, \bibinfo {author} {\bibfnamefont
  {Z.}~\bibnamefont {Huang}}, \bibinfo {author} {\bibfnamefont
  {P.}~\bibnamefont {Hundertmakr}}, \bibinfo {author} {\bibfnamefont {J.~W.}\
  \bibnamefont {Lynn}}, \bibinfo {author} {\bibfnamefont {N.~C.}\ \bibnamefont
  {Maliszewskyj}}, \bibinfo {author} {\bibfnamefont {J.}~\bibnamefont {Moyer}},
  \bibinfo {author} {\bibfnamefont {J.}~\bibnamefont {Orndorff}}, \bibinfo
  {author} {\bibfnamefont {D.}~\bibnamefont {Pierce}}, \bibinfo {author}
  {\bibfnamefont {T.~D.}\ \bibnamefont {Pike}}, \bibinfo {author}
  {\bibfnamefont {G.}~\bibnamefont {Scharfstein}}, \bibinfo {author}
  {\bibfnamefont {S.~A.}\ \bibnamefont {Smee}}, \ and\ \bibinfo {author}
  {\bibfnamefont {R.}~\bibnamefont {Vilaseca}},\ }\href {\doibase
  10.1088/0957-0233/19/3/034023} {\bibfield  {journal} {\bibinfo  {journal}
  {Meas. Sci. Technol.}\ }\textbf {\bibinfo {volume} {19}},\ \bibinfo {pages}
  {034023} (\bibinfo {year} {2008})}\BibitemShut {NoStop}%
\bibitem [{\citenamefont {des Cloizeaux}\ and\ \citenamefont
  {Pearson}(1962)}]{Cloizeaux1962128}%
  \BibitemOpen
  \bibfield  {author} {\bibinfo {author} {\bibfnamefont {J.}~\bibnamefont {des
  Cloizeaux}}\ and\ \bibinfo {author} {\bibfnamefont {J.~J.}\ \bibnamefont
  {Pearson}},\ }\href {\doibase 10.1103/PhysRev.128.2131} {\bibfield  {journal}
  {\bibinfo  {journal} {Phys. Rev.}\ }\textbf {\bibinfo {volume} {128}},\
  \bibinfo {pages} {2131} (\bibinfo {year} {1962})}\BibitemShut {NoStop}%
\bibitem [{\citenamefont {M\"uller}\ \emph {et~al.}(1981)\citenamefont
  {M\"uller}, \citenamefont {Thomas}, \citenamefont {Beck},\ and\ \citenamefont
  {Bonner}}]{Muller198124}%
  \BibitemOpen
  \bibfield  {author} {\bibinfo {author} {\bibfnamefont {G.}~\bibnamefont
  {M\"uller}}, \bibinfo {author} {\bibfnamefont {H.}~\bibnamefont {Thomas}},
  \bibinfo {author} {\bibfnamefont {H.}~\bibnamefont {Beck}}, \ and\ \bibinfo
  {author} {\bibfnamefont {J.~C.}\ \bibnamefont {Bonner}},\ }\href {\doibase
  10.1103/PhysRevB.24.1429} {\bibfield  {journal} {\bibinfo  {journal} {Phys.
  Rev. B}\ }\textbf {\bibinfo {volume} {24}},\ \bibinfo {pages} {1429}
  (\bibinfo {year} {1981})}\BibitemShut {NoStop}%
\bibitem [{\citenamefont {Endoh}\ \emph {et~al.}(1974)\citenamefont {Endoh},
  \citenamefont {Shirane}, \citenamefont {Birgeneau}, \citenamefont
  {Richards},\ and\ \citenamefont {Holt}}]{Endoh197432}%
  \BibitemOpen
  \bibfield  {author} {\bibinfo {author} {\bibfnamefont {Y.}~\bibnamefont
  {Endoh}}, \bibinfo {author} {\bibfnamefont {G.}~\bibnamefont {Shirane}},
  \bibinfo {author} {\bibfnamefont {R.~J.}\ \bibnamefont {Birgeneau}}, \bibinfo
  {author} {\bibfnamefont {P.~M.}\ \bibnamefont {Richards}}, \ and\ \bibinfo
  {author} {\bibfnamefont {S.~L.}\ \bibnamefont {Holt}},\ }\href {\doibase
  10.1103/PhysRevLett.32.170} {\bibfield  {journal} {\bibinfo  {journal} {Phys.
  Rev. Lett.}\ }\textbf {\bibinfo {volume} {32}},\ \bibinfo {pages} {170}
  (\bibinfo {year} {1974})}\BibitemShut {NoStop}%
\bibitem [{\citenamefont {Coldea}\ \emph {et~al.}(2003)\citenamefont {Coldea},
  \citenamefont {Tennant},\ and\ \citenamefont {Tylczynski}}]{coldea200368}%
  \BibitemOpen
  \bibfield  {author} {\bibinfo {author} {\bibfnamefont {R.}~\bibnamefont
  {Coldea}}, \bibinfo {author} {\bibfnamefont {D.~A.}\ \bibnamefont {Tennant}},
  \ and\ \bibinfo {author} {\bibfnamefont {Z.}~\bibnamefont {Tylczynski}},\
  }\href {\doibase 10.1103/PhysRevB.68.134424} {\bibfield  {journal} {\bibinfo
  {journal} {Phys. Rev. B}\ }\textbf {\bibinfo {volume} {68}},\ \bibinfo
  {pages} {134424} (\bibinfo {year} {2003})}\BibitemShut {NoStop}%
\bibitem [{\citenamefont {Tennant}\ \emph {et~al.}(1993)\citenamefont
  {Tennant}, \citenamefont {Perring}, \citenamefont {Cowley},\ and\
  \citenamefont {Nagler}}]{Tennant199370}%
  \BibitemOpen
  \bibfield  {author} {\bibinfo {author} {\bibfnamefont {D.~A.}\ \bibnamefont
  {Tennant}}, \bibinfo {author} {\bibfnamefont {T.~G.}\ \bibnamefont
  {Perring}}, \bibinfo {author} {\bibfnamefont {R.~A.}\ \bibnamefont {Cowley}},
  \ and\ \bibinfo {author} {\bibfnamefont {S.~E.}\ \bibnamefont {Nagler}},\
  }\href {\doibase 10.1103/PhysRevLett.70.4003} {\bibfield  {journal} {\bibinfo
   {journal} {Phys. Rev. Lett.}\ }\textbf {\bibinfo {volume} {70}},\ \bibinfo
  {pages} {4003} (\bibinfo {year} {1993})}\BibitemShut {NoStop}%
\bibitem [{\citenamefont {Nagler}\ \emph {et~al.}(1991)\citenamefont {Nagler},
  \citenamefont {Tennant}, \citenamefont {Cowley}, \citenamefont {Perring},\
  and\ \citenamefont {Satija}}]{Nagler199144}%
  \BibitemOpen
  \bibfield  {author} {\bibinfo {author} {\bibfnamefont {S.~E.}\ \bibnamefont
  {Nagler}}, \bibinfo {author} {\bibfnamefont {D.~A.}\ \bibnamefont {Tennant}},
  \bibinfo {author} {\bibfnamefont {R.~A.}\ \bibnamefont {Cowley}}, \bibinfo
  {author} {\bibfnamefont {T.~G.}\ \bibnamefont {Perring}}, \ and\ \bibinfo
  {author} {\bibfnamefont {S.~K.}\ \bibnamefont {Satija}},\ }\href {\doibase
  10.1103/PhysRevB.44.12361} {\bibfield  {journal} {\bibinfo  {journal} {Phys.
  Rev. B}\ }\textbf {\bibinfo {volume} {44}},\ \bibinfo {pages} {12361}
  (\bibinfo {year} {1991})}\BibitemShut {NoStop}%
\bibitem [{\citenamefont {Lake}\ \emph {et~al.}(2013)\citenamefont {Lake},
  \citenamefont {Tennant}, \citenamefont {Caux}, \citenamefont {Barthel},
  \citenamefont {Schollw\"ock}, \citenamefont {Nagler},\ and\ \citenamefont
  {Frost}}]{Lake2013111}%
  \BibitemOpen
  \bibfield  {author} {\bibinfo {author} {\bibfnamefont {B.}~\bibnamefont
  {Lake}}, \bibinfo {author} {\bibfnamefont {D.~A.}\ \bibnamefont {Tennant}},
  \bibinfo {author} {\bibfnamefont {J.-S.}\ \bibnamefont {Caux}}, \bibinfo
  {author} {\bibfnamefont {T.}~\bibnamefont {Barthel}}, \bibinfo {author}
  {\bibfnamefont {U.}~\bibnamefont {Schollw\"ock}}, \bibinfo {author}
  {\bibfnamefont {S.~E.}\ \bibnamefont {Nagler}}, \ and\ \bibinfo {author}
  {\bibfnamefont {C.~D.}\ \bibnamefont {Frost}},\ }\href {\doibase
  10.1103/PhysRevLett.111.137205} {\bibfield  {journal} {\bibinfo  {journal}
  {Phys. Rev. Lett.}\ }\textbf {\bibinfo {volume} {111}},\ \bibinfo {pages}
  {137205} (\bibinfo {year} {2013})}\BibitemShut {NoStop}%
\bibitem [{\citenamefont {Stone}\ \emph {et~al.}(2003)\citenamefont {Stone},
  \citenamefont {Reich}, \citenamefont {Broholm}, \citenamefont {Lefmann},
  \citenamefont {Rischel}, \citenamefont {Landee},\ and\ \citenamefont
  {Turnbull}}]{Stone200391}%
  \BibitemOpen
  \bibfield  {author} {\bibinfo {author} {\bibfnamefont {M.~B.}\ \bibnamefont
  {Stone}}, \bibinfo {author} {\bibfnamefont {D.~H.}\ \bibnamefont {Reich}},
  \bibinfo {author} {\bibfnamefont {C.}~\bibnamefont {Broholm}}, \bibinfo
  {author} {\bibfnamefont {K.}~\bibnamefont {Lefmann}}, \bibinfo {author}
  {\bibfnamefont {C.}~\bibnamefont {Rischel}}, \bibinfo {author} {\bibfnamefont
  {C.~P.}\ \bibnamefont {Landee}}, \ and\ \bibinfo {author} {\bibfnamefont
  {M.~M.}\ \bibnamefont {Turnbull}},\ }\href {\doibase
  10.1103/PhysRevLett.91.037205} {\bibfield  {journal} {\bibinfo  {journal}
  {Phys. Rev. Lett.}\ }\textbf {\bibinfo {volume} {91}},\ \bibinfo {pages}
  {037205} (\bibinfo {year} {2003})}\BibitemShut {NoStop}%
\bibitem [{\citenamefont {Mourigal}\ \emph {et~al.}(2013)\citenamefont
  {Mourigal}, \citenamefont {Enderle}, \citenamefont {Kloepperpieper},
  \citenamefont {Caux}, \citenamefont {Stunault},\ and\ \citenamefont
  {Ronnow}}]{Mourigal20139}%
  \BibitemOpen
  \bibfield  {author} {\bibinfo {author} {\bibfnamefont {M.}~\bibnamefont
  {Mourigal}}, \bibinfo {author} {\bibfnamefont {M.}~\bibnamefont {Enderle}},
  \bibinfo {author} {\bibfnamefont {A.}~\bibnamefont {Kloepperpieper}},
  \bibinfo {author} {\bibfnamefont {J.-S.}\ \bibnamefont {Caux}}, \bibinfo
  {author} {\bibfnamefont {A.}~\bibnamefont {Stunault}}, \ and\ \bibinfo
  {author} {\bibfnamefont {H.~M.}\ \bibnamefont {Ronnow}},\ }\href {\doibase
  10.1038/NPHYS2652} {\bibfield  {journal} {\bibinfo  {journal} {Nat. Phys.}\
  }\textbf {\bibinfo {volume} {9}},\ \bibinfo {pages} {435} (\bibinfo {year}
  {2013})}\BibitemShut {NoStop}%
\bibitem [{\citenamefont {Heilmann}\ \emph {et~al.}(1981)\citenamefont
  {Heilmann}, \citenamefont {Kjems}, \citenamefont {Endoh}, \citenamefont
  {Reiter}, \citenamefont {Shirane},\ and\ \citenamefont
  {Birgeneau}}]{heilmann198124}%
  \BibitemOpen
  \bibfield  {author} {\bibinfo {author} {\bibfnamefont {I.~U.}\ \bibnamefont
  {Heilmann}}, \bibinfo {author} {\bibfnamefont {J.~K.}\ \bibnamefont {Kjems}},
  \bibinfo {author} {\bibfnamefont {Y.}~\bibnamefont {Endoh}}, \bibinfo
  {author} {\bibfnamefont {G.~F.}\ \bibnamefont {Reiter}}, \bibinfo {author}
  {\bibfnamefont {G.}~\bibnamefont {Shirane}}, \ and\ \bibinfo {author}
  {\bibfnamefont {R.~J.}\ \bibnamefont {Birgeneau}},\ }\href {\doibase
  10.1103/PhysRevB.24.3939} {\bibfield  {journal} {\bibinfo  {journal} {Phys.
  Rev. B}\ }\textbf {\bibinfo {volume} {24}},\ \bibinfo {pages} {3939}
  (\bibinfo {year} {1981})}\BibitemShut {NoStop}%
\bibitem [{\citenamefont {Huberman}\ \emph {et~al.}(2005)\citenamefont
  {Huberman}, \citenamefont {Coldea}, \citenamefont {Cowley}, \citenamefont
  {Tennant}, \citenamefont {Leheny}, \citenamefont {Christianson},\ and\
  \citenamefont {Frost}}]{huberman200572}%
  \BibitemOpen
  \bibfield  {author} {\bibinfo {author} {\bibfnamefont {T.}~\bibnamefont
  {Huberman}}, \bibinfo {author} {\bibfnamefont {R.}~\bibnamefont {Coldea}},
  \bibinfo {author} {\bibfnamefont {R.~A.}\ \bibnamefont {Cowley}}, \bibinfo
  {author} {\bibfnamefont {D.~A.}\ \bibnamefont {Tennant}}, \bibinfo {author}
  {\bibfnamefont {R.~L.}\ \bibnamefont {Leheny}}, \bibinfo {author}
  {\bibfnamefont {R.~J.}\ \bibnamefont {Christianson}}, \ and\ \bibinfo
  {author} {\bibfnamefont {C.~D.}\ \bibnamefont {Frost}},\ }\href {\doibase
  10.1103/PhysRevB.72.014413} {\bibfield  {journal} {\bibinfo  {journal} {Phys.
  Rev. B}\ }\textbf {\bibinfo {volume} {72}},\ \bibinfo {pages} {014413}
  (\bibinfo {year} {2005})}\BibitemShut {NoStop}%
\bibitem [{\citenamefont {Songvilay}\ \emph {et~al.}(2018)\citenamefont
  {Songvilay}, \citenamefont {Rodriguez}, \citenamefont {Lindsay},
  \citenamefont {Green}, \citenamefont {Walker}, \citenamefont
  {Rodriguez-Rivera},\ and\ \citenamefont {Stock}}]{songvilay2018}%
  \BibitemOpen
  \bibfield  {author} {\bibinfo {author} {\bibfnamefont {M.}~\bibnamefont
  {Songvilay}}, \bibinfo {author} {\bibfnamefont {E.}~\bibnamefont
  {Rodriguez}}, \bibinfo {author} {\bibfnamefont {R.}~\bibnamefont {Lindsay}},
  \bibinfo {author} {\bibfnamefont {M.}~\bibnamefont {Green}}, \bibinfo
  {author} {\bibfnamefont {H.}~\bibnamefont {Walker}}, \bibinfo {author}
  {\bibfnamefont {J.}~\bibnamefont {Rodriguez-Rivera}}, \ and\ \bibinfo
  {author} {\bibfnamefont {C.}~\bibnamefont {Stock}},\ }\href {\doibase
  10.1103/PhysRevLett.121.087201} {\bibfield  {journal} {\bibinfo  {journal}
  {Physical Review Letters}\ }\textbf {\bibinfo {volume} {121}},\ \bibinfo
  {pages} {087201} (\bibinfo {year} {2018})}\BibitemShut {NoStop}%
\bibitem [{\citenamefont {Hammar}\ \emph {et~al.}(1998)\citenamefont {Hammar},
  \citenamefont {Reich}, \citenamefont {Broholm},\ and\ \citenamefont
  {Trouw}}]{hammar199857}%
  \BibitemOpen
  \bibfield  {author} {\bibinfo {author} {\bibfnamefont {P.~R.}\ \bibnamefont
  {Hammar}}, \bibinfo {author} {\bibfnamefont {D.~H.}\ \bibnamefont {Reich}},
  \bibinfo {author} {\bibfnamefont {C.}~\bibnamefont {Broholm}}, \ and\
  \bibinfo {author} {\bibfnamefont {F.}~\bibnamefont {Trouw}},\ }\href
  {\doibase 10.1103/PhysRevB.57.7846} {\bibfield  {journal} {\bibinfo
  {journal} {Phys. Rev. B}\ }\textbf {\bibinfo {volume} {57}},\ \bibinfo
  {pages} {7846} (\bibinfo {year} {1998})}\BibitemShut {NoStop}%
\bibitem [{\citenamefont {Xu}\ \emph {et~al.}(2013)\citenamefont {Xu},
  \citenamefont {Xu},\ and\ \citenamefont {Tranquada}}]{xu201384}%
  \BibitemOpen
  \bibfield  {author} {\bibinfo {author} {\bibfnamefont {G.}~\bibnamefont
  {Xu}}, \bibinfo {author} {\bibfnamefont {Z.}~\bibnamefont {Xu}}, \ and\
  \bibinfo {author} {\bibfnamefont {J.~M.}\ \bibnamefont {Tranquada}},\ }\href
  {\doibase 10.1063/1.4818323} {\bibfield  {journal} {\bibinfo  {journal} {Rev.
  Sci. Instrum.}\ }\textbf {\bibinfo {volume} {84}},\ \bibinfo {pages} {083906}
  (\bibinfo {year} {2013})}\BibitemShut {NoStop}%
\bibitem [{\citenamefont {Nakatsuji}\ \emph {et~al.}(2012)\citenamefont
  {Nakatsuji}, \citenamefont {Kuga}, \citenamefont {Kimura}, \citenamefont
  {Satake}, \citenamefont {Katayama}, \citenamefont {Nishibori}, \citenamefont
  {Sawa}, \citenamefont {Ishii}, \citenamefont {Hagiwara}, \citenamefont
  {Bridges}, \citenamefont {Ito}, \citenamefont {Higemoto}, \citenamefont
  {Karaki}, \citenamefont {Halim}, \citenamefont {Nugroho}, \citenamefont
  {Rodriguez-Rivera}, \citenamefont {Green},\ and\ \citenamefont
  {Broholm}}]{Nakatsuji2012339}%
  \BibitemOpen
  \bibfield  {author} {\bibinfo {author} {\bibfnamefont {S.}~\bibnamefont
  {Nakatsuji}}, \bibinfo {author} {\bibfnamefont {K.}~\bibnamefont {Kuga}},
  \bibinfo {author} {\bibfnamefont {K.}~\bibnamefont {Kimura}}, \bibinfo
  {author} {\bibfnamefont {R.}~\bibnamefont {Satake}}, \bibinfo {author}
  {\bibfnamefont {N.}~\bibnamefont {Katayama}}, \bibinfo {author}
  {\bibfnamefont {E.}~\bibnamefont {Nishibori}}, \bibinfo {author}
  {\bibfnamefont {H.}~\bibnamefont {Sawa}}, \bibinfo {author} {\bibfnamefont
  {R.}~\bibnamefont {Ishii}}, \bibinfo {author} {\bibfnamefont
  {M.}~\bibnamefont {Hagiwara}}, \bibinfo {author} {\bibfnamefont
  {F.}~\bibnamefont {Bridges}}, \bibinfo {author} {\bibfnamefont {T.~U.}\
  \bibnamefont {Ito}}, \bibinfo {author} {\bibfnamefont {W.}~\bibnamefont
  {Higemoto}}, \bibinfo {author} {\bibfnamefont {Y.}~\bibnamefont {Karaki}},
  \bibinfo {author} {\bibfnamefont {M.}~\bibnamefont {Halim}}, \bibinfo
  {author} {\bibfnamefont {A.~A.}\ \bibnamefont {Nugroho}}, \bibinfo {author}
  {\bibfnamefont {J.~A.}\ \bibnamefont {Rodriguez-Rivera}}, \bibinfo {author}
  {\bibfnamefont {M.~A.}\ \bibnamefont {Green}}, \ and\ \bibinfo {author}
  {\bibfnamefont {C.}~\bibnamefont {Broholm}},\ }\href {\doibase
  10.1126/science.1212154} {\bibfield  {journal} {\bibinfo  {journal}
  {Science}\ }\textbf {\bibinfo {volume} {339}},\ \bibinfo {pages} {559}
  (\bibinfo {year} {2012})}\BibitemShut {NoStop}%
\bibitem [{\citenamefont {Sears}(1992)}]{sears19923}%
  \BibitemOpen
  \bibfield  {author} {\bibinfo {author} {\bibfnamefont {V.~F.}\ \bibnamefont
  {Sears}},\ }\href {\doibase 10.1080/10448639208218770} {\bibfield  {journal}
  {\bibinfo  {journal} {Neutron News}\ }\textbf {\bibinfo {volume} {3}},\
  \bibinfo {pages} {26} (\bibinfo {year} {1992})}\BibitemShut {NoStop}%
\bibitem [{\citenamefont {Zaliznyak}\ and\ \citenamefont
  {Lee}(2005)}]{zaliznyak2005}%
  \BibitemOpen
  \bibfield  {author} {\bibinfo {author} {\bibfnamefont {I.~A.}\ \bibnamefont
  {Zaliznyak}}\ and\ \bibinfo {author} {\bibfnamefont {S.-H.}\ \bibnamefont
  {Lee}},\ }in\ \href {\doibase 10.1007/0-387-23395-4_1} {\emph {\bibinfo
  {booktitle} {Modern {Techniques} for {Characterizing} {Magnetic}
  {Materials}}}},\ \bibinfo {editor} {edited by\ \bibinfo {editor}
  {\bibfnamefont {Y.}~\bibnamefont {Zhu}}}\ (\bibinfo  {publisher} {Springer},\
  \bibinfo {address} {Boston, MA},\ \bibinfo {year} {2005})\ pp.\ \bibinfo
  {pages} {3--64}\BibitemShut {NoStop}%
\bibitem [{\citenamefont {Sarte}(2019)}]{sarte2019a}%
  \BibitemOpen
  \bibfield  {author} {\bibinfo {author} {\bibfnamefont {P.~M.}\ \bibnamefont
  {Sarte}},\ }\emph {\bibinfo {title} {Neutron Inelastic Scattering Studies of
  Effective Spin-1/2 Magnets}},\ \href {https://era.ed.ac.uk/handle/1842/35553}
  {Ph.D. thesis} (\bibinfo {year} {2019})\BibitemShut {NoStop}%
\bibitem [{\citenamefont {Sarte}\ \emph {et~al.}(2018)\citenamefont {Sarte},
  \citenamefont {{Ar{\'e}valo-L{\'o}pez}}, \citenamefont {Songvilay},
  \citenamefont {Le}, \citenamefont {Guidi}, \citenamefont
  {{Garc{\'i}a-Sakai}}, \citenamefont {Mukhopadhyay}, \citenamefont {Capelli},
  \citenamefont {Ratcliff}, \citenamefont {Hong}, \citenamefont {McNally},
  \citenamefont {Pachoud}, \citenamefont {Attfield},\ and\ \citenamefont
  {Stock}}]{sarte201898}%
  \BibitemOpen
  \bibfield  {author} {\bibinfo {author} {\bibfnamefont {P.~M.}\ \bibnamefont
  {Sarte}}, \bibinfo {author} {\bibfnamefont {A.~M.}\ \bibnamefont
  {{Ar{\'e}valo-L{\'o}pez}}}, \bibinfo {author} {\bibfnamefont
  {M.}~\bibnamefont {Songvilay}}, \bibinfo {author} {\bibfnamefont
  {D.}~\bibnamefont {Le}}, \bibinfo {author} {\bibfnamefont {T.}~\bibnamefont
  {Guidi}}, \bibinfo {author} {\bibfnamefont {V.}~\bibnamefont
  {{Garc{\'i}a-Sakai}}}, \bibinfo {author} {\bibfnamefont {S.}~\bibnamefont
  {Mukhopadhyay}}, \bibinfo {author} {\bibfnamefont {S.~C.}\ \bibnamefont
  {Capelli}}, \bibinfo {author} {\bibfnamefont {W.~D.}\ \bibnamefont
  {Ratcliff}}, \bibinfo {author} {\bibfnamefont {K.~H.}\ \bibnamefont {Hong}},
  \bibinfo {author} {\bibfnamefont {G.~M.}\ \bibnamefont {McNally}}, \bibinfo
  {author} {\bibfnamefont {E.}~\bibnamefont {Pachoud}}, \bibinfo {author}
  {\bibfnamefont {J.~P.}\ \bibnamefont {Attfield}}, \ and\ \bibinfo {author}
  {\bibfnamefont {C.}~\bibnamefont {Stock}},\ }\href {\doibase
  10.1103/PhysRevB.98.224410} {\bibfield  {journal} {\bibinfo  {journal} {Phys.
  Rev. B}\ }\textbf {\bibinfo {volume} {98}},\ \bibinfo {pages} {224410}
  (\bibinfo {year} {2018})}\BibitemShut {NoStop}%
\bibitem [{\citenamefont {Stock}\ \emph {et~al.}(2009)\citenamefont {Stock},
  \citenamefont {Chapon}, \citenamefont {Adamopoulos}, \citenamefont {Lappas},
  \citenamefont {Giot}, \citenamefont {Taylor}, \citenamefont {Green},
  \citenamefont {Brown},\ and\ \citenamefont {Radaelli}}]{stock2009103}%
  \BibitemOpen
  \bibfield  {author} {\bibinfo {author} {\bibfnamefont {C.}~\bibnamefont
  {Stock}}, \bibinfo {author} {\bibfnamefont {L.~C.}\ \bibnamefont {Chapon}},
  \bibinfo {author} {\bibfnamefont {O.}~\bibnamefont {Adamopoulos}}, \bibinfo
  {author} {\bibfnamefont {A.}~\bibnamefont {Lappas}}, \bibinfo {author}
  {\bibfnamefont {M.}~\bibnamefont {Giot}}, \bibinfo {author} {\bibfnamefont
  {J.~W.}\ \bibnamefont {Taylor}}, \bibinfo {author} {\bibfnamefont {M.~A.}\
  \bibnamefont {Green}}, \bibinfo {author} {\bibfnamefont {C.~M.}\ \bibnamefont
  {Brown}}, \ and\ \bibinfo {author} {\bibfnamefont {P.~G.}\ \bibnamefont
  {Radaelli}},\ }\href {\doibase 10.1103/PhysRevLett.103.077202} {\bibfield
  {journal} {\bibinfo  {journal} {Phys. Rev. Lett.}\ }\textbf {\bibinfo
  {volume} {103}},\ \bibinfo {pages} {077202} (\bibinfo {year}
  {2009})}\BibitemShut {NoStop}%
\bibitem [{\citenamefont {Stone}\ \emph {et~al.}(2002)\citenamefont {Stone},
  \citenamefont {Chen}, \citenamefont {Rittner}, \citenamefont {Yardimci},
  \citenamefont {Reich}, \citenamefont {Broholm}, \citenamefont {Ferraris},\
  and\ \citenamefont {Lectka}}]{stone200265}%
  \BibitemOpen
  \bibfield  {author} {\bibinfo {author} {\bibfnamefont {M.~B.}\ \bibnamefont
  {Stone}}, \bibinfo {author} {\bibfnamefont {Y.}~\bibnamefont {Chen}},
  \bibinfo {author} {\bibfnamefont {J.}~\bibnamefont {Rittner}}, \bibinfo
  {author} {\bibfnamefont {H.}~\bibnamefont {Yardimci}}, \bibinfo {author}
  {\bibfnamefont {D.~H.}\ \bibnamefont {Reich}}, \bibinfo {author}
  {\bibfnamefont {C.}~\bibnamefont {Broholm}}, \bibinfo {author} {\bibfnamefont
  {D.~V.}\ \bibnamefont {Ferraris}}, \ and\ \bibinfo {author} {\bibfnamefont
  {T.}~\bibnamefont {Lectka}},\ }\href {\doibase 10.1103/PhysRevB.65.064423}
  {\bibfield  {journal} {\bibinfo  {journal} {Phys. Rev. B}\ }\textbf {\bibinfo
  {volume} {65}},\ \bibinfo {pages} {064423} (\bibinfo {year}
  {2002})}\BibitemShut {NoStop}%
\bibitem [{\citenamefont {Matsumoto}(2021)}]{Matsumoto21:90}%
  \BibitemOpen
  \bibfield  {author} {\bibinfo {author} {\bibfnamefont {M.}~\bibnamefont
  {Matsumoto}},\ }\href {\doibase 10.7566/JPSJ.90.014701} {\bibfield  {journal}
  {\bibinfo  {journal} {J. Phys. Soc. Jpn.}\ }\textbf {\bibinfo {volume}
  {90}},\ \bibinfo {pages} {014701} (\bibinfo {year} {2021})}\BibitemShut
  {NoStop}%
\bibitem [{\citenamefont {Songvilay}\ \emph {et~al.}(2020)\citenamefont
  {Songvilay}, \citenamefont {Robert}, \citenamefont {Petit}, \citenamefont
  {Rodriguez-Rivera}, \citenamefont {Ratcliff}, \citenamefont {Damay},
  \citenamefont {Balédent}, \citenamefont {Jiménez-Ruiz}, \citenamefont
  {Lejay}, \citenamefont {Pachoud}, \citenamefont {Hadj-Azzem}, \citenamefont
  {Simonet},\ and\ \citenamefont {Stock}}]{songvilay2020102}%
  \BibitemOpen
  \bibfield  {author} {\bibinfo {author} {\bibfnamefont {M.}~\bibnamefont
  {Songvilay}}, \bibinfo {author} {\bibfnamefont {J.}~\bibnamefont {Robert}},
  \bibinfo {author} {\bibfnamefont {S.}~\bibnamefont {Petit}}, \bibinfo
  {author} {\bibfnamefont {J.~A.}\ \bibnamefont {Rodriguez-Rivera}}, \bibinfo
  {author} {\bibfnamefont {W.~D.}\ \bibnamefont {Ratcliff}}, \bibinfo {author}
  {\bibfnamefont {F.}~\bibnamefont {Damay}}, \bibinfo {author} {\bibfnamefont
  {V.}~\bibnamefont {Balédent}}, \bibinfo {author} {\bibfnamefont
  {M.}~\bibnamefont {Jiménez-Ruiz}}, \bibinfo {author} {\bibfnamefont
  {P.}~\bibnamefont {Lejay}}, \bibinfo {author} {\bibfnamefont
  {E.}~\bibnamefont {Pachoud}}, \bibinfo {author} {\bibfnamefont
  {A.}~\bibnamefont {Hadj-Azzem}}, \bibinfo {author} {\bibfnamefont
  {V.}~\bibnamefont {Simonet}}, \ and\ \bibinfo {author} {\bibfnamefont
  {C.}~\bibnamefont {Stock}},\ }\href {\doibase 10.1103/PhysRevB.102.224429}
  {\bibfield  {journal} {\bibinfo  {journal} {Phys. Rev. B}\ }\textbf {\bibinfo
  {volume} {102}},\ \bibinfo {pages} {224429} (\bibinfo {year}
  {2020})}\BibitemShut {NoStop}%
\bibitem [{\citenamefont {Songvilay}\ \emph {et~al.}(2021)\citenamefont
  {Songvilay}, \citenamefont {Petit}, \citenamefont {Damay}, \citenamefont
  {Roux}, \citenamefont {Qureshi}, \citenamefont {Walker}, \citenamefont
  {{Rodriguez-Rivera}}, \citenamefont {Gao}, \citenamefont {Cheong},\ and\
  \citenamefont {Stock}}]{songvilay2021126}%
  \BibitemOpen
  \bibfield  {author} {\bibinfo {author} {\bibfnamefont {M.}~\bibnamefont
  {Songvilay}}, \bibinfo {author} {\bibfnamefont {S.}~\bibnamefont {Petit}},
  \bibinfo {author} {\bibfnamefont {F.}~\bibnamefont {Damay}}, \bibinfo
  {author} {\bibfnamefont {G.}~\bibnamefont {Roux}}, \bibinfo {author}
  {\bibfnamefont {N.}~\bibnamefont {Qureshi}}, \bibinfo {author} {\bibfnamefont
  {H.~C.}\ \bibnamefont {Walker}}, \bibinfo {author} {\bibfnamefont {J.~A.}\
  \bibnamefont {{Rodriguez-Rivera}}}, \bibinfo {author} {\bibfnamefont
  {B.}~\bibnamefont {Gao}}, \bibinfo {author} {\bibfnamefont {S.-W.}\
  \bibnamefont {Cheong}}, \ and\ \bibinfo {author} {\bibfnamefont
  {C.}~\bibnamefont {Stock}},\ }\href {\doibase 10.1103/PhysRevLett.126.017201}
  {\bibfield  {journal} {\bibinfo  {journal} {Phys. Rev. Lett.}\ }\textbf
  {\bibinfo {volume} {126}},\ \bibinfo {pages} {017201} (\bibinfo {year}
  {2021})}\BibitemShut {NoStop}%
\bibitem [{\citenamefont {Cowley}(2003)}]{cowley200315}%
  \BibitemOpen
  \bibfield  {author} {\bibinfo {author} {\bibfnamefont {R.~A.}\ \bibnamefont
  {Cowley}},\ }\href {\doibase 10.1088/0953-8984/15/24/308} {\bibfield
  {journal} {\bibinfo  {journal} {J. Phys.: Condens. Matter}\ }\textbf
  {\bibinfo {volume} {15}},\ \bibinfo {pages} {4143} (\bibinfo {year}
  {2003})}\BibitemShut {NoStop}%
\bibitem [{\citenamefont {Stock}\ \emph {et~al.}(2010)\citenamefont {Stock},
  \citenamefont {Cowley}, \citenamefont {Taylor},\ and\ \citenamefont
  {Bennington}}]{Stock201081}%
  \BibitemOpen
  \bibfield  {author} {\bibinfo {author} {\bibfnamefont {C.}~\bibnamefont
  {Stock}}, \bibinfo {author} {\bibfnamefont {R.~A.}\ \bibnamefont {Cowley}},
  \bibinfo {author} {\bibfnamefont {J.~W.}\ \bibnamefont {Taylor}}, \ and\
  \bibinfo {author} {\bibfnamefont {S.~M.}\ \bibnamefont {Bennington}},\ }\href
  {\doibase 10.1103/PhysRevB.81.024303} {\bibfield  {journal} {\bibinfo
  {journal} {Phys. Rev. B}\ }\textbf {\bibinfo {volume} {81}},\ \bibinfo
  {pages} {024303} (\bibinfo {year} {2010})}\BibitemShut {NoStop}%
\bibitem [{\citenamefont {Toth}\ and\ \citenamefont {Lake}(2015)}]{toth201527}%
  \BibitemOpen
  \bibfield  {author} {\bibinfo {author} {\bibfnamefont {S.}~\bibnamefont
  {Toth}}\ and\ \bibinfo {author} {\bibfnamefont {B.}~\bibnamefont {Lake}},\
  }\href {\doibase 10.1088/0953-8984/27/16/166002} {\bibfield  {journal}
  {\bibinfo  {journal} {J. Phys. Condens. Matter}\ }\textbf {\bibinfo {volume}
  {27}},\ \bibinfo {pages} {166002} (\bibinfo {year} {2015})}\BibitemShut
  {NoStop}%
\bibitem [{\citenamefont {Smart}(1966)}]{smart1966}%
  \BibitemOpen
  \bibfield  {author} {\bibinfo {author} {\bibfnamefont {J.~S.}\ \bibnamefont
  {Smart}},\ }\href@noop {} {\emph {\bibinfo {title} {Effective field theories
  of magnetism}}}\ (\bibinfo  {publisher} {Saunders},\ \bibinfo {address}
  {Philadelphia},\ \bibinfo {year} {1966})\BibitemShut {NoStop}%
\bibitem [{\citenamefont {Sarte}\ \emph {et~al.}(2019)\citenamefont {Sarte},
  \citenamefont {Songvilay}, \citenamefont {Pachoud}, \citenamefont {Ewings},
  \citenamefont {Frost}, \citenamefont {Prabhakaran}, \citenamefont {Hong},
  \citenamefont {Browne}, \citenamefont {Yamani}, \citenamefont {Attfield},
  \citenamefont {Rodriguez}, \citenamefont {Wilson},\ and\ \citenamefont
  {Stock}}]{sarte2019100}%
  \BibitemOpen
  \bibfield  {author} {\bibinfo {author} {\bibfnamefont {P.~M.}\ \bibnamefont
  {Sarte}}, \bibinfo {author} {\bibfnamefont {M.}~\bibnamefont {Songvilay}},
  \bibinfo {author} {\bibfnamefont {E.}~\bibnamefont {Pachoud}}, \bibinfo
  {author} {\bibfnamefont {R.~A.}\ \bibnamefont {Ewings}}, \bibinfo {author}
  {\bibfnamefont {C.~D.}\ \bibnamefont {Frost}}, \bibinfo {author}
  {\bibfnamefont {D.}~\bibnamefont {Prabhakaran}}, \bibinfo {author}
  {\bibfnamefont {K.~H.}\ \bibnamefont {Hong}}, \bibinfo {author}
  {\bibfnamefont {A.~J.}\ \bibnamefont {Browne}}, \bibinfo {author}
  {\bibfnamefont {Z.}~\bibnamefont {Yamani}}, \bibinfo {author} {\bibfnamefont
  {J.~P.}\ \bibnamefont {Attfield}}, \bibinfo {author} {\bibfnamefont {E.~E.}\
  \bibnamefont {Rodriguez}}, \bibinfo {author} {\bibfnamefont {S.~D.}\
  \bibnamefont {Wilson}}, \ and\ \bibinfo {author} {\bibfnamefont
  {C.}~\bibnamefont {Stock}},\ }\href {\doibase 10.1103/PhysRevB.100.075143}
  {\bibfield  {journal} {\bibinfo  {journal} {Phys. Rev. B}\ }\textbf {\bibinfo
  {volume} {100}},\ \bibinfo {pages} {075143} (\bibinfo {year}
  {2019})}\BibitemShut {NoStop}%
\bibitem [{\citenamefont {Lane}\ \emph
  {et~al.}(2021{\natexlab{a}})\citenamefont {Lane}, \citenamefont {Pachoud},
  \citenamefont {Rodriguez-Rivera}, \citenamefont {Songvilay}, \citenamefont
  {Xu}, \citenamefont {Gehring}, \citenamefont {Attfield}, \citenamefont
  {Ewings},\ and\ \citenamefont {Stock}}]{Lane2021a}%
  \BibitemOpen
  \bibfield  {author} {\bibinfo {author} {\bibfnamefont {H.}~\bibnamefont
  {Lane}}, \bibinfo {author} {\bibfnamefont {E.}~\bibnamefont {Pachoud}},
  \bibinfo {author} {\bibfnamefont {J.~A.}\ \bibnamefont {Rodriguez-Rivera}},
  \bibinfo {author} {\bibfnamefont {M.}~\bibnamefont {Songvilay}}, \bibinfo
  {author} {\bibfnamefont {G.}~\bibnamefont {Xu}}, \bibinfo {author}
  {\bibfnamefont {P.~M.}\ \bibnamefont {Gehring}}, \bibinfo {author}
  {\bibfnamefont {J.~P.}\ \bibnamefont {Attfield}}, \bibinfo {author}
  {\bibfnamefont {R.~A.}\ \bibnamefont {Ewings}}, \ and\ \bibinfo {author}
  {\bibfnamefont {C.}~\bibnamefont {Stock}},\ }\href {\doibase
  10.1103/PhysRevB.104.L020411} {\bibfield  {journal} {\bibinfo  {journal}
  {Phys. Rev. B}\ }\textbf {\bibinfo {volume} {104}},\ \bibinfo {pages}
  {L020411} (\bibinfo {year} {2021}{\natexlab{a}})}\BibitemShut {NoStop}%
\bibitem [{\citenamefont {Lane}\ \emph
  {et~al.}(2021{\natexlab{b}})\citenamefont {Lane}, \citenamefont {Rodriguez},
  \citenamefont {Walker}, \citenamefont {Niedermayer}, \citenamefont {Stuhr},
  \citenamefont {Bewley}, \citenamefont {Voneshen}, \citenamefont {Green},
  \citenamefont {Rodriguez-Rivera}, \citenamefont {Fouquet}, \citenamefont
  {Cheong}, \citenamefont {Attfield}, \citenamefont {Ewings},\ and\
  \citenamefont {Stock}}]{Lane2021}%
  \BibitemOpen
  \bibfield  {author} {\bibinfo {author} {\bibfnamefont {H.}~\bibnamefont
  {Lane}}, \bibinfo {author} {\bibfnamefont {E.~E.}\ \bibnamefont {Rodriguez}},
  \bibinfo {author} {\bibfnamefont {H.~C.}\ \bibnamefont {Walker}}, \bibinfo
  {author} {\bibfnamefont {C.}~\bibnamefont {Niedermayer}}, \bibinfo {author}
  {\bibfnamefont {U.}~\bibnamefont {Stuhr}}, \bibinfo {author} {\bibfnamefont
  {R.~I.}\ \bibnamefont {Bewley}}, \bibinfo {author} {\bibfnamefont {D.~J.}\
  \bibnamefont {Voneshen}}, \bibinfo {author} {\bibfnamefont {M.~A.}\
  \bibnamefont {Green}}, \bibinfo {author} {\bibfnamefont {J.~A.}\ \bibnamefont
  {Rodriguez-Rivera}}, \bibinfo {author} {\bibfnamefont {P.}~\bibnamefont
  {Fouquet}}, \bibinfo {author} {\bibfnamefont {S.-W.}\ \bibnamefont {Cheong}},
  \bibinfo {author} {\bibfnamefont {J.~P.}\ \bibnamefont {Attfield}}, \bibinfo
  {author} {\bibfnamefont {R.~A.}\ \bibnamefont {Ewings}}, \ and\ \bibinfo
  {author} {\bibfnamefont {C.}~\bibnamefont {Stock}},\ }\href {\doibase
  10.1103/PhysRevB.104.104404} {\bibfield  {journal} {\bibinfo  {journal}
  {Phys. Rev. B}\ }\textbf {\bibinfo {volume} {104}},\ \bibinfo {pages}
  {104404} (\bibinfo {year} {2021}{\natexlab{b}})}\BibitemShut {NoStop}%
\bibitem [{\citenamefont {Lane}\ \emph {et~al.}(2022)\citenamefont {Lane},
  \citenamefont {Songvilay}, \citenamefont {Ewings},\ and\ \citenamefont
  {Stock}}]{Lane22:106}%
  \BibitemOpen
  \bibfield  {author} {\bibinfo {author} {\bibfnamefont {H.}~\bibnamefont
  {Lane}}, \bibinfo {author} {\bibfnamefont {M.}~\bibnamefont {Songvilay}},
  \bibinfo {author} {\bibfnamefont {R.~A.}\ \bibnamefont {Ewings}}, \ and\
  \bibinfo {author} {\bibfnamefont {C.}~\bibnamefont {Stock}},\ }\href
  {\doibase 10.1103/PhysRevB.106.054431} {\bibfield  {journal} {\bibinfo
  {journal} {Phys. Rev. B}\ }\textbf {\bibinfo {volume} {106}},\ \bibinfo
  {pages} {054431} (\bibinfo {year} {2022})}\BibitemShut {NoStop}%
\bibitem [{\citenamefont {Buyers}\ \emph {et~al.}(1975)\citenamefont {Buyers},
  \citenamefont {Holden},\ and\ \citenamefont {Perreault}}]{Buyers197511}%
  \BibitemOpen
  \bibfield  {author} {\bibinfo {author} {\bibfnamefont {W.~J.~L.}\
  \bibnamefont {Buyers}}, \bibinfo {author} {\bibfnamefont {T.~M.}\
  \bibnamefont {Holden}}, \ and\ \bibinfo {author} {\bibfnamefont
  {A.}~\bibnamefont {Perreault}},\ }\href {\doibase 10.1103/PhysRevB.11.266}
  {\bibfield  {journal} {\bibinfo  {journal} {Phys. Rev. B}\ }\textbf {\bibinfo
  {volume} {11}},\ \bibinfo {pages} {266} (\bibinfo {year} {1975})}\BibitemShut
  {NoStop}%
\bibitem [{\citenamefont {Cooke}(1973)}]{Cooke19737}%
  \BibitemOpen
  \bibfield  {author} {\bibinfo {author} {\bibfnamefont {J.~F.}\ \bibnamefont
  {Cooke}},\ }\href {\doibase 10.1103/PhysRevB.7.1108} {\bibfield  {journal}
  {\bibinfo  {journal} {Phys. Rev. B}\ }\textbf {\bibinfo {volume} {7}},\
  \bibinfo {pages} {1108} (\bibinfo {year} {1973})}\BibitemShut {NoStop}%
\bibitem [{\citenamefont {Sarte}\ \emph {et~al.}(2020)\citenamefont {Sarte},
  \citenamefont {Stock}, \citenamefont {Ortiz}, \citenamefont {Hong},\ and\
  \citenamefont {Wilson}}]{Sarte2020102}%
  \BibitemOpen
  \bibfield  {author} {\bibinfo {author} {\bibfnamefont {P.~M.}\ \bibnamefont
  {Sarte}}, \bibinfo {author} {\bibfnamefont {C.}~\bibnamefont {Stock}},
  \bibinfo {author} {\bibfnamefont {B.~R.}\ \bibnamefont {Ortiz}}, \bibinfo
  {author} {\bibfnamefont {K.~H.}\ \bibnamefont {Hong}}, \ and\ \bibinfo
  {author} {\bibfnamefont {S.~D.}\ \bibnamefont {Wilson}},\ }\href {\doibase
  10.1103/PhysRevB.102.245119} {\bibfield  {journal} {\bibinfo  {journal}
  {Phys. Rev. B}\ }\textbf {\bibinfo {volume} {102}},\ \bibinfo {pages}
  {245119} (\bibinfo {year} {2020})}\BibitemShut {NoStop}%
\end{thebibliography}
\end{document}